\documentclass[10pt,aps,pra,twocolumn,nofootinbib]{revtex4-1}   % style for Physical Review B and AJP are similar

\usepackage[dvips]{epsfig}
\usepackage{amsmath}    % need for subequations
\usepackage{amsfonts}  %note how statements can be commented out
\usepackage{amssymb}
\usepackage{graphicx}   % for figures
\usepackage{mathrsfs}
\usepackage{ulem}
\usepackage{soul}
\usepackage[usenames,dvipsnames]{color}

\usepackage[colorlinks]{hyperref}

\usepackage{colortbl}

\definecolor{gray(x11gray)}{rgb}{0.75, 0.75, 0.75}

\definecolor{darksienna}{rgb}{0.24, 0.08, 0.08}

\definecolor{darkblue}{rgb}{0.0, 0.13, 0.35}

\newcommand{\ket}[1]{| #1 \rangle}

\newcommand{\ignore}[1]{}

\AtBeginDocument{%
    \hypersetup{
      urlcolor     = blue, %Color for external hyperlinks
      filecolor   = red,
      linkcolor    = blue, %Color of internal links
      citecolor   = blue, %Color of citations
    }
} % end  \AtBeginDocument

\begin{document}

\title{Optimal Mach-Zehnder phase sensitivity with Gaussian states}

\author{Stefan~Ataman}
\affiliation{Extreme Light Infrastructure - Nuclear Physics (ELI-NP), `Horia Hulubei' National R\&D Institute for Physics and Nuclear Engineering (IFIN-HH), 30 Reactorului Street, 077125 M\u{a}gurele, jud. Ilfov, Romania}
\email{stefan.ataman@eli-np.ro}

\date{\today}

% ---------------------------------------------------------
% ----------------------- ABSTRACT ------------------------
% ---------------------------------------------------------

\begin{abstract}
We address in this work the phase sensitivity of a Mach-Zehnder interferometer with Gaussian input states. A squeezed-coherent plus squeezed vacuum input state allows us to unambiguously determine the optimal phase-matching conditions in order to maximize the quantum Fisher information. Realistic detection schemes are described and their performance compared in respect with the quantum Cram\'er-Rao bound. The core of this paper discusses in detail the most general Gaussian input state, without any apriori parameter restrictions. Prioritizing the maximization of various terms in the quantum Fisher information has the consequence of imposing the input phase-matching conditions. We discuss in detail when each scenario yields an optimal performance. Realistic detection scenarios are also considered and their performance compared to the theoretical optimum. The impact of the beam splitter types employed on the optimum phase-matching conditions is also discussed. We find a number of potentially interesting advantages of these states over the coherent plus squeezed vacuum input case.
\end{abstract}

\maketitle

% ---------------------------------------------------------
% ------------------- INTRODUCTION ------------------------
% ---------------------------------------------------------
\section{Introduction}
\label{sec:introduction}

Interferometric phase sensitivity is an ongoing research topic benefiting from a high interest from the research community \cite{LIGO13,Tay13,Gar17,Li14,Lan13,Lan14,API18,Tak17,Pre19,Mic18,Dem15}. The theoretic works \cite{Lan13,Lan14,Gar17,Li14,Tak17,API18,Pre19,Mic18} are paralleled by the practical interest from the gravitational-wave detection \cite{LIGO13,Ace14,Oel14,Meh18,Vah18} and quantum technology \cite{Lan14,Gio12} communities.

% ------- GO BELOW SHOT NOISE
The shot-noise limited single coherent input interferometer has been long ago shown to be surpassed by the use of non-classical states of light \cite{Cav81,Xia87,Tay13,Gio04}. The coherent plus squeezed vacuum input state \cite{Pez08,Jar12,Wu19} became a popular choice, also due to its good performance in the low- as well as in the high-power regimes \cite{Gar17,API18,Wu19}. Recently, the squeezing technique has been shown to reduce laser power fluctuations \cite{Vah18}, detect mechanical motion of an oscillator \cite{Bur19} or help the search for axion-like particles \cite{Mal19}. For a recent review on the applications of squeezed states, see reference \cite{Xu19}.

% -------- GW  with SQUEEZING----
After the first round of observations, the gravitational-wave observatories enhanced their sensitivities by employing squeezed states of light \cite{Gro13}. Boosted by these needs, the generation of squeezed light became a mature technology \cite{Vah16,Sch17} delivering ever increasing squeezing factors \cite{Meh18}.

% --- phase sensitivity of MZI

The phase sensitivity of a Mach-Zehnder interferometer (MZI) is generally not constant over a wide range of total internal phase shifts \cite{Yur86,Dem15,API18} and it depends on the detection scheme employed \cite{Gar17,Dem15,Mic18,Wu19}. Although for some states, workarounds to extend this range are known \cite{Pez07}, it is generally preferred to operate the interferometer at or near the optimum working point, $\varphi_\textrm{opt}$ (sometimes also called the ``sweet spot''). For a difference-intensity detection scheme the optimum working point is generally at $\varphi_\textrm{opt}=\pi/2$. This is true for a wide class of input states including the single coherent, coherent plus squeezed vacuum as well as the squeezed-coherent plus squeezed vacuum states \cite{Par95,API18,Lan14,Pre19}. In references \cite{API18} and \cite{Pre19} it has been shown that generally $\varphi_\textrm{opt}\neq\pi/2$ for a double coherent input. In this paper we will show that this is also the case for the most general input Gaussian state, namely the squeezed-coherent plus squeezed coherent input. Other realistic detection schemes yield other optimum internal phase shifts. For example a single-mode intensity and the homodyne detection schemes have the optimum working point $\varphi_\textrm{opt}\approx\pi$ for a large class of input states \citep{Gar17,API18}.

% === FISHER INFORMATION -----------
The quantum Fisher information (QFI) and its associated quantum Cram\'er-Rao bound (QCRB) \cite{Bra94,Dem12,Dem15,Pez15} has been shown to be a powerful tool in setting upper performance bounds in phase estimation. We will employ a two-parameter Fisher information calculation \cite{Jar12,Lan13,Pez15} in order to avoid accounting fictitious resources that are actually unavailable \cite{Jar12,Tak17,Dem15}. 

The QFI approach was applied to single coherent, dual coherent and coherent plus squeezed vacuum input scenarios \cite{Pez08,API18,Lan13,Jar12}, thus providing ways to evaluate the sub-optimality of realistic detection schemes \cite{API18,Gar17}. The most general squeezed-coherent plus squeezed-coherent input was considered in the literature \cite{Spa15,Spa16} with a single-parameter Fisher estimation technique. This approach yielded over-optimistic results by counting resources that are actually not available. The origin of this type of discrepancy was discussed by Jarzyna \emph{et al.} \cite{Jar12}. Moreover, the coherent sources as well as one squeezing operator were assumed to have zero phase \cite{Spa15,Spa16}. In this paper we reconsider this input state, however we insist on not a priori limiting any input parameter.

% --- phase mis-match => PMC

The effect of input phase matching (i.e. the relative phases of the various input sources) has been discussed in the literature \cite{Jar12,Liu13,Pre19}. In \cite{Pre19}, this problem was thoroughly analyzed for an unbalanced interferometer. Generally assumed phase matching conditions set all input phases to zero \cite{Jar12,Liu13,Par95}. As shown in reference \cite{Pre19}, this is not always the optimal choice. In this paper the input phase matching conditions  (PMCs) will be a central point in the discussion. They will prove to be of paramount importance in the characterization of the squeezed-coherent plus squeezed-coherent input scenario. As we will show, the three phase matching conditions that appear will be consequences of the maximization of the Fisher information and not apriori assumptions.

% ----------- LOSSES

Losses adversely affect the phase sensitivity and we can distinguish between internal losses (photon absorption, decoherence etc.) \cite{Dor09,Dem09,Ono10} and the non-ideality of the photo-detectors \cite{Kim99,Spa16}. In this work we only consider the latter and evaluate their impact on the interferometric phase sensitivity performance.

% -------- TYPE of BS used

An often ignored problem is also considered in this paper, namely the impact of the types beam splitters used on the optimum PMCs. Indeed, two main types of beam splitters are used today and this also divides the works in this field: the ones employing symmetrical beam splitters \cite{Liu13,Lan13,Lan14,API18,Pre19} and the ones employing cube beam splitters \cite{Gar17,Par95,Spa15,Spa16}. As we will show, this choice is not without consequences on the optimum PMCs, sometimes giving the impression that different papers give different accounts for the same input state.

This paper is structured as follows. In Section \ref{sec:MZI_setup_and_sensitivities} we formalize our tools used throughout this paper. Among them we introduce two functions, we specify the field operator transformations for our interferometer, and define the Cram\'er-Rao bound as well as the realistic detection schemes considered. The squeezed-coherent plus squeezed vacuum input with all its consequences is considered in Section \ref{sec:sqz_coh_plus_squeezed_vacuum}. The most general case involving Gaussian states \emph{i. e.} the squeezed-coherent plus squeezed coherent input is thoroughly analyzed in Section \ref{sec:sqz_coh_plus_sqz_coh}. The impact of the types of beam splitter used on the optimum input phase-matching conditions is discussed in Section \ref{sec:impact_BS_on_PMC}. Finally, conclusions are drawn in Section \ref{sec:conclusions}.

% ---------------------------------------------------------
% ------------------ MZI SETUP and SENSITIVITIES ----------
% ---------------------------------------------------------
\section{MZI setup: detection sensitivities}
\label{sec:MZI_setup_and_sensitivities}

% ---------------------------------------------------------
% -------------------- PARAMETER ESTIMATION ---------------
% ---------------------------------------------------------
\subsection{Parameter estimation: A short introduction}
\label{subsec:theo_introduction}
We briefly discuss the problem of parameter estimation in quantum mechanics. Longer introductions are available in the literature \cite{API18,Dem15,Pez15,Par09}.

We assume an experimentally accessible Hermitian operator $\hat{A}$ that depends on a parameter $\varphi$. In our case this parameter is the internal phase shift in a Mach-Zehnder interferometer. The fact that $\varphi$ may or may not be an observable makes no difference in our case since we estimate it through the observable $\hat{A}$. The average of this operator is $\langle\hat{A}\left(\varphi\right)\rangle=\langle\psi\vert\hat{A}\left(\varphi\right)\vert\psi\rangle$ where $\vert\psi\rangle$ is the wavefunction of the system. The {\em sensitivity}, $\Delta\varphi$ is defined by \cite{Dem15,Par09,API18,Pre19}
\begin{equation}
\label{eq:Delta_varphi_DEFINITION}
\Delta\varphi=\frac{\Delta\hat{A}}{\big\vert\frac{\partial}{\partial\varphi}\langle\hat{A}\rangle\big\vert}
\end{equation}
where the standard deviation is defined as $\Delta\hat{A}= \sqrt{\Delta^2\hat{A}}$ and the variance is $\Delta^2\hat{A}= \langle\psi\vert\hat{A}^2\vert\psi\rangle- \langle\hat{A}\rangle^2$.

Throughout this paper, the explicit dependence on $\varphi$ of various averages and variances is not necessarily emphasized \emph{i. e.} for simplicity we write $\langle\hat{N}\rangle$ instead of $\langle\hat{N}(\varphi)\rangle$ etc.

% ---------------------------------------------------------
% ------------ FIELD OPERATOR TRANSFORMATIONS -------------
% ---------------------------------------------------------
\subsection{Transformations of the field operators}
\label{subsec:MZI_field_operator_transf}
\begin{figure}
\centering
\includegraphics[scale=0.6]{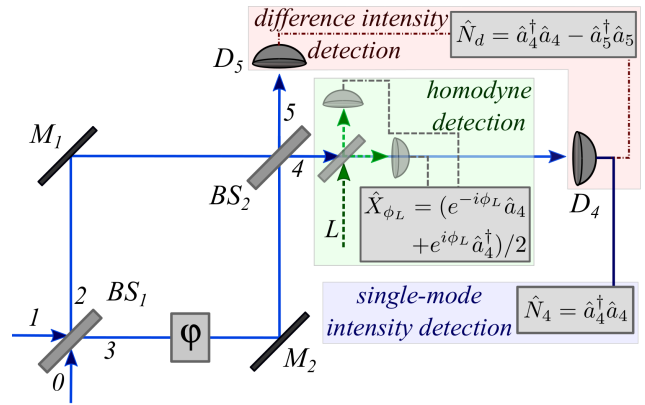}
\caption{The detection schemes and their observables for the MZI we analyze here. The parameter to be estimated via a suitable observable is the phase difference $\varphi$ between the two arms of the MZI.}\label{fig:MZI_2D_single_diff}
\end{figure}

We consider a balanced Mach-Zehnder interferometer (see. Fig.~\ref{fig:MZI_2D_single_diff}). It is composed of two symmetrical beam splitters (BS). We have the well-known field operator transformations e. g. for the first BS we have \cite{GerryKnight,Yur86}
\begin{equation}
\label{eq:field_op_transf_BS_sym}
\left\{
\begin{array}{l}
\hat{a}_0^\dagger=\frac{1}{\sqrt{2}}\hat{a}_2^\dagger+\frac{i}{\sqrt{2}}\hat{a}_3^\dagger\\
\hat{a}_1^\dagger=\frac{i}{\sqrt{2}}\hat{a}_2^\dagger\frac{1}{\sqrt{2}}\hat{a}_3^\dagger
\end{array}
\right.
\end{equation}
where $\hat{a}_k$ ($\hat{a}_k^\dagger$) denotes the annihilation (creation) operator on port $k$ . The two input (output) ports are denoted by $0$ and $1$ ($4$ and $5$). The input-output field operator transformations for the MZI are given by
\begin{equation}
\label{eq:field_op_transf_MZI}
\left\{
\begin{array}{l}
\hat{a}_4^\dagger=-\sin\left(\frac{\varphi}{2}\right)\hat{a}_0^\dagger+\cos\left(\frac{\varphi}{2}\right)\hat{a}_1^\dagger\\
\hat{a}_5^\dagger=\mbox{ }\cos\left(\frac{\varphi}{2}\right)\hat{a}_0^\dagger+\sin\left(\frac{\varphi}{2}\right)\hat{a}_1^\dagger
\end{array}
\right.
\end{equation}
and we ignored global phases. Unless specified otherwise, we assume that the output ports $4$ and $5$ are connected to ideal photo-detectors. No losses inside the MZI are considered throughout this paper.

% ---------------------------------------------------------
% ----------------------------------- FIGURE --- FIGURE ---
% ---------------------------------------------------------

In the following we denote by $\varphi$ the total phase shift inside the interferometer. It is composed of two parts: (i) the experimentally-controllable part $\varphi_\textrm{exp}$ and (ii) the unknown phase shift $\varphi_s$, which is the quantity we want to measure. We have:
\begin{equation}
\label{eq:total_internal_phase_shift}
\varphi= \varphi_s + \varphi_\textrm{exp}
\end{equation}
In all realistic detection scenarios an \emph{optimum total internal phase shift} $\varphi_{\mathrm{opt}}$ (i. e. ``working point'' or ``sweet spot'') exists. It is desirable to have $\varphi$ as close as possible to $\varphi_{\mathrm{opt}}$. If $\vert\varphi_s\vert \ll |\varphi|$, this is generally possible by adjusting the experimentally available parameter $\varphi_\textrm{exp}$.

When speaking of a phase sensitivity at any given total internal phase shift \eqref{eq:total_internal_phase_shift} we will denote it with $\Delta\varphi$ and it it is generally a function of $\varphi$. When speaking about the phase sensitivity \emph{at} the optimum working point (\emph{i. e.} when the total internal phase shift is $\varphi_{\mathrm{opt}}$), we will denote it by $\Delta\tilde{\varphi}$.

% ---------------------------------------------------------
% ------------- OUTPUT OBSERVABLES --- GENERAL ------------
% ---------------------------------------------------------
\subsection{Output observables}
\label{subsec:output_observables}
We consider three realistic detection schemes, each one having an associated observable.

In the difference intensity detection scheme we calculate 
the difference between the output photo-currents (i.e., at the detectors $D_4$ and $D_5$, see Fig.~\ref{fig:MZI_2D_single_diff}). This is a very popular setup, especially at low intensities \cite{Pez07}. Thus, the observable conveying information about the phase $\varphi$ is
\begin{equation}
\label{eq:N_d_operator_DEFINITION}
\hat{N}_d\left(\varphi\right)=\hat{n}_4-\hat{n}_5
\end{equation}
where $\hat{n}_k=\hat{a}^\dagger_k\hat{a}_k$ is the number operator for mode $k$. Using the field operator transformations eqs.~\eqref{eq:field_op_transf_MZI} we have
\begin{eqnarray}
\label{eq:Nd_average}
\langle\hat{N}_d\rangle=\cos\varphi\left(\langle{\hat{a}_1^\dagger\hat{a}_1}\rangle-\langle{\hat{a}_0^\dagger\hat{a}_0}\rangle\right)
%\nonumber\\
-\sin\varphi\left(\langle{\hat{a}_0}{\hat{a}_1^\dagger}\rangle
+\langle{\hat{a}_0^\dagger}{\hat{a}_1}\rangle\right)
%-----------
\nonumber\\
% ------------ written with Re
=\cos\varphi\left(\langle{\hat{n}_1}\rangle
-\langle{\hat{n}_0}\rangle\right)
-2\sin\varphi\Re\langle{\hat{a}_0}{\hat{a}_1^\dagger}\rangle
\qquad\qquad\qquad\,\,
\end{eqnarray}
where the expectation values are calculated w.r.t.~the input state $\ket{\psi_{in}}$ and $\Re$ denotes the real part. To estimate the phase sensitivity in equation~\eqref{eq:Delta_varphi_DEFINITION} we need the absolute value of the derivative
\begin{eqnarray}
\label{eq:del_Nd_average_over_del_phi}
\bigg\vert\frac{\partial\langle\hat{N}_d\rangle}{\partial\varphi}\bigg\vert
=\vert\sin\varphi(\langle\hat{n}_0\rangle-\langle\hat{n}_1\rangle)
%\nonumber\\
-2\cos\varphi\Re\langle\hat{a}_0\hat{a}_1^\dagger\rangle\vert.
\end{eqnarray}
The calculation of the standard deviation ${\Delta\hat{N}_d}$ needed in equation \eqref{eq:Delta_varphi_DEFINITION} is detailed in Appendix  \ref{subsec:app:variance_calculation_diff_det}.

% ---------------------------------------------------------
% -------------------- SINGLE DETECTOR --------------------
% ---------------------------------------------------------
In the single-mode intensity detection scheme we have only one detector coupled at the output port $4$, see Fig.~\ref{fig:MZI_2D_single_diff}. Thus the operator of interest is $\hat{N}_4=\hat{a}_4^\dagger\hat{a}_4$. From equation~\eqref{eq:field_op_transf_MZI} we have
\begin{eqnarray}
\label{eq:N_4_average_GENERAL}
\langle\hat{N}_4\rangle=\sin^2\left(\frac{\varphi}{2}\right)\langle\hat{n}_0\rangle+\cos^2\left(\frac{\varphi}{2}\right)\langle\hat{n}_1\rangle
%\nonumber\\
-\sin\varphi\Re\langle\hat{a}_0\hat{a}_1^\dagger\rangle
\quad
\end{eqnarray}
and the absolute value of its derivative w.r.t. $\varphi$ is
\begin{eqnarray}
\label{eq:N4_derivative_average}
\bigg\vert\frac{\partial\langle\hat{N}_4\rangle}{\partial\varphi}\bigg\vert
=\frac{1}{2}\vert\sin\varphi\left(\langle\hat{n}_0\rangle-\langle\hat{n}_1\rangle\right)
%\nonumber\\
-2\cos\varphi\Re\langle\hat{a}_0\hat{a}_1^\dagger\rangle
\vert
\end{eqnarray}
Similar to the difference-intensity detection scheme, the calculation of the standard deviation $\Delta\hat{N}_4$ is detailed Appendix \ref{subsec:app:single_intensity_detection}.

If we assume a (balanced) homodyne detection at the output port $4$, the operator modeling this  detection scheme is given by $\hat{X}_{\phi_L}=(e^{-i\phi_L}\hat{a}_4+e^{i\phi_L}\hat{a}_4^\dagger)/2$ and, expressed in respect with the input field operators,
\begin{eqnarray}
\label{eq:Obervable_Homodyne}
\hat{X}_{\phi_L}
%=\frac{e^{-i\phi_L}\hat{a}_4+e^{i\phi_L}\hat{a}_4^\dagger}{2}
=-\sin\left(\frac{\varphi}{2}\right)\frac{e^{-i\phi_L}{\hat{a}_0}+e^{i\phi_L}{\hat{a}_0^\dagger}}{2}
\nonumber\\
+\cos\left(\frac{\varphi}{2}\right)\frac{e^{-i\phi_L}{\hat{a}_1}+e^{i\phi_L}{\hat{a}_1^\dagger}}{2}
\end{eqnarray}
where $\phi_L$ is the phase of the local coherent source (assumed fixed and adjustable in respect with $\theta_\alpha$).

% ---------------------------------------------------------
% --------- ESTIMATION via FISHER INFORMATION -------------
% ---------------------------------------------------------
\subsection{Fisher information and the Cram\'er-Rao bound}
\label{subsec:Fisher_and_QCRB}
The Fisher information is a powerful approach to find the best-case solution of parameter estimation \cite{Bra94,Dem15,Par09,Lan13,Pez15}. The lower bound for the estimation of a parameter $\varphi$ is given by the Cram\'er-Rao bound (CRB) \cite{Dem15,Par09}. The Fisher information is maximised by the QFI $\mathcal{F}\left(\varphi\right)$ \cite{Bra94,Dem15} and this leads to the QCRB,
\begin{equation}
\label{eq:Quantum_Cramer_Rao_Bound}
\Delta\varphi_{QCRB}=\frac{1}{\sqrt{\mathcal{F}}}
\end{equation}
Since we will be interested in the difference phase shift sensitivity only (see details in Appendix \ref{sec:app:Fisher_information}), we define the QFI as
\begin{equation}
\label{eq:Fisher_definition}
\mathcal{F}=\mathcal{F}_{dd}-\frac{(\mathcal{F}_{sd})^2}{\mathcal{F}_{ss}}.
\end{equation}
Similar to reference \cite{Pre19}, we will not consider $\mathcal{F}\approx\mathcal{F}_{dd}$, as done by many authors \cite{Lan13,Lan14}. Although some input states justify this approximation (for example the coherent plus squeezed vacuum input), in our case the Fisher matrix coefficient $\mathcal{F}_{sd}$ from equation \eqref{eq:app:F_sd_GENERIC_FINAL_balanced} will play an important role in the discussion from Section \ref{sec:sqz_coh_plus_sqz_coh}.

% ------------------------ FIGURE -------------------
\begin{figure}
\includegraphics[scale=0.47]{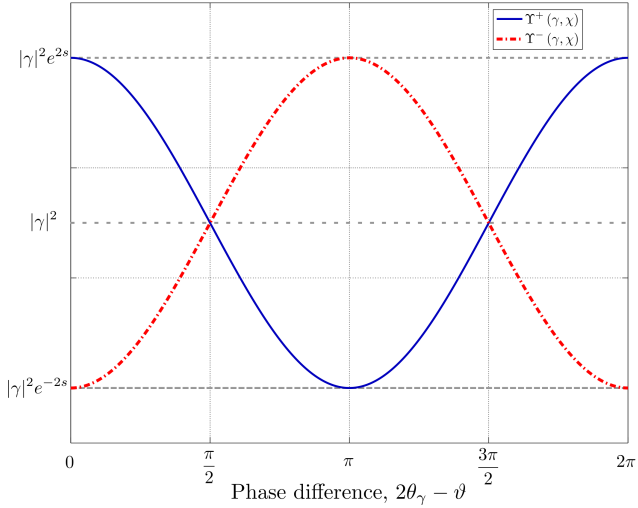}
\caption{The functions $\Upsilon^+\left(\gamma,\chi\right)$ and $\Upsilon^-\left(\gamma,\chi\right)$. For the input phase-matching condition ${2\theta_\gamma-\vartheta=0}$ we have ${\Upsilon^+\left(\gamma,\chi\right)=\vert\gamma\vert^2e^{2s}}$ and $\Upsilon^-\left(\gamma,\chi\right)=\vert\gamma\vert^2e^{-2s}$.}
\label{fig:upsilon_functions}
\end{figure}

For $N$ repeated experiments we have a scaling $\Delta\varphi_{QCRB}=1/\sqrt{N\mathcal{F}}$ \cite{Bra94,Dem15} and the same $1/\sqrt{N}$ applies to $\Delta\varphi$ from equation \eqref{eq:Delta_varphi_DEFINITION}. For simplicity, we consider $N=1$ throughout our discussion.

% ---------------------------------------------------------
% --------- THE UPSILON FUNCTIONS -------------------------
% ---------------------------------------------------------
\subsection{Two useful functions} 
\label{subsec:Upsilon_functions}
We now introduce two functions that will repeatedly appear in our calculations, allowing a more compact writing of the obtained results. We define the function
\begin{equation}
\label{eq:function:Upsilon_Plus_Definition}
\Upsilon^+\left(\gamma,\chi\right)={\vert\gamma\vert^2}\left(\cosh2s
+\sinh2s\cos(2\theta_\gamma-\vartheta)\right)
\end{equation}
where both arguments are complex and we have ${\gamma=\vert\gamma\vert e^{i\theta_\gamma}}$ and ${\chi=se^{i\vartheta}}$ with $s\in\mathbb{R}^+$, $\theta_\gamma,\vartheta\in[0,2\pi]$. Similarly, we introduce the function
\begin{equation}
\label{eq:function:Upsilon_Minus_Definition}
\Upsilon^-\left(\gamma,\chi\right)=\vert\gamma\vert^2\left(\cosh{{2s}}
-\sinh{{2s}}\cos(2{\theta_\gamma}-{\vartheta})\right)
\end{equation}
We plot these functions in Fig.~\ref{fig:upsilon_functions}. 
In our context, these functions will model the squeezing-induced ($\chi$) enhancement/reduction of the coherent source's ($\gamma$) fluctuations. %These functions represent the fluctuations sensitive to the input phase mis-match. 
Indeed, one can see that  if we impose the phase-matching condition (PMC) ${2{\theta_\gamma}-{\vartheta}=0}$ (${2{\theta_\gamma}-{\vartheta}=\pi}$), we have ${\Upsilon^+\left(\gamma,\chi\right)={\vert\gamma\vert^2}e^{2s}}$ (${\Upsilon^+\left({\gamma},\chi\right)=\vert\gamma\vert^2e^{-2s}}$) and ${\Upsilon^-\left(\gamma,\chi\right)=\vert\gamma\vert^2e^{-2s}}$ (${\Upsilon^-\left(\gamma,\chi\right)={\vert\gamma\vert^2}e^{2s}}$) \emph{i. e.} the fluctuations are enhanced/reduced.

We can also connect these functions to the quadrature measurement at a given angle $\hat{X}_{\theta_{q}}$ (\cite{GerryKnight} equation (7.7), Section $7.1$). Thus, the function $\Upsilon^+\left(\gamma,\chi\right)$ is proportional to the fluctuations of quadrature measurement on the $\theta_{L}=2{\theta_\gamma}-{\vartheta}$ axis while $\Upsilon^-\left({\gamma},{\chi}\right)$ is proportional to the measurement on the $\theta_{L}=2{\theta_\gamma}-{\vartheta}+\pi/2$ axis. More precisely, we have
\begin{equation}
\left\{
\begin{array}{l}
\Upsilon^+\left(\gamma,\chi\right)=4{\vert\gamma\vert^2}\Delta^2\hat{X}_{2{\theta_\gamma}-{\vartheta}}\\
\Upsilon^-\left(\gamma,\chi\right)=4{\vert\gamma\vert^2}\Delta^2\hat{X}_{2{\theta_\gamma}-{\vartheta}+\frac{\pi}{2}}
\end{array}
\right.
\end{equation}
For $\theta_{L}=0$ ($\theta_{L}=\pi/2$) we have a measurement on the $X_1$ ($X_2$) quadrature (sometimes called $X$ and $Y$, e. g. \cite{Loudon}, equations (5.6.7)-(5.6.8), Section $5.6$).

% ---------------------------------------------------------
% ----------- SQUEEZED-COHERENT plus SQUEEZED VACUUM ------
% ---------------------------------------------------------
\section{Squeezed-coherent plus squeezed vacuum input}
\label{sec:sqz_coh_plus_squeezed_vacuum}
In the first scenario we consider a squeezed-coherent plus squeezed vacuum \cite{Pre19} input state,
\begin{equation}
\label{eq:psi_in_squeezed_coherent_plus_squeezed_vac}
\vert\psi_{in}\rangle=\vert(\alpha\zeta)_1\xi_0\rangle=\hat{D}_1\left(\alpha\right)\hat{S}_1\left(\zeta\right)\hat{S}_0\left(\xi\right)\vert0\rangle
\end{equation}
applied to input of the interferometer. A squeezed vacuum state is obtained by applying the squeezing operator \cite{GerryKnight,Yue76}
\begin{equation}
\label{eq:Squeezing_operator}
\hat{S}_m\left(\chi\right)=e^{[\chi^*\hat{a}_m^2-\chi(\hat{a}_m^\dagger)^2]/2}
\end{equation}
with $\chi=se^{i\vartheta}$ to a port $m$ previously found in the vacuum state, $\vert0\rangle$. We call $s\in\mathbb{R}^{+}$ the squeezing factor and ${\vartheta}$ is the phase of the squeezed state. For the input state \eqref{eq:psi_in_squeezed_coherent_plus_squeezed_vac} we use a squeezed state with $\xi=re^{i\theta}$ ($\zeta=ze^{i\phi}$) applied to the input port $0$  ($1$). The displacement operator \cite{GerryKnight,MandelWolf,Aga12} for a port $k$ is defined by 
\begin{equation}
\label{eq:displacement_operator_def}
\hat{D}_k\left(\alpha\right)=e^{\alpha\hat{a}_k^\dagger-\alpha^*\hat{a}_k}.
\end{equation}
Please note that in the input state from equation \eqref{eq:psi_in_squeezed_coherent_plus_squeezed_vac} we first \emph{squeeze} and then \emph{displace} the vacuum from input port $1$.

We also note that our state is separable (non-entangled), i. e. we can write equation \eqref{eq:psi_in_squeezed_coherent_plus_squeezed_vac} as $\vert\psi_{in}\rangle=\vert(\alpha\zeta)_1\rangle\otimes\vert\xi_0\rangle$. The same remark will apply to the input state discussed in Section \ref{sec:sqz_coh_plus_sqz_coh}. Firstly, this state of facts corresponds to the experimental reality. Second, by forbidding entanglement at the input of our interferometer, we avoid pathologies connected to the Fisher information (see e. g. \cite{Lan14} an references within).

% ----------------- SINGLE COHERENT INPUT -----------------
% ----------------- DIFFERENTIAL DETECTION ----------------
% ---------------------------------------------------------
\subsection{Fisher information and the Cram\'er-Rao bound}
\label{subsec:sqz_coherent_sqz_vac_input_Fisher}
Our input state  \eqref{eq:psi_in_squeezed_coherent_plus_squeezed_vac} applied to equation \eqref{eq:app:F_dd_FINAL_FORM_GENERAL_balanced} yields the difference-difference Fisher matrix element \cite{Pre19}
\begin{eqnarray}
\label{eq:app:F_dd_coh_sqz_vac_sqz_vac_Upsilon}
\mathcal{F}_{dd}
=\Upsilon^+\left(\alpha,\xi\right)
+\frac{\cosh2r\cosh2z}{2}
\nonumber\\
-\frac{\sinh2r\sinh2z\cos(\theta-\phi)+1}{2}.
%-\frac{1}{2}
\end{eqnarray}
Since $\mathcal{F}_{sd}=0$ for this input state, we have $\mathcal{F}=\mathcal{F}_{dd}$. The function $\Upsilon^+\left(\alpha,\xi\right)$ reaches its maximum value of $\vert\alpha\vert^2e^{2r}$ if we impose the input PMC
\begin{equation}
\label{eq:phase_matching_cond_coh_plus_sqz_vac}
2\theta_\alpha-\theta=0
\end{equation}
and this is the same constraint already reported and discussed in the literature for the coherent plus squeezed vacuum input \cite{Pez08,Liu13,API18}. In order to maximize the last term from equation \eqref{eq:app:F_dd_coh_sqz_vac_sqz_vac_Upsilon} we have to impose the supplementary input phase-matching condition 
\begin{equation}
\label{eq:phase_matching_cond_sqz_coh_plus_sqz_vac}
\theta-\phi=\pm\pi, 
\end{equation}
yielding the optimum QFI
\begin{equation}
\label{eq:Fisher_sqz-coh_plus_sqz_vac_MAXIMAL}
\mathcal{F}=\vert\alpha\vert^2e^{2r}
%+\frac{\cosh(2\left(r+z\right))-1}{2}
+\sinh^2\left(r+z\right).
\end{equation}
and thus the QCRB for the input state \eqref{eq:psi_in_squeezed_coherent_plus_squeezed_vac},
\begin{equation}
\label{eq:delta_varphi_QCRB_sqz_coh_sqz_vac}
\Delta\varphi_{QCRB}=\frac{1}{\sqrt{\vert\alpha\vert^2e^{2r}+\sinh^2\left(r+z\right)}}.
\end{equation}

% ---------------------------------------------------------
% ---------------- DIFFERENTIAL DETECTION -----------------
% ---------------------------------------------------------
\subsection{Difference intensity detection scheme}
\label{subsec:SqueezedCoherent_SqueezedVac_differential_det}
The input state \eqref{eq:psi_in_squeezed_coherent_plus_squeezed_vac} applied to equation \eqref{eq:Nd_average} gives
\begin{equation}
\label{eq:Nd_average_sqz-coherent_squeezed-vac}
\langle\hat{N}_d\rangle=\cos\varphi\left(\vert\alpha\vert^2+\sinh^2z
-\sinh^2r\right)
\end{equation}
For the variance (see details in Appendix \ref{subsec:app:coh_sqz_plus_sqz_vac_DIFF_DET}) we obtain
\begin{eqnarray}
\label{eq:VARIANCE_N_d_sqz-coh_sqz_vac_FINAL}
\Delta^2\hat{N}_d
=\cos^2\varphi\left(
{\frac{\sinh^22r}{2}}
+\frac{\sinh^22z}{2}
+\Upsilon^-\left(\alpha,\zeta\right)\right)
\nonumber\\
+\sin^2\varphi\bigg(
\Upsilon^-\left(\alpha,\xi\right)
%\right.
\nonumber\\
%\left.
+\frac{\cosh2r\cosh2z+\sinh2r\sinh2z
\cos\left({\phi}-{\theta}\right)-1}{2}
\bigg)
\quad
\end{eqnarray}
and the phase sensitivity is given by equation \eqref{eq:Delta_pdi_szq-coh_sqz_vac_GENERAL}. The best sensitivity is achieved for the optimum total internal phase shift $\varphi_\textrm{opt}=\pi/2$. Introducing this result in equation \eqref{eq:Delta_pdi_szq-coh_sqz_vac_GENERAL} takes us to the best achievable phase sensitivity
\begin{eqnarray}
\label{eq:app:Delta_phi_N_d_sqz-coh_sqz_vac_pi_over_2}
\Delta\tilde{\varphi}_\mathrm{df}
=\frac{\sqrt{\Upsilon^-\left({\alpha},{\xi}\right)
+\frac{\cosh{2r}\cosh{2z}+\sinh{2r}\sinh{2z}
\cos\left({\phi}-{\theta}\right)-1}{2}}}
{\vert\vert\alpha\vert^2+\sinh^2z-\sinh^2r\vert}
\quad\:\:
\end{eqnarray}
The phase sensitivity is further improved by imposing the phase matching conditions \eqref{eq:phase_matching_cond_coh_plus_sqz_vac} and \eqref{eq:phase_matching_cond_sqz_coh_plus_sqz_vac} yielding
\begin{eqnarray}
\label{eq:Delta_varphi_Sqz-coh_Sqz_Vac_DIFF_Best}
\Delta\tilde{\varphi}_\mathrm{df}
=\frac{\sqrt{\vert\alpha\vert^2e^{-2r}+\sinh^2(r-z)}}
{\vert\vert\alpha\vert^2+\sinh^2z-\sinh^2r\vert}
\end{eqnarray}

% ---------------------------------------------------------
% ----------------- SINGLE DETECTOR -----------------------
% ---------------------------------------------------------
\subsection{Single-mode intensity detection scheme}
\label{subsec:Sqz_coherent_plus_squeezedVac_single_det}
For the input state \eqref{eq:psi_in_squeezed_coherent_plus_squeezed_vac} the average number of photons for a single-intensity detection scheme is
\begin{equation}
\label{eq:N4_average_coh-squeezed_sqz-vacuum}
\langle\hat{N}_4\rangle=\sin^2\left(\frac{\varphi}{2}\right)\sinh^2r+\cos^2\left(\frac{\varphi}{2}\right)\left({\vert\alpha\vert^2+\sinh^2z}\right)
\end{equation}
The variance is found to be (see details in Appendix \ref{sec:app:sqz_coh_plus_sqz_vac})
\begin{eqnarray}
\label{eq:Variance_N_4_coh_sqz_sqz_vac}
% --------------------------------------
\Delta^2\hat{N}_4
=\cos^4\left(\frac{\varphi}{2}\right)\left(\frac{\sinh^22z}{2}+\Upsilon^-\left({\alpha},{\zeta}\right)\right)
% -----%-----
\nonumber\\
+\sin^4\left(\frac{\varphi}{2}\right)\frac{\sinh^22r}{2}
+\frac{\sin^2\varphi}{4}\bigg(
\Upsilon^-\left(\alpha,\xi\right)
\qquad
%\right.
% -----%-----
\nonumber\\
%\left.
+\frac{\cosh2r\cosh2z
+\sinh2r\sinh2z\cos\left(\theta-\phi\right)-1}
{2}
\bigg)
\:
\end{eqnarray}
and the phase sensitivity $\Delta\varphi_\mathrm{sg}$ in the single-intensity detection scenario is given by equation \eqref{eq:Delta_phi_sqz_coh__sqz_vac_SINGLE_DET_Upsilon}. We can find an optimum internal phase shift,
\begin{equation}
\label{eq:varphi_opt_sqz-coh_plus_sqz_vac_sing_det}
\varphi_\textrm{opt}
%=2\arctan\sqrt[4]{\frac{A}{B}}
=\pm2\arctan\sqrt[4]{\frac{{\sinh^22z}+2\Upsilon^-\left({\alpha},{\zeta}\right)}{{\sinh^22r}}}
\end{equation}
and introducing this result into equation \eqref{eq:Delta_phi_sqz_coh__sqz_vac_SINGLE_DET_Upsilon} yields the best achievable phase sensitivity $\Delta\tilde{\varphi}_\mathrm{sg}$ from equation \eqref{eq:Delta_varphi_Sqz-coh_Sqz_Vac_SING_Best_GENERAL}. This sensitivity can be further improved by imposing the input phase matching conditions \eqref{eq:phase_matching_cond_coh_plus_sqz_vac} and \eqref{eq:phase_matching_cond_sqz_coh_plus_sqz_vac} yielding the result from equation \eqref{eq:app:Delta_varphi_Sqz-coh_Sqz_Vac_SING_Best}.

% ---------------------------------------------------------
% ----------------- HOMODYNE DETECTION --------------------
% ---------------------------------------------------------
\subsection{Homodyne detection scheme}
\label{subsec:Sqz_coherent_plus_squeezedVac_homodyne_det}
Using the results from Appendix \ref{subsec:app:homodyne_detection} we find
\begin{equation}
\vert\partial_\varphi\langle\hat{X}_{\phi_L}\rangle\vert
=\frac{1}{2}\Big\vert\sin\left(\frac{\varphi}{2}\right)\vert\alpha\vert\cos(\phi_L-\theta_\alpha)\Big\vert
\end{equation}
We impose $\cos(\phi_L-\theta_\alpha)=1$ thus $\phi_L=\theta_\alpha$ i. e. the local oscillator should be in phase with the coherent source. We find the variance
\begin{eqnarray}
\label{eq:Homodyne_variance_sqz_coh_plus_sqz_vac}
\Delta^2\hat{X}_{\phi_L}
=\frac{\cos^2\left(\frac{\varphi}{2}\right)\Upsilon^-\left(\alpha,\zeta\right)+\sin^2\left(\frac{\varphi}{2}\right)\Upsilon^-\left(\alpha,\xi\right)}{4\vert\alpha\vert^2}
\quad
\end{eqnarray}
thus yielding a phase sensitivity
\begin{equation}
\label{eq:Delta_varphi_scq_coh_plus_sqz_vac_Homodyne}
\Delta\varphi_\mathrm{hom}
=\frac{\sqrt{\Upsilon^-(\alpha,\xi)
+\cot^2\left(\frac{\varphi}{2}\right)\Upsilon^-(\alpha,\zeta)}}
{\vert\alpha\vert^2}
\end{equation}
At the optimum angle $\varphi_\textrm{opt}=\pi$ the sensitivity becomes 
\begin{equation}
\label{eq:Delta_varphi_scq_coh_plus_sqz_vac_Homodyne_BEST}
\Delta\tilde{\varphi}_\mathrm{hom}=\frac{\sqrt{\Upsilon^-(\alpha,\xi)}}{\vert\alpha\vert^2}
\end{equation}
and further imposing the PMC \eqref{eq:phase_matching_cond_coh_plus_sqz_vac} yields  $\Delta\tilde{\varphi}_\mathrm{hom}=e^{-r}/\vert\alpha\vert$, a result identical to the one with a coherent plus squeezed vacuum input \cite{Gar17}. 

% ---------------------------------------------------------
% --------------- DISCUSSION ------------------------------
% ---------------------------------------------------------
\subsection{Discussion}
\label{subsec:Sqz_coherent_plus_squeezedVac_DISCUSSION}

% ---------------- HEISENBERG SCALING -----------------------
\subsubsection{Analysis of the obtained results}

Using the PMCs given by Eqs.~\eqref{eq:phase_matching_cond_coh_plus_sqz_vac} and \eqref{eq:phase_matching_cond_sqz_coh_plus_sqz_vac}, if we compare the best achievable sensitivities at the optimal working point, we actually have
\begin{equation}
\Delta\tilde{\varphi}_\textrm{sg}\geq\Delta\tilde{\varphi}_\textrm{df}\geq\Delta\varphi_{QCRB}\textrm{ and }
\Delta\tilde{\varphi}_\textrm{hom}\geq\Delta\varphi_{QCRB}
\end{equation}
showing that all discussed schemes are slightly suboptimal (optimality for $r=z=0$). We note that in the case with equal squeezing in both inputs ($r=z$) we find the interesting result
\begin{equation}
%\label{eq:Delta_varphi_scq_coh_plus_sqz_vac_Homodyne}
\Delta\tilde{\varphi}_\mathrm{hom}=
\Delta\tilde{\varphi}_\mathrm{df}=\frac{e^{-r}}{\vert\alpha\vert}
\end{equation}
In the high-$\vert\alpha\vert$ regime (i. e. when $\vert\alpha\vert^2\gg\{\sinh^2r,\:\sinh^2z\}$) we have $\Delta\tilde{\varphi}_\textrm{sg}\approx\Delta\tilde{\varphi}_\textrm{df}\approx\Delta\tilde{\varphi}_\textrm{hom}\approx\Delta\varphi_{QCRB}\approx e^{-r}/\vert\alpha\vert$.

% ----------------------- FIGURE --------------- LOW-ALPHA ----
\begin{figure}
\centering
\includegraphics[scale=0.46]{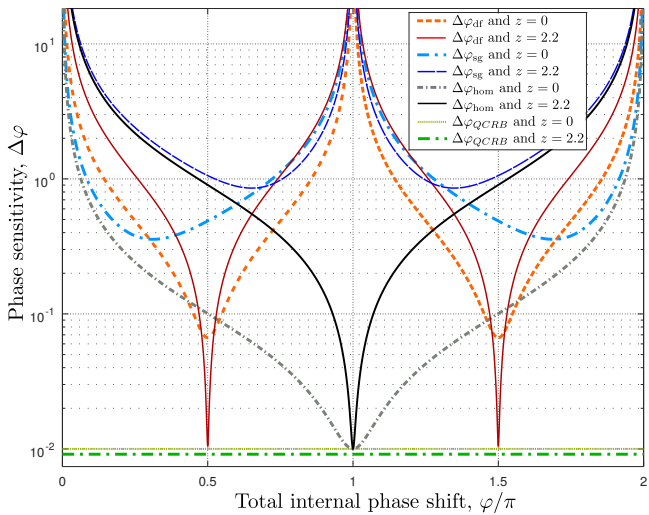}
\caption{The phase sensitivity $\Delta\varphi$ versus the phase shift in the low-$\vert\alpha\vert$ regime. Adding squeezing does not noticeably enhance the quantum Cram\'er-Rao bound, however it substantially enhances the performance of a difference-intensity detection scheme. Parameters used: $\vert\alpha\vert=10$, $r=2.3$, $\theta=0$, $\theta_\alpha=0$, $\phi=\pi$.}
\label{fig:coh_sqz_plus_sqz_vac_versus_varphi}
\end{figure}

In Fig.~\ref{fig:coh_sqz_plus_sqz_vac_versus_varphi} we plot the realistic detection schemes against the QCRB in the low-$\vert\alpha\vert$ limit for two scenarios: $z=0$ (\emph{i. e.} no squeezing in port $1$) and $z=2.2$. One notes the swift enhancement in sensitivity  in the case of a difference intensity detection scheme (solid red and dashed orange curves), due to the supplementary squeezing. For the single-mode intensity detection scheme (dashed blue and dash-dotted cyan curves), on the contrary: the supplementary squeezing simply degrades the performance in respect with no squeezing in port $1$.

The best sensitivity reached by a homodyne detection, with or without squeezing in port $1$ (dash-dotted gray and solid black curves) reaches the same value: $\Delta\tilde{\varphi}_\mathrm{hom}=e^{-r}/\vert\alpha\vert$. In this case, too, the second squeezing ($\zeta$) brings no benefit, quite on the contrary: without squeezing in port $1$ the sensitivity degrades slower when $\varphi\neq\pi/2$. From equation \eqref{eq:Delta_varphi_scq_coh_plus_sqz_vac_Homodyne} we immediately find the culprit: the $\Upsilon^-(\alpha,\zeta)$ term. Indeed, employing the PMC from equation \eqref{eq:phase_matching_cond_sqz_coh_plus_sqz_vac} maximized this term to $\vert\alpha\vert^2e^{2z}$. We return shortly to this problem.

From Fig.~\ref{fig:coh_sqz_plus_sqz_vac_versus_varphi} it is apparent that the gain we found for the difference intensity detection scenario is rather fragile. Indeed, if the internal phase shift drifts from $\varphi_\textrm{opt}=(2k+1)\pi/2$ (with $k\in\mathbb{Z}$), the performance quickly degrades. Tracing back this issue in $\Delta^2\hat{N}_d$ from equation \eqref{eq:VARIANCE_N_d_sqz-coh_sqz_vac_FINAL}, one notes that for the phase matching conditions given by equations \eqref{eq:phase_matching_cond_coh_plus_sqz_vac} and \eqref{eq:phase_matching_cond_sqz_coh_plus_sqz_vac} we have $\Upsilon^-\left(\alpha,\xi\right)$ minimized to $\vert\alpha\vert^2e^{-2r}$ however $\Upsilon^-\left(\alpha,\zeta\right)$ is maximized again to $\vert\alpha\vert^2e^{2z}$. Thus, as soon as $\cos\varphi\neq0$, the contribution of $\Upsilon^-\left(\alpha,\zeta\right)$ is far from negligible, hence the swift degradation in phase sensitivity. %The same problem plagues the phase sensitivity $\Delta\varphi_\mathrm{hom}$ from equation \eqref{eq:Delta_varphi_scq_coh_plus_sqz_vac_Homodyne} if $\varphi\neq k\pi$ (with $k\in\mathbb{Z}$). 
We might conclude at this point that, with the exception of the difference-intensity detection scheme, adding a second squeezing actually worsens the overall performance. 

Nonetheless, if we relax our restrictions on the phase matching conditions leading to the optimal QFI from equation \eqref{eq:Fisher_sqz-coh_plus_sqz_vac_MAXIMAL}, many useful advantages will arise from adding a second squeezing. We thus alter now the second phase matching condition given by equation \eqref{eq:phase_matching_cond_sqz_coh_plus_sqz_vac} to
\begin{equation}
\label{eq:phase_matching_cond_sqz_coh_plus_sqz_vac_BIS}
\theta-\phi=0
\end{equation}
This new PMC, when replacing the optimal $\theta-\phi=\pm\pi$ constraint, decreases the QFI from the value given in equation \eqref{eq:Fisher_sqz-coh_plus_sqz_vac_MAXIMAL} to 
\begin{equation}
\label{eq:Fisher_sqz_coh_plus_sqz_vac_PMC_phi_is_zero}
{\mathcal{F}=\vert\alpha\vert^2e^{2r}+\sinh^2\left(r-z\right)}. 
\end{equation}
But if we are in experimentally interesting high-$\vert\alpha\vert$ regime, the effect is small. 
The question is now if this sub-optimal PMC has any practical advantage.

\begin{figure}
\centering
\includegraphics[scale=0.49]{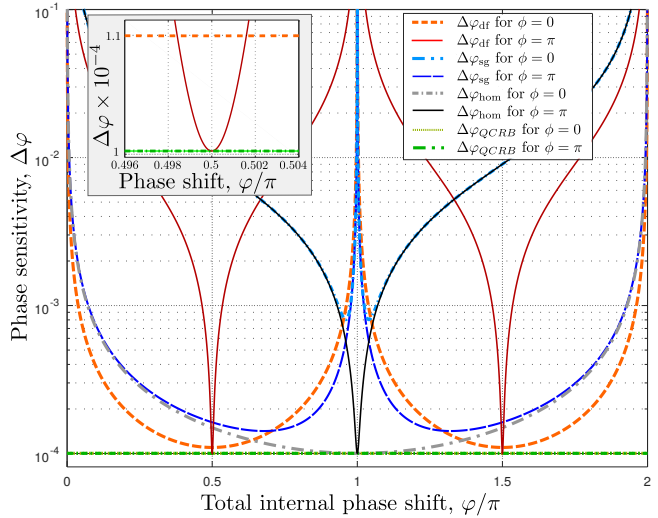}
\caption{The phase sensitivity $\Delta\varphi$ versus the phase shift in the high-$\vert\alpha\vert$ regime ($\vert\alpha\vert^2\gg\{\sinh^2{\xi},\:\sinh^2\zeta\}$). The phase difference between the squeezers $\theta-\phi$ has a negligible impact on the QCRB but an important impact on the performance of realistic detection schemes. Parameters used: $\vert\alpha\vert=10^3$, $\theta_\alpha=0$, $r=2.3$, $z=2.2$, $\theta=0$. Inset: zoom around $\varphi=\pi/2$ showing that the phase sensitivity in the difference-intensity detection scheme is indeed maximized for $\phi=\pi$.}
\label{fig:Delta_phi_coh_sqz_sqz_vac_vs_theta_alpha1k_phi_0_pi}
\end{figure}

In order to answer this question, the $\theta-\phi=0/\pi$ scenarios are depicted in Fig.~\ref{fig:Delta_phi_coh_sqz_sqz_vac_vs_theta_alpha1k_phi_0_pi}. One immediately notes that with the PMC from equation \eqref{eq:phase_matching_cond_sqz_coh_plus_sqz_vac_BIS}, $\Delta\varphi_\mathrm{hom}$, $\Delta\varphi_\mathrm{df}$ and $\Delta\varphi_\mathrm{sg}$ are much more insensitive to the variation of $\varphi$. This is an experimental advantage, since a wider range of total internal phase shifts can be measured more accurately. Moreover, the single-intensity detection scheme shows much better results, this time the squeezing from port $1$ showing its positive effect. This improvement can be traced back to the term $\Upsilon^-\left(\alpha,\zeta\right)$ that has been minimized this time to $\vert\alpha\vert^2e^{-2z}$.

% ------------- FIGURE --------- versus VARPHI --- HIGH ALPHA ---

\begin{figure}
\centering
\includegraphics[scale=0.45]{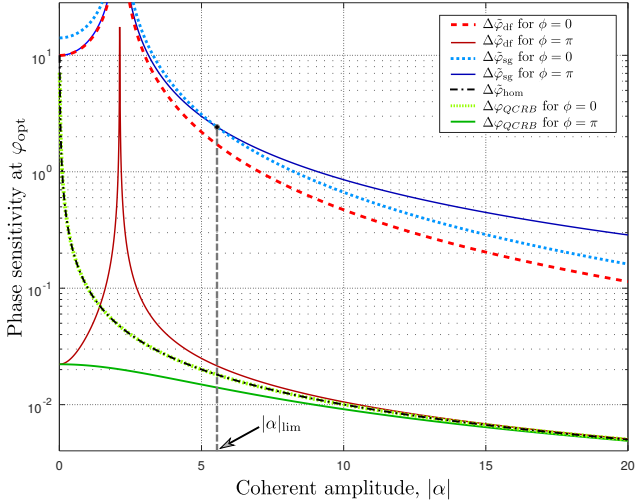}
\caption{The phase sensitivities $\Delta\tilde{\varphi}_\mathrm{df}$, $\Delta\tilde{\varphi}_\mathrm{sg}$, $\Delta\tilde{\varphi}_\mathrm{hom}$ and $\Delta\varphi_{QCRB}$ versus the coherent amplitude $\vert\alpha\vert$. Parameters used: $\theta_\alpha=0$, $r=2.3$, $z=2.2$, $\theta=0$.}
\label{fig:Delta_phi_coh_squeezed_plus_squeezed_vac_versus_alpha_BIS}
\end{figure}

In Fig.~\ref{fig:Delta_phi_coh_squeezed_plus_squeezed_vac_versus_alpha_BIS} we discuss the same $\theta-\phi=0/\pi$ problem at the optimal phase shift ($\varphi_\textrm{opt}$) for all considered detection schemes versus the QCRB. One notices that at low $\vert\alpha\vert$ the phase matching condition $\theta-\phi=0$ brings a significant penalty on both $\Delta\tilde{\varphi}_\mathrm{df}$ (red curves) and $\Delta\varphi_{QCRB}$ (green curves). The phase matching conditions \eqref{eq:phase_matching_cond_coh_plus_sqz_vac} and \eqref{eq:phase_matching_cond_sqz_coh_plus_sqz_vac} yield the best optimal phase sensitivity $\Delta\tilde{\varphi}_\mathrm{df}$ throughout the whole range of $\vert\alpha\vert$. This is not true anymore for a single-mode intensity detection scheme (blue curves). Indeed, the aforementioned PMCs yield the best sensitivity only for $\vert\alpha\vert<\vert\alpha\vert_\mathrm{lim}$, where we define
\begin{equation}
\label{eq:alpha_lim_sqz_coh_plus_sqz_vac}
\vert\alpha\vert_\mathrm{lim}=\frac{\sqrt{\cosh2z+\sqrt{4\cosh^22z-3}}}{2}
\end{equation}
and using the value of $z$ used throughout this paper ($z=2.2$) we find $\vert\alpha\vert_\mathrm{lim}\approx5.5$. For $\vert\alpha\vert>\vert\alpha\vert_\mathrm{lim}$ the optimum PMCs are given by equations \eqref{eq:phase_matching_cond_coh_plus_sqz_vac} and \eqref{eq:phase_matching_cond_sqz_coh_plus_sqz_vac_BIS} and the best achievable sensitivity for a single-mode intensity detection scheme is given by equation  \eqref{eq:app:Delta_varphi_Sqz-coh_Sqz_Vac_SING_Best_phi_is_zero}. For even higher values of $\vert\alpha\vert$ the gap between the two performances increases in the favor of the PMC $\theta-\phi=0$. Although not optimal, this detection scheme is experimentally interesting due to its simplicity and because the output port is ``dark'', thus extremely sensitive p-i-n photo-diodes can be used. We can also point to the results of reference \cite{Wu19}, where it has been shown that photon-number-resolving detection of only a small number of photons in the dark port can achieve the QCRB.

Except in the regime where $\vert\alpha\vert\ll\{\sinh^2r,\sinh^2z\}$ (where the difference-intensity detection scheme yields better results), the homodyne outperforms the other detection schemes considered. We mention that we have a single curve for $\Delta\tilde{\varphi}_\mathrm{hom}$ in Fig.~\ref{fig:Delta_phi_coh_squeezed_plus_squeezed_vac_versus_alpha_BIS} because the phase sensitivity for a homodyne detection scheme at the optimum working point  does not depend on the phase $\phi$.

% ------------- FIGURE --------- versus ALPHA --------------

In the coherent plus squeezed vacuum scenario (\emph{i. e.} for $z=0$) the optimal PMC \eqref{eq:phase_matching_cond_coh_plus_sqz_vac} is indisputable \cite{API18,Pre19,Liu13}, most authors using it by default \cite{Pez08,Dem15,Lan13}. Adding squeezing to the coherent source from the input port $1$ brought forward two scenarios. Indeed, the optimality given by the PMC from equation \eqref{eq:phase_matching_cond_sqz_coh_plus_sqz_vac} is to be taken with a grain of salt. If one chases the ultimate bound on sensitivity, then the QFI from equation \eqref{eq:Fisher_sqz-coh_plus_sqz_vac_MAXIMAL} and the related QCRB from equation \eqref{eq:delta_varphi_QCRB_sqz_coh_sqz_vac} are the answer. If one is more interested in a wider range of $\varphi$ while keeping a good sensitivity, then the PMCs \eqref{eq:phase_matching_cond_coh_plus_sqz_vac} and \eqref{eq:phase_matching_cond_sqz_coh_plus_sqz_vac_BIS} are more appropriate.

% --------------------- FIGURE ------------ LOSSES
\begin{figure}
\includegraphics[scale=0.48]{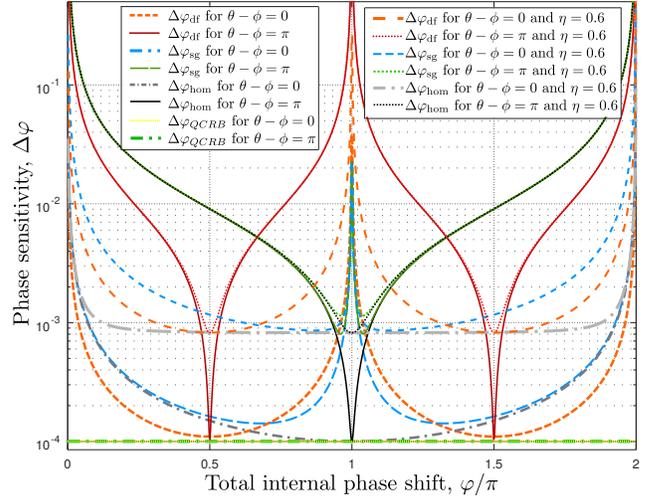}
\caption{The effect of non-unit photo-detection efficiency on the phase sensitivity. While losses degrade the performance of all realistic scenarios, the PMC $\theta-\phi=0$ still retains a better overall performance. Wherever not specified it is implied that $\eta=1$.  Parameters used: $\vert\alpha\vert=10^3$, $r=2.3$, $z=2.2$, $\theta_\alpha=\theta=0$.}
\label{fig:Delta_varphi_versus_varphi_sqz-coh_sqz-coh_LOSSES}
\end{figure}

%----------------------------------------------------------
% --------------------- EFFECT LOSSES ---------------------
\subsubsection{Non-unit photo detection efficiency}

Losses inside the interferometer and due to the coupling with the environment are outside the scope of this paper. We point the interested reader to the available literature \cite{Dor09,Dem09,Ono10}.

In Appendix \ref{sec:app:LOSSES} we briefly describe how to account for losses caused by non-unit photo-detection efficiency (assumed identical to all detectors and modelled by the parameter $\eta\leq1$, the ideal case implying $\eta=1$). We begin with a single-mode intensity detection scheme and using equation \eqref{eq:app:Variance_nk_LOSSY} we arrive at the phase sensitivity
\begin{equation}
\label{eq:delta_varphi_LOSSY}
\Delta\varphi'_\mathrm{sg}=\frac{\sqrt{\Delta^2\hat{n}_4+\frac{1-\eta}{\eta}\langle\hat{n}_4\rangle}}{\vert\partial_\varphi\langle\hat{n}_4\rangle\vert}
\end{equation}
The numerator variance has the supplementary term $(1-\eta)/\eta\langle\hat{n}_4\rangle$. For a shot-noise limited detection (i. e. $\Delta^2\hat{n}_4=\langle\hat{n}_4\rangle$) equation \eqref{eq:delta_varphi_LOSSY} reduces to $\Delta\varphi'=\Delta\hat{n}_4/(\sqrt{\eta}\vert\partial_\varphi\langle\hat{n}_4\rangle\vert)$, a well-known result \cite{Ono10}. However the whole interest of squeezed states is to have $\Delta^2\hat{n}_4<\langle\hat{n}_4\rangle$ and thus the addition of the $\langle\hat{n}_4\rangle$ term is a clear degradation of the performance.

In Fig.~\ref{fig:Delta_varphi_versus_varphi_sqz-coh_sqz-coh_LOSSES} we plot the phase sensitivity for the ideal case ($\eta=1$) as well as for the one including losses ($\eta=0.6)$. Although a swift degradation of the phase sensitivity in the case of PMC \eqref{eq:phase_matching_cond_sqz_coh_plus_sqz_vac_BIS} is found, this setting yields still a better overall performance compared to the PMC \eqref{eq:phase_matching_cond_sqz_coh_plus_sqz_vac}.

The losses affect all considered realistic detection schemes. A general pattern emerges: the peak performance is the most affected and whatever the internal phase shift $\varphi$ is we have $\Delta\varphi\leq\Delta\varphi'$. The experimentally preferable detection scheme emerges the homodyne detection due to its higher immunity to losses over a large range of internal phase shifts.

% -------------- PHASE-SPACE REPRESENTATION ----------------
\subsubsection{Phase-space representation and some physical insights}

\begin{figure}
\centering
\includegraphics[scale=0.98]{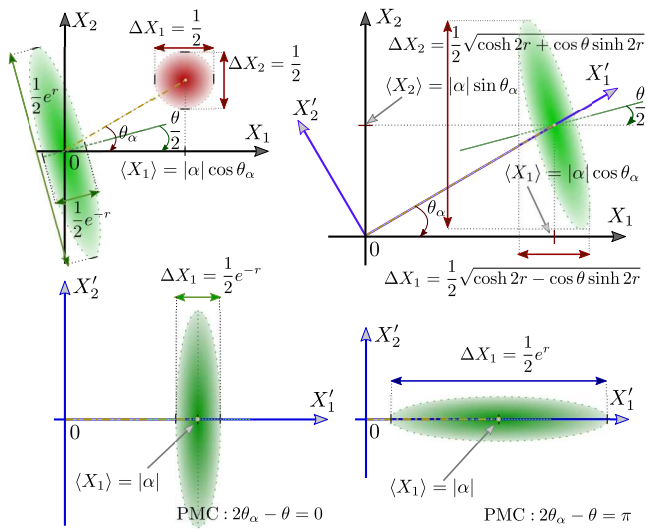}
\caption{Top left graphic: phase-space representation of a coherent (red circle) and squeezed vacuum (green ellipse) states. Top right graphic: standard phase-space representation of a squeezed-coherent state (see e. g. \cite{Loudon}, Section 5.6.). However, in the case of our interferometer, the phase reference is the coherent source $\alpha$, therefore we rotate our measurement axis following $\theta_\alpha$ (see blue axis, $X_1'$ and $X_2'$). Bottom left: phase-space representation of a squeezed-coherent state in the rotated frame with PMC  $2\theta_\alpha-\theta=0$. Bottom right: phase-space representation of a squeezed-coherent state in the rotated frame with PMC  $2\theta_\alpha-\theta=\pi$.}
\label{fig:phase_space_representation}
\end{figure}

We give now a qualitative phase-space representation and some physical insights concerning the obtained results, especially the PMCs. In Fig.~\ref{fig:phase_space_representation} (top left graphic) we have a standard phase-space representation of a coherent state (red circle) and a squeezed vacuum (green ellipse). Please note that the angle of rotation for the squeezed state is $\theta/2$ i. e. a rotation of $\theta=\pi$ brings the ellipse to a perpendicular position wrt its original state \cite{GerryKnight,Loudon}. The standard representation of a squeezed coherent state is given in Fig.~\ref{fig:phase_space_representation} (top right graphic).

However, in our interferometer, the coherent source $\alpha$ acts as a phase reference, therefore we have to rotate the phase space with $\theta_\alpha$ (see blue axis in Fig.~\ref{fig:phase_space_representation}, top right graphic).
%
%
%
%
% ------------- FIGURE --------- versus ALPHA --------------
This is what actually happens in the homodyne detection technique.  The most important term determining the performance of $\Delta\hat{X}_{\phi_L}=\sqrt{\Delta^2\hat{X}_{\phi_L}}$ from equation \eqref{eq:Homodyne_variance_sqz_coh_plus_sqz_vac} is $\Upsilon^-(\alpha,\xi)$. The fact that $2\theta_\alpha-\theta=0$ effectively squeezes the variance of $\Upsilon^-(\alpha,\xi)$, thus the average measured value is more accurate (see Fig.~\ref{fig:phase_space_representation}, bottom, left graphic). 

We can extend the discussion to the difference- and single-mode detection techniques. The same term $\Upsilon^-(\alpha,\xi)$ is present in both equations \eqref{eq:VARIANCE_N_d_sqz-coh_sqz_vac_FINAL} and \eqref{eq:Variance_N_4_coh_sqz_sqz_vac}, thus the same PMC from equation \eqref{eq:phase_matching_cond_coh_plus_sqz_vac} minimizes the respective variances, thus optimizing the phase sensitivity.

The fact that the squeezers have to be in anti-phase was explained in the literature \cite{Lan14} (see also the discussion from Appendix \ref{sec:app:optimization_two_squeezers}). Indeed, the optimal input state with two equal squeezed vacuums in anti-phase ($\zeta=-\xi$) is an eigenvector of the beam splitter i. e. the input state \eqref{eq:psi_in_squeezed_coherent_plus_squeezed_vac} becomes $\vert\psi'\rangle=\hat{D}_2(i\alpha/\sqrt{2})\hat{D}_3(\alpha/\sqrt{2})\hat{S}_2(-\xi)\hat{S}_3(\xi)\vert0\rangle$ after $BS_1$, result also reported in \cite{Kim96}. Thus, the same, un-attenuated squeezing coefficient is available inside the interferometer.

At the same time with minimizing $\Upsilon^-(\alpha,\xi)$, one notes that $\Upsilon^+(\alpha,\xi)$ is maximized for $2\theta_\alpha-\theta=0$. This fact is reassuring, since the Fisher information from equation \eqref{eq:app:F_dd_coh_sqz_vac_sqz_vac_Upsilon} contains this term and we wish to have it maximized. Similarly, having the squeezers in anti-phase changes the sign of the last term from equation \eqref{eq:app:F_dd_coh_sqz_vac_sqz_vac_Upsilon}, thus maximizing the Fisher information.

In the high-$\alpha$ regime, if we drop the insistence on having the squeezers in anti-phase and the MZI is not operating at the optimum working point, another term, namely $\Upsilon^-(\alpha,\zeta)$ has to be minimized. Using the same arguments as before, leads us to the condition $\theta-\phi=0$. This gives a physical explanation for the PMC from equation \eqref{eq:phase_matching_cond_sqz_coh_plus_sqz_vac_BIS}.

% ---------------------------------------------------------
% ---------------- HEISENBERG SCALING ---------------------
\subsubsection{Heisenberg scaling}

We ponder now if a Heisenberg scaling \cite{Dem15,Ou96} can be reached by an interferometer fed by the input state \eqref{eq:psi_in_squeezed_coherent_plus_squeezed_vac} \emph{i. e.} if we can reach
\begin{equation}
\label{eq:delta_varphi_HL_coh_sqz_vac}
\Delta\varphi_{HL}\approx\frac{1}{\langle{N}_\textrm{tot}\rangle}
\end{equation}
where $\langle{N}_\textrm{tot}\rangle=\vert\alpha\vert^2+\sinh^2r+\sinh^2z$. Pezz\'e \& Smerzi \cite{Pez08} showed that the scaling \eqref{eq:delta_varphi_HL_coh_sqz_vac} can be reached by a coherent plus squeezed vacuum input if we consider $\vert\alpha\vert^2\approx\sinh^2r\approx\langle{N}_\textrm{tot}\rangle/2\gg1$. We  denote $f_\alpha=\vert\alpha\vert^2/\langle{N}_\textrm{tot}\rangle$, $f_r=\sinh^2r/\langle{N}_\textrm{tot}\rangle$, $f_z=\sinh^2z/\langle{N}_\textrm{tot}\rangle$ and assume $\{\vert\alpha\vert^2,\:\sinh^2r,\:\sinh^2z\}\gg1$. From equation \eqref{eq:Fisher_sqz-coh_plus_sqz_vac_MAXIMAL} we obtain the Fisher information $\mathcal{F}\approx4\langle{N}_\textrm{tot}\rangle^2f_r(f_\alpha+f_z)$, hence the scaling
\begin{equation}
\label{eq:delta_varphi_HL_coh_sqz_vac_scaling_beta1_beta2}
\Delta\varphi_{HL}\approx
%\frac{1}{\sqrt{4\langle{N}_\textrm{tot}\rangle^2f_r(f_\alpha+f_z)}}
%=
\frac{1}{\sqrt{4\langle{N}_\textrm{tot}\rangle^2f_r(1-f_r)}}.
\end{equation}
The optimum is obtained when $f_r=1/2$ yielding the QFI $\mathcal{F}=\langle{N}_\textrm{tot}\rangle^2$ and thus Heisenberg scaling from equation \eqref{eq:delta_varphi_HL_coh_sqz_vac}. Similar to the  result from reference \cite{Pez08}, half of the input power has to go into the squeezing from port $0$ in order to guarantee a Heisenberg scaling. However, surprisingly, the limit \eqref{eq:delta_varphi_HL_coh_sqz_vac_scaling_beta1_beta2} does not depend on $f_\alpha$ and $f_z$. Thus, the experimenter is free to choose the fraction of squeezed/coherent power in input port $1$ at its own will (as long as $f_\alpha+f_z=1/2$), while being guaranteed to reach a Heisenberg scaling.

% ---------------------------------------------------------
% ------- FIGURE --- FISHER 2D plot versus alpha and beta -
% ---------------------------------------------------------
\begin{figure}
\includegraphics[scale=0.5]{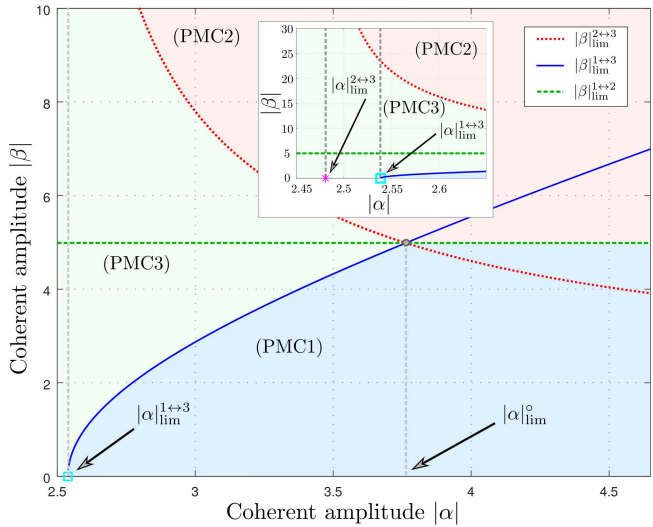}
\caption{All the optimum PMCs versus the input coherent amplitudes $\vert\alpha\vert$ and $\vert\beta\vert$ summarized in a graphical manner. The low-$\vert\alpha\vert$ coherent amplitude regime is depicted in the inset. With the squeezing factors $r=2.3$ and $z=2.2$ we get the limit values: $\vert\alpha\vert^{1\leftrightarrow3}_\textrm{lim}\approx2.54$, $\vert\alpha\vert^{2\leftrightarrow3}_\textrm{lim}\approx2.48$, $\vert\alpha\vert^\circ_\textrm{lim}\approx3.76$ and $\vert\beta\vert^{1\leftrightarrow2}_\textrm{lim}\approx4.98$.}
\label{fig:alpha_LIM_beta_LIM_sqzcoh_sqzcoh}
\end{figure}
% ---------------------------------------------------------
% ------- END FIGURE --- 2D plot versus alpha and beta ----
% ---------------------------------------------------------

The aforementioned Heisenberg scaling assumed the PMCs given by equations \eqref{eq:phase_matching_cond_coh_plus_sqz_vac} and \eqref{eq:phase_matching_cond_sqz_coh_plus_sqz_vac}. If we use instead the constraint \eqref{eq:phase_matching_cond_sqz_coh_plus_sqz_vac_BIS}, we arrive at the QFI given by equation  \eqref{eq:Fisher_sqz_coh_plus_sqz_vac_PMC_phi_is_zero} therefore $\mathcal{F}\approx4\langle{N}_\textrm{tot}\rangle^2f_\alpha f_r$. This time the Heisenberg scaling \eqref{eq:delta_varphi_HL_coh_sqz_vac} imposes $f_\alpha=f_r\to1/2$ (and consequently $f_z\to0$), thus the optimum is found for a coherent plus squeezed vacuum input with half of the power denoted to squeezing \cite{Pez08}.

% ---------------------------------------------------------
% --------------- COHERENT plus SQUEEZED VACUUM -----------
% ---------------------------------------------------------
\section{Squeezed-coherent plus squeezed-coherent input state}
\label{sec:sqz_coh_plus_sqz_coh}
In this scenario we consider the most general Gaussian input state, namely a squeezed-coherent plus squeezed-coherent input,
\begin{equation}
\label{eq:psi_in_squeezed_coherent_plus_squeezed_coherent}
\vert\psi_\mathrm{in}\rangle=\vert(\alpha\zeta)_1(\beta\xi)_0\rangle=\hat{D}_1\left(\alpha\right)\hat{S}_1\left(\zeta\right)\hat{D}_0\left(\beta\right)\hat{S}_0\left(\xi\right)\vert0\rangle
\end{equation}
We impose for the time being no restriction on any of the parameters involved in this state.

Due to the number of variables, this scenario is more difficult to tackle. We start our discussion with the QFI and use it as a guide in order to be able to evaluate how well realistic detection schemes behave.

% ----------------- SINGLE COHERENT INPUT -----------------
% ----------------- DIFFERENTIAL DETECTION ----------------
% ---------------------------------------------------------
\subsection{Fisher information and the Cram\'er-Rao bound}
\label{subsec:sqz_coherent_sqz_coh_input_Fisher}
The Fisher matrix coefficients $\mathcal{F}_{ss}$, $\mathcal{F}_{dd}$ and $\mathcal{F}_{sd}$ are computed in Section  \ref{subsec:app:fisher_sqz_coh_plus_sqz_coh}. In order to minimize the QCRB, one wishes to maximize the QFI given by equation \eqref{eq:Fisher_definition}. However, this time the problem is less trivial. In Section \ref{sec:sqz_coh_plus_squeezed_vacuum} the maximization of the QFI gave the phase-matching conditions \eqref{eq:phase_matching_cond_coh_plus_sqz_vac} and \eqref{eq:phase_matching_cond_sqz_coh_plus_sqz_vac}. Moreover, the phase-matching conditions did not depend on the values of the parameters involved \emph{i. e.} the amplitude of the coherent sources and the squeezing factor(s). This assertion is no longer true in the squeezed-coherent plus squeezed-coherent scenario.

% ---------------------------------------------------------
% ----------- FIGURE --- FISHER versus BETA --- ALPHA=2.5 -
% ---------------------------------------------------------
\begin{figure}
	\centering
		\includegraphics[scale=0.49]{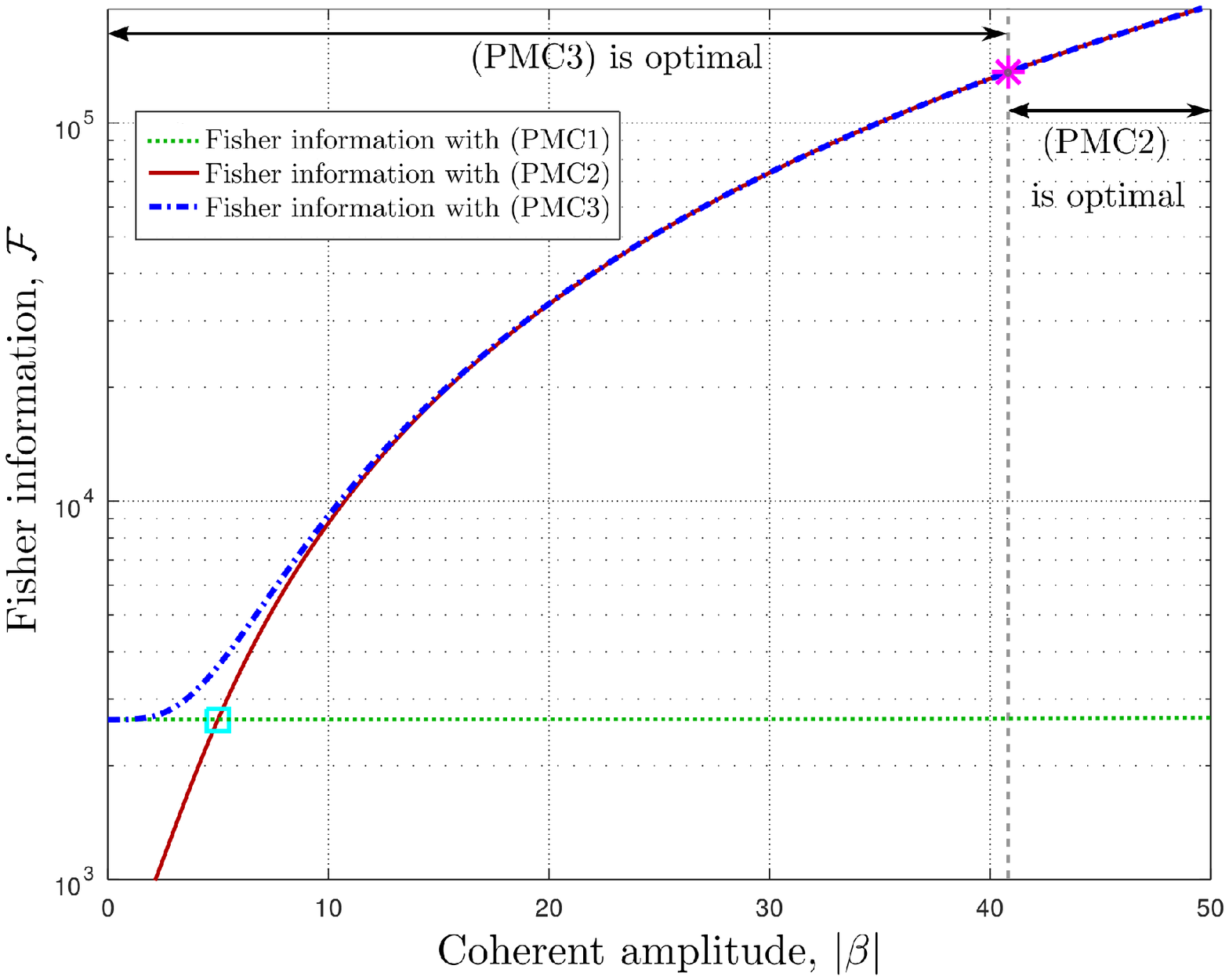}
	\caption{The QFI versus the coherent amplitude $\vert\beta\vert$ for $\vert\alpha\vert=2.5$. Since $\vert\alpha\vert\in[\vert\alpha\vert^{2\leftrightarrow3}_\textrm{lim},\vert\alpha\vert^{1\leftrightarrow3}_\textrm{lim}]$, the optimal input scenario is (PMC3)  for $\vert\beta\vert$ smaller than $\vert\beta\vert^{2\leftrightarrow3}_\textrm{lim}$ (magenta star) and (PMC2) thereafter. Parameters used: $r=2.3$, $z=2.2$.}
	\label{fig:F_sqz-coh_sqz-coh_versus_beta_alpha_2_5}
\end{figure}
% ---------------------------------------------------------
% ------- END FIGURE --- FISHER versus BETA --- ALPHA=2.5 -
% ---------------------------------------------------------

Throughout our discussion, without loss of generality, we consider the coherent source $\vert\alpha\vert$ the primary one, thus, if necessary, the maximization of the coefficient $\Upsilon^+\left(\alpha,\xi\right)$ is privileged over the maximization of $\Upsilon^+\left(\beta,\zeta\right)$. The discussion is, of course, symmetric and one can switch $\alpha\leftrightarrow\beta$ and conduct a similar analysis. In this section we first present the phase-matching conditions leading to the optimal QFI and later we will justify them (see also the discussion from Appendix \ref{subsec:app:PMC_for_fisher_sqz_coh_plus_sqz_coh}). Throughout the discussion we fix the squeezing factors $r$ and $z$ and vary the coherent amplitudes. Thus, all the ``limit values'' ($\vert\alpha\vert_\textrm{lim}$ and $\vert\beta\vert_\textrm{lim}$) that will appear in our analysis will be functions of $r$ and $z$. Please note that the values of $r$ and $z$ are in no way constrained.

Intense computer simulations showed that the Fisher information is maximized only by the PMCs $\theta_\alpha-\theta_\beta=n\pi/2$, $2\theta_\alpha-\theta=n'\pi$ and $2\theta_\beta-\phi=n''\pi$ with $n,n',n''\in\mathbb{Z}$. This result constrained substantially our search for the optimum input PMCs.

We start from the PMCs \eqref{eq:phase_matching_cond_coh_plus_sqz_vac} and \eqref{eq:phase_matching_cond_sqz_coh_plus_sqz_vac} from Section \ref{sec:sqz_coh_plus_squeezed_vacuum} and add the condition \eqref{eq:phase_matcing_theta_alpha_equal_theta_beta} on $\theta_\beta$. We obtain the first set of input phase-matching conditions,
\begin{equation}
\label{eq:phase_matching_array_0_0_phi_is_pi}
\textrm{(PMC1) }
\left\{
\begin{array}{l}
2\theta_\alpha-\theta=0\\
\phi-\theta=\pm\pi\\
\theta_\alpha-\theta_\beta=0
\end{array}
\right.
\end{equation}
These PMCs applied to equation \eqref{eq:Fisher_sqzcoh_sqz_coh_PMC_theta_0_phi_pi_thetabeta_free} give the QFI
\begin{eqnarray}
\label{eq:Fisher_sqzcoh_sqz_coh_PMC_theta_0_phi_pi_thetabeta_0}
\mathcal{F}
=\vert\alpha\vert^2e^{2r}
+\vert\beta\vert^2e^{-2z}
+\sinh^2\left(r+z\right).
\end{eqnarray}
For $\{\vert\alpha\vert^2,\:\sinh^2r,\:\sinh^2z\}\gg\vert\beta\vert^2$ this QFI is obviously optimal (see also Fig.~\ref{fig:alpha_LIM_beta_LIM_sqzcoh_sqzcoh}, blue shaded area). However, when $\vert\beta\vert$ becomes comparable to $\vert\alpha\vert$, this is clearly not the case anymore. We thus impose a different set of phase matching conditions when  $\{\vert\alpha\vert^2,\:\vert\beta\vert^2\}\gg\{\sinh^2r,\:\sinh^2z\}$,
\begin{equation}
\label{eq:phase_matching_array_0_0_0}
\textrm{(PMC2) }
\left\{
\begin{array}{l}
2\theta_\alpha-\theta=0\\
\phi-\theta=0\\
\theta_\alpha-\theta_\beta=0
\end{array}
\right.
\end{equation}
that applied to equation \eqref{eq:Fisher_sqzcoh_sqz_coh_PMC_theta_0_phi_pi_thetabeta_free} give the QFI
\begin{eqnarray}
\label{eq:Fisher_sqzcoh_sqz_coh_PMC_theta_0_phi_0_thetabeta_0}
\mathcal{F}
=\vert\alpha\vert^2e^{2r}
+\vert\beta\vert^2e^{2z}
+\sinh^2\left(r-z\right).
\end{eqnarray}
This time the gain in the second term is accompanied by a less important contribution from the two squeezers. Comparing equations \eqref{eq:Fisher_sqzcoh_sqz_coh_PMC_theta_0_phi_pi_thetabeta_0} and \eqref{eq:Fisher_sqzcoh_sqz_coh_PMC_theta_0_phi_0_thetabeta_0} yields the limit value of $\vert\beta\vert$,
\begin{equation}
\label{eq:beta_mod_LIMIT_coh-sqz_plus_coh-sqz}
\vert\beta\vert_\textrm{lim}^{1\leftrightarrow2}=\sqrt{\frac{\cosh2r}{2}}
\end{equation}
Thus, above a certain limit value of $\vert\alpha\vert$ (denoted $\vert\alpha\vert^\circ_\mathrm{lim}$, to be specified shortly, see equation \eqref{eq:csqz_csqz_alpha_LIM_triple_point} and also Fig.~\ref{fig:alpha_LIM_beta_LIM_sqzcoh_sqzcoh}), if ${\vert\beta\vert<\vert\beta\vert_\textrm{lim}^{1\leftrightarrow2}}$, the QFI from equation \eqref{eq:Fisher_sqzcoh_sqz_coh_PMC_theta_0_phi_pi_thetabeta_0} is maximal and the optimal PMCs are given by equation \eqref{eq:phase_matching_array_0_0_phi_is_pi}. If ${\vert\beta\vert>\vert\beta\vert_\textrm{lim}^{1\leftrightarrow2}}$, the QFI from equation \eqref{eq:Fisher_sqzcoh_sqz_coh_PMC_theta_0_phi_0_thetabeta_0} is maximal and equation \eqref{eq:phase_matching_array_0_0_0} gives the optimum PMCs (see also Fig.~\ref{fig:alpha_LIM_beta_LIM_sqzcoh_sqzcoh}, red shaded area).

% ---------------------------------------------------------
% ----------- FIGURE --- FISHER versus BETA --- ALPHA=2.8 -
% ---------------------------------------------------------
\begin{figure}
	\centering
	\includegraphics[scale=0.48]{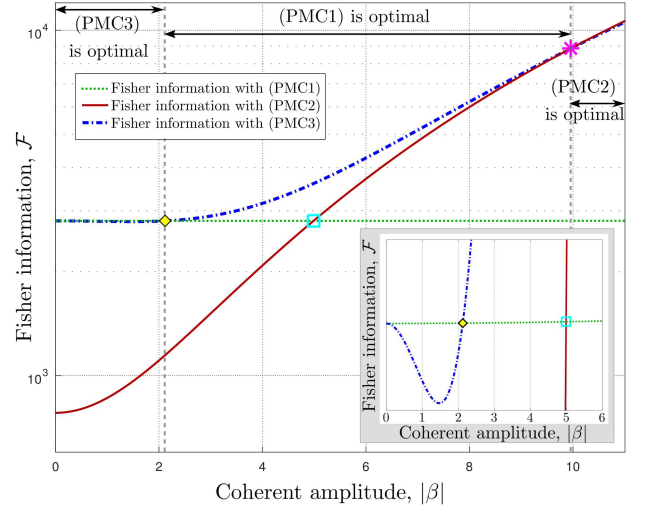}
	\caption{The QFI versus the coherent amplitude $\vert\beta\vert$ for $\vert\alpha\vert=2.8$. For $\vert\beta\vert$ smaller than $\vert\beta\vert_\textrm{lim}^{1\leftrightarrow3}$ (yellow diamond) (PMC3) is optimal while for $\vert\beta\vert$ bigger than $\vert\beta\vert_\textrm{lim}^{2\leftrightarrow3}$ (magenta star) (PMC2) yields the best performance. In between these values (PMC1) is optimal. Inset: zoom on the range $\vert\beta\vert\in[0,6]$. Parameters used: $r=2.3$ and $z=2.2$.}
	\label{fig:F_sqz-coh_sqz-coh_versus_beta_alpha_2_8}
\end{figure}

We introduce now the third and final set of phase-matching conditions,
\begin{equation}
\label{eq:phase_matching_array_0_0_pi_over_2}
\textrm{({PMC3}) }
\left\{
\begin{array}{l}
2\theta_\alpha-\theta=0\\
2\theta_\beta-\phi=0\\
\theta_\alpha-\theta_\beta=\frac{\pi}{2}
\end{array}
\right.
\end{equation}
that applied to equation \eqref{eq:Fisher_sqzcoh_sqz_coh_PMC_theta_0_phi_pi_thetabeta_free} give the Fisher information
\begin{eqnarray}
\label{eq:Fisher_sqzcoh_sqz_coh_PMC_theta_0_phi_pi_thetabeta_pi2}
\mathcal{F}
=\vert\alpha\vert^2e^{2r}
+\vert\beta\vert^2e^{2z}
+\sinh^2\left(r+z\right)
\nonumber\\
-\frac{\vert\alpha\beta\vert^2\left(e^{2r}+e^{2z}\right)^2}{\frac{\sinh^22r}{2}
+\vert\beta\vert^2e^{2r}
+\frac{\sinh^22z}{2}
+\vert\alpha\vert^2e^{2z}}
\end{eqnarray}
As we will show, (PMC3) is optimal in the limit $\{\sinh^2r,\:\sinh^2z\}\gg\{\vert\alpha\vert^2,\:\vert\beta\vert^2\}$ (see also Fig.~\ref{fig:alpha_LIM_beta_LIM_sqzcoh_sqzcoh}, green shaded area).

We need to find now the limit values of $\vert\alpha\vert$ and $\vert\beta\vert$ (themselves functions of the squeezing parameters $r$ and $z$) that make the transition from one PMC to another. Without loss of generality, we fix the values $\vert\alpha\vert_\textrm{lim}$ and write $\vert\beta\vert_\textrm{lim}$ as functions of $\vert\alpha\vert_\textrm{lim}$. If a value $\vert\alpha\vert_\textrm{lim}$ makes the transition between e. g. (PMC1) and (PMC2), it will be denoted by $\vert\alpha\vert^{1\leftrightarrow2}_\textrm{lim}$ etc.

We first consider the low-$\vert\alpha\vert$ regime. Imposing equal Fisher information to equations \eqref{eq:Fisher_sqzcoh_sqz_coh_PMC_theta_0_phi_pi_thetabeta_0} and \eqref{eq:Fisher_sqzcoh_sqz_coh_PMC_theta_0_phi_pi_thetabeta_pi2} (as $\vert\beta\vert\to0$), we get the limit amplitude
\begin{equation}
\label{eq:alpha_LIM_circ_sqz-coh_plus_sqz-coh}
\vert\alpha\vert^{1\leftrightarrow3}_\textrm{lim}
=\sqrt{\frac{2S\sinh{2z}}
{e^{2r}(e^{2r}+2e^{2z})+1}}
\end{equation}
where $S=(\sinh^22r+\sinh^22z)/2$. With the values taken throughout this paper ($r=2.3$, $z=2.2$), we obtain the limit value $\vert\alpha\vert^{1\leftrightarrow3}_\textrm{lim}\approx2.54$. 

% ---------------------------------------------------------
% ----------- FIGURE --- FISHER versus BETA --- ALPHA=4 ---
% ---------------------------------------------------------
\begin{figure}
	\centering
		\includegraphics[scale=0.48]{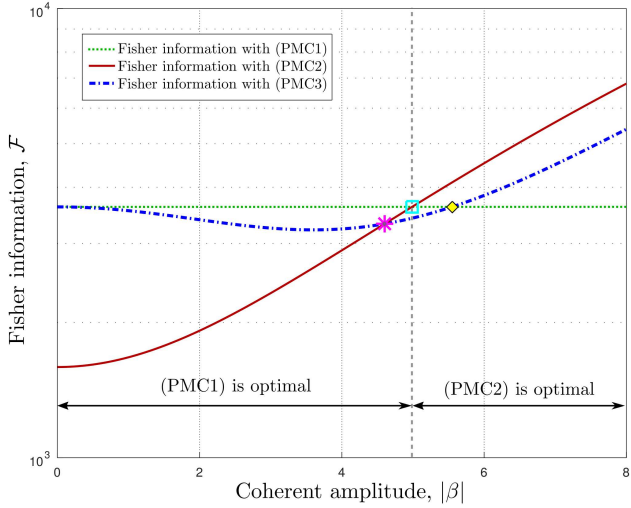}
	\caption{The QFI versus the coherent amplitude $\vert\beta\vert$ for $\vert\alpha\vert=4$. 
For 	$\vert\beta\vert\leq\vert\beta\vert_\textrm{lim}^{1\leftrightarrow2}$/ $\vert\beta\vert>\vert\beta\vert_\textrm{lim}^{1\leftrightarrow2}$ (PMC1)/(PMC2) is optimal.
	The cyan square marks $\vert\beta\vert_\textrm{lim}^{1\leftrightarrow2}$ given by equation \eqref{eq:beta_mod_LIMIT_coh-sqz_plus_coh-sqz}. Parameters used: $r=2.3$, $z=2.2$.}
	\label{fig:F_sqz-coh_sqz-coh_versus_beta_alpha_4}
\end{figure}
% ---------------------------------------------------------
% ------- END FIGURE --- FISHER versus BETA --- ALPHA=4 ---
% ---------------------------------------------------------

As $\vert\alpha\vert$ increases, we cannot disregard the scenario employing (PMC2) from equation \eqref{eq:phase_matching_array_0_0_0}. Comparing the QFI from equations \eqref{eq:Fisher_sqzcoh_sqz_coh_PMC_theta_0_phi_0_thetabeta_0} and \eqref{eq:Fisher_sqzcoh_sqz_coh_PMC_theta_0_phi_pi_thetabeta_pi2} we arrive at the  limit value,
\begin{equation}
\label{eq:alpha_LIM_star_sqz-coh_plus_sqz-coh}
\vert\alpha\vert^{2\leftrightarrow3}_\textrm{lim}=e^{-z}\frac{\sqrt{\sinh2r\sinh2z}}{2\cosh(r-z)}
\end{equation}
With the above values of $r$ and $z$, we find $\vert\alpha\vert^{2\leftrightarrow3}_\textrm{lim}\approx2.48$.

Since we have now the two limit values $\vert\alpha\vert^{1\leftrightarrow3}_\textrm{lim}$ and $\vert\alpha\vert^{2\leftrightarrow3}_\textrm{lim}$, we can vary of the parameter $\vert\alpha\vert$ from zero to arbitrary large values and search for the limit values of $\vert\beta\vert$. Thus, if $\vert\alpha\vert\in[\vert\alpha\vert^{2\leftrightarrow3}_\textrm{lim},\vert\alpha\vert^{1\leftrightarrow3}_\textrm{lim}]$ we introduce the limit value of $\vert\beta\vert$,
\begin{equation}
\label{eq:beta_LIM_star_sqz-coh_plus_sqz-coh}
\vert\beta\vert^{2\leftrightarrow3}_\textrm{lim}
=\sqrt{
\frac{e^{2r}\sinh{2r}\sinh{2z}(\vert\alpha\vert^2e^{2z}+S)}{4\vert\alpha\vert^2e^{2z}\cosh^2(r-z)-\sinh{2r}\sinh{2z}
}}
\end{equation}
We recall that $\vert{\beta}\vert^{2\leftrightarrow3}_\textrm{lim}$ is a function of $\vert\alpha\vert$ (see also Fig.~ \ref{fig:alpha_LIM_beta_LIM_sqzcoh_sqzcoh}). For $\vert{\beta}\vert\leq\vert\beta\vert^{2\leftrightarrow3}_\textrm{lim}$ the optimal QFI is obtained by imposing (PMC3) given by equation \eqref{eq:phase_matching_array_0_0_pi_over_2}. If $\vert\beta\vert>\vert\beta\vert^{2\leftrightarrow3}_\textrm{lim}$ the optimum is (PMC2) given by equation \eqref{eq:phase_matching_array_0_0_0}. This scenario is depicted in Fig.~\ref{fig:F_sqz-coh_sqz-coh_versus_beta_alpha_2_5} (see also the inset from Fig.~\ref{fig:alpha_LIM_beta_LIM_sqzcoh_sqzcoh}).

% ---------------------------------------------------------
% ----------- FIGURE --- FISHER versus BETA --- ALPHA=500 -
% ---------------------------------------------------------
\begin{figure}
	\centering
	\includegraphics[scale=0.48]{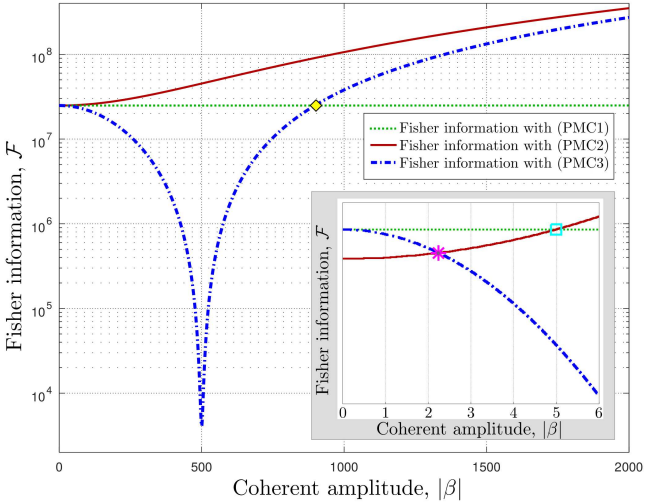}
	\caption{The QFI versus the coherent amplitude $\vert\beta\vert$ for $\vert\alpha\vert=500$. (PMC1) is optimal for $\vert\beta\vert$ below $\vert\beta\vert_\textrm{lim}^{1\leftrightarrow2}$ (cyan square) while above this value (PMC2) is preferred. Inset: zoom on the range $\vert\beta\vert\in[0,6]$. Parameters used $r=2.3$, $z=2.2$.}
	\label{fig:F_sqz-coh_sqz-coh_versus_beta_alpha_500}
\end{figure}

Satisfying the condition $\vert\alpha\vert\leq\vert\alpha\vert^{1\leftrightarrow3}_\textrm{lim}$ guarantees that for very small $\vert\beta\vert$, (PMC3) from equation \eqref{eq:phase_matching_array_0_0_pi_over_2} is always optimal. If $\vert\alpha\vert>\vert\alpha\vert^{1\leftrightarrow3}_\textrm{lim}$, this is no longer true. We introduce now the limit value,
\begin{equation}
\label{eq:beta_LIM_circ_sqz-coh_plus_sqz-coh}
\vert\beta\vert^{1\leftrightarrow3}_\textrm{lim}
=e^{z-r}\sqrt{\vert\alpha\vert^2\left(\frac{2e^{2r}\cosh^2(r-z)}{\sinh2z}-1\right)-\frac{S}{e^{2z}}}
\end{equation}
For  $\vert\alpha\vert>\vert\alpha\vert^{1\leftrightarrow3}_\textrm{lim}$ as $\vert\beta\vert$ starts to grow from $0$, (PMC1) from equation \eqref{eq:phase_matching_array_0_0_phi_is_pi} will yield the maximum Fisher information until $\vert\beta\vert=\vert\beta\vert^{1\leftrightarrow3}_\textrm{lim}$. At this point both scenarios yield the same Fisher information.

It can be shown that there exists a limit value $\vert\alpha\vert=\vert\alpha\vert_\textrm{lim}^\circ$ \emph{s. t.} $\vert\beta\vert_\textrm{lim}^{1\leftrightarrow2}=\vert\beta\vert_\textrm{lim}^{1\leftrightarrow3}=\vert\beta\vert_\textrm{lim}^{2\leftrightarrow3}$ and we find
\begin{equation}
\label{eq:csqz_csqz_alpha_LIM_triple_point}
\vert\alpha\vert_\textrm{lim}^\circ=\sqrt{\frac{e^{2r}\cosh2r\sinh2z+2S}{e^{2r}\left(e^{2r}+2e^{2z}\right)+1}}.
\end{equation}
For the parameters used we have $\vert\alpha\vert_\textrm{lim}^\circ\approx3.76$. For $\vert\alpha\vert>\vert\alpha\vert_\textrm{lim}^\circ$ we are in a strong coherent regime. The optimum PMCs are to be chosen between (PMC1) and (PMC2) with a threshold given by $\vert\beta\vert_\textrm{lim}^{1\leftrightarrow2}$, as mentioned before.

% ------------------- FIGURE ---------- ALPHA=0.2 BETA=0.25 ----
\begin{figure}
\centering
\includegraphics[scale=0.49]{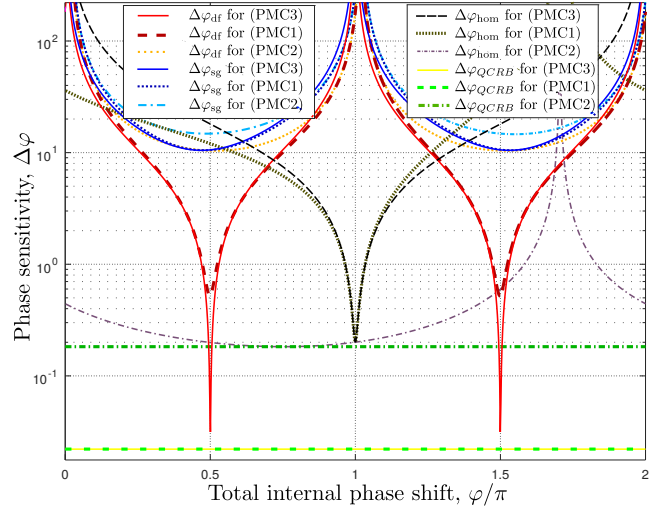}
\caption{Phase sensitivity versus the phase shift at low coherent amplitudes. Parameters used: $\vert\alpha\vert=0.5$, $\vert\beta\vert=0.25$, $r=2.3$, and $z=2.2$. As expected, for the low-intensity regime, (PMC3) is optimal. All realistic detection schemes are suboptimal, with the difference-intensity detection scheme yielding the best performance.}
\label{fig:Delta_varphi_versus_varphi_alpha1_beta05}
\end{figure}

Thus, if $\vert\alpha\vert<\vert\alpha\vert_\textrm{lim}^\circ$ we have $\vert\beta\vert^{1\leftrightarrow3}_\textrm{lim}<\vert\beta\vert^{1\leftrightarrow2}_\textrm{lim}<\vert\beta\vert^{2\leftrightarrow3}_\textrm{lim}$, see Fig.~\ref{fig:alpha_LIM_beta_LIM_sqzcoh_sqzcoh}. This scenario is also depicted in Fig.~\ref{fig:F_sqz-coh_sqz-coh_versus_beta_alpha_2_8}. For $\vert\beta\vert\leq\vert\beta\vert^{1\leftrightarrow3}_\textrm{lim}$ we have the optimum QFI given by (PMC1) from equation \eqref{eq:phase_matching_array_0_0_phi_is_pi}. For $\vert\beta\vert\in [\vert\beta\vert^{1\leftrightarrow3}_\textrm{lim},\vert\beta\vert^{2\leftrightarrow3}_\textrm{lim}]$ the optimum is given by (PMC3) from equation \eqref{eq:phase_matching_array_0_0_pi_over_2} while for $\vert\beta\vert>\vert\beta\vert^{2\leftrightarrow3}_\textrm{lim}$ the optimum is given by (PMC2) from equation \eqref{eq:phase_matching_array_0_0_0}.

For $\vert\alpha\vert\geq\vert\alpha\vert_\textrm{lim}^\circ$ we have $\vert\beta\vert^{2\leftrightarrow3}_\textrm{lim}\leq\vert\beta\vert^{1\leftrightarrow2}_\textrm{lim}\leq\vert\beta\vert^{1\leftrightarrow3}_\textrm{lim}$ (see also Fig.~\ref{fig:alpha_LIM_beta_LIM_sqzcoh_sqzcoh}). This scenario is depicted in Fig.~\ref{fig:F_sqz-coh_sqz-coh_versus_beta_alpha_4}. It corresponds to a higher power regime for the coherent sources w.r.t. the squeezing, therefore it is expected that (PMC3) from equation \eqref{eq:phase_matching_array_0_0_pi_over_2} will lose the upper hand. Indeed for $\vert\beta\vert<\vert\beta\vert^{1\leftrightarrow2}_\textrm{lim}$ (PMC1) yields the maximum QFI while for $\vert\beta\vert>\vert\beta\vert^{1\leftrightarrow2}_\textrm{lim}$ (PMC2) is optimal.

% -------------- FIGURE ----------- ALPHA=1000 BETA=800 ----
\begin{figure}
\centering
\includegraphics[scale=0.49]{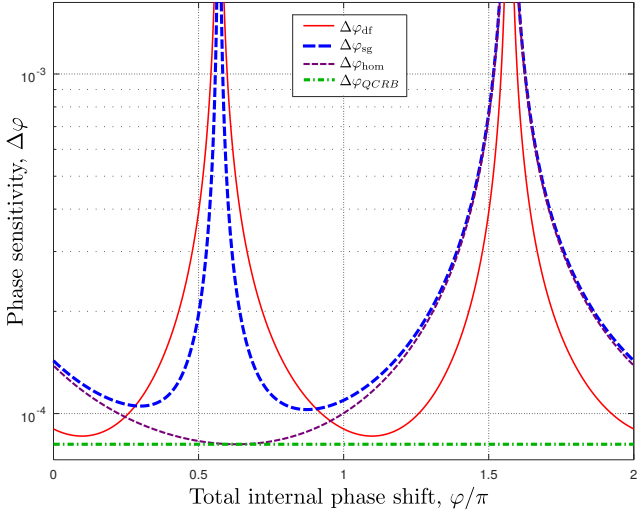}
\caption{Phase sensitivity for a squeezed-coherent plus squeezed-coherent input versus the phase at high coherent amplitudes with (PMC2). Parameters used: $\vert\alpha\vert=1000$,  $\vert\beta\vert=800$,  $r=2.3$,  $z=2.2$.}
\label{fig:Delta_varphi_versus_varphi_alpha1000_beta800_noPMC3}
\end{figure}

This state of fact does not alter for high values of the coherent amplitudes. In Fig.~\ref{fig:F_sqz-coh_sqz-coh_versus_beta_alpha_500}, the QFI is depicted for three input phase-matching scenarios. There is no qualitative difference between the behavior in this case and in the one depicted in Fig.~\ref{fig:F_sqz-coh_sqz-coh_versus_beta_alpha_4}. The remarkable dip in the Fisher information corresponding to (PMC3) for  $\vert\beta\vert\approx\vert\alpha\vert$ can be explained using the double-coherent scenario discussed in references \citep{API18} and \cite{Pre19}. Indeed, since $\{\vert\alpha\vert,\vert\beta\vert\}\gg\{\sinh{r},\sinh{z}\}$, we can approximate this situation with a double coherent input with PMC $\theta_\alpha-\theta_\beta=\pi/2$, yielding the minimal QFI (see \emph{e. g.} equation (12) in reference \cite{Pre19}).

We are able now to briefly discuss the case $\vert\alpha\vert^{2\leftrightarrow3}_\textrm{lim}>\vert\alpha\vert^{1\leftrightarrow3}_\textrm{lim}$. Indeed, considering the inset of Fig.~\ref{fig:alpha_LIM_beta_LIM_sqzcoh_sqzcoh}, for $\vert\alpha\vert\in[\vert\alpha\vert^{1\leftrightarrow3}_\textrm{lim},\vert\alpha\vert^{2\leftrightarrow3}_\textrm{lim}]$ one can see that instead of having the optimal (PMC3) for $\vert\beta\vert\leq\vert\beta\vert_\textrm{lim}^{2\leftrightarrow3}$, we have (PMC1) optimal for $\vert\beta\vert\leq\vert\beta\vert_\textrm{lim}^{1\leftrightarrow3}$ and (PMC3) for $\vert\beta\vert>\vert\beta\vert_\textrm{lim}^{1\leftrightarrow3}$. The rest of the discussion does not change.

To conclude, in a low-coherent scenario (i. e. when  $\{\vert\alpha\vert,\vert\beta\vert\}\ll\{\sinh{r},\sinh{z}\}$), (PMC3) yields the optimum QFI. Intuitively this can be explained by the fact that the most important term  from equation \eqref{eq:app:F_dd_sqz-coh_plus_sqz_coh_FINAL_FORM} in this regime is the third one. (PMC3) ensures that it is maximized and it manages to maximize the other two terms. This happens however with the price of having $\mathcal{F}_{sd}\neq0$. Why (PMC3) is still optimal for $\vert\alpha\vert$ small but $\vert\beta\vert$ arbitrarily large can be explained by rewriting the phase-matching conditions as $2\theta_\beta-\phi=0$ and $\theta-\phi=\pm\pi$ ($\vert\alpha\vert$ being small, $\theta_\alpha$ is disregarded). We recognize here the squeezed-coherent plus squeezed vacuum scenario discussed in Section \ref{sec:sqz_coh_plus_squeezed_vacuum}, however with the input ports inverted.

As $\vert\alpha\vert$ grows there is a transition regime with various interplays between (PMC1) and (PMC3) for low $\vert\beta\vert$. For  high $\vert\beta\vert$,  (PMC2) is optimal. 

In the high $\vert\alpha\vert$ regime (i. e. for $\vert\alpha\vert\geq\vert\alpha\vert^\circ_\textrm{lim}$), for very low $\vert\beta\vert$, (PMC1) shortly dominates but as $\vert\beta\vert$ increases (PMC2) takes over.

% ---------------------------------------------------------
% ---------------- DIFFERENTIAL DETECTION -----------------
% ---------------------------------------------------------
\subsection{Difference intensity detection scheme}
\label{subsec:SqueezedCoherent_SqueezedCoherent_differential_det}
From equation \eqref{eq:Nd_average} and using the input state \eqref{eq:psi_in_squeezed_coherent_plus_squeezed_coherent} we have the average of the observable $\hat{N}_d$,
\begin{eqnarray}
\label{eq:Nd_average_sqz-coherent_sqz-coh}
\langle\hat{N}_d\rangle=\cos\varphi\left(\vert\alpha\vert^2-\vert\beta\vert^2+\sinh^2z-\sinh^2r\right)
\nonumber\\
-2\sin\varphi\vert\alpha\beta\vert\cos(\theta_\alpha-\theta_\beta)
\end{eqnarray}
The variance of $\hat{N}_d$ has been calculated in Appendix \ref{sec:app:sqz_coh_plus_sqz_sqz_coh}, equation~\eqref{eq:VARIANCE_N_d_FINAL_sqz_coh_sqz_coh_02_Upsilon} and the phase sensitivity is given by equation \eqref{eq:app:Delta_varphi_sqz-coh_plus_sqz-coh_Diff_det}.

Similar to the previous sections an optimal total internal phase shift $\Delta\tilde{\varphi}_\textrm{df}$ can be computed and is formally given in equation \eqref{eq:app:_delta_varphi_OPT_sqz-coh_plus_sqz_coh_Diff_det}.

% ---------------------------------------------------------
% ----------------- SINGLE DETECTOR -----------------------
% ---------------------------------------------------------
\subsection{Single-mode intensity detection scheme}
\label{subsec:Sqz_coherent_plus_Sqz_coherent_single_det}
The average value of the operator $\hat{N}_4$ with the input state \eqref{eq:psi_in_squeezed_coherent_plus_squeezed_coherent} is found to be
\begin{eqnarray}
\label{eq:N_4_average_coh-sqz_plus_coh-sqz}
\langle\hat{N}_4\rangle=\sin^2\left(\frac{\varphi}{2}\right)\left(\vert\beta\vert^2+\sinh^2r\right)
+\cos^2\left(\frac{\varphi}{2}\right)\left(\vert\alpha\vert^2+\sinh^2z\right)
%\qquad\qquad
\nonumber\\
-\sin\varphi\vert\alpha\beta\vert\cos\left(\theta_\alpha-\theta_\beta\right)
\qquad
\end{eqnarray}
and the absolute value of its derivative w.r.t. $\varphi$ is
\begin{eqnarray}
\label{eq:N4_derivative_average_coh-sqz_plus_coh-sqz}
\bigg\vert\frac{\partial\langle\hat{N}_4\rangle}{\partial\varphi}\bigg\vert
=\Big\vert\frac{1}{2}\sin\varphi\left(\vert\alpha\vert^2+\sinh^2z-\vert\beta\vert^2-\sinh^2r\right)
\nonumber\\
+\cos\varphi\vert\alpha\beta\vert\cos\left(\theta_\alpha-\theta_\beta\right)\Big\vert
\quad
\end{eqnarray}
The variance $\Delta^2\hat{N}_4$ is calculated and given in equation \eqref{eq:app:Variance_N_4_sqz_coh_sqz_coh_FINAL}. The phase sensitivity $\Delta\varphi$ is also computed and given by equation \eqref{eq:app:Delta_varphi_sqz-coh_plus_sqz-coh_Sing_det}.

% --------------------- FIGURE ------------ HIGH COHERENT versus VARPHI --
\begin{figure}
\includegraphics[scale=0.48]{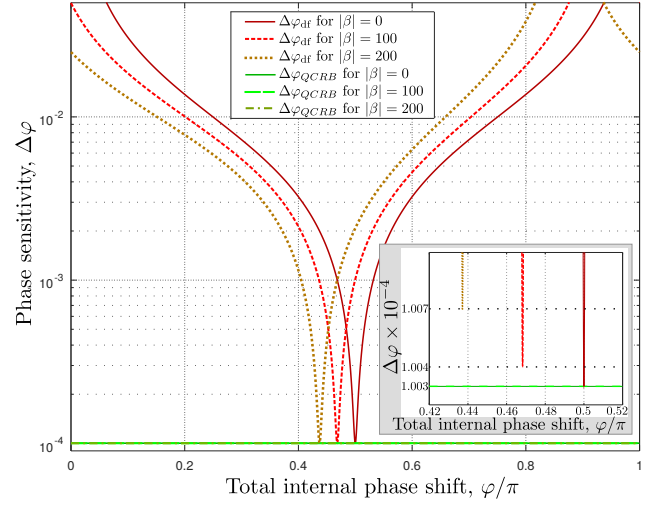}
\caption{Phase sensitivity for $\vert\alpha\vert=10^3$ and a small $\vert\beta\vert$. The (PMC1) constraints are used. Inset: extreme zoom in the region $\varphi\in[0.42\pi, 0.52\pi]$. Parameters used: $r=2.3$, $z=2.2$.}
\label{fig:Delta_varphi_versus_varphi_sqz-coh_sqz-coh_alpha1000_beta100_phi_pi}
\end{figure}

% ---------------------------------------------------------
% ----------------- HOMODYNE DETECTION --------------------
% ---------------------------------------------------------
\subsection{Homodyne detection scheme}
\label{subsec:Sqz_coherent_plus_Sqz_coherent_homodyne_det}
Using equation \eqref{eq:Homodyne_del_X_del_varphi} and setting again $\phi_L-\theta_\alpha=0$ we have
\begin{equation}
\Big\vert\frac{\partial}{\partial\varphi}\langle\hat{X}_{\phi_L}\rangle\Big\vert
=\frac{1}{2}\bigg\vert\cos\left(\frac{\varphi}{2}\right)\vert\beta\vert\cos(\theta_\beta-\theta_\alpha)
+\sin\left(\frac{\varphi}{2}\right)\vert\alpha\vert\bigg\vert
\end{equation}
The variance of $\hat{X}_{\phi_L}$ is computed using equation \eqref{eq:Homodyne_Variance_GENERIC} and yields the same result from equation \eqref{eq:Homodyne_variance_sqz_coh_plus_sqz_vac}. The phase sensitivity is thus given by
\begin{equation}
\label{eq:Delta_varphi_sqz_coh_sqz_coh_Homodyne}
\Delta\varphi_\mathrm{hom}
=\frac{\sqrt{\cot^2\left(\frac{\varphi}{2}\right)\Upsilon^-\left(\alpha,\zeta\right)+\Upsilon^-\left(\alpha,\xi\right)}}
{\vert\alpha\vert\big\vert\cot\left(\frac{\varphi}{2}\right)\vert\beta\vert\cos(\theta_\beta-\theta_\alpha)
+\vert\alpha\vert\big\vert}
\end{equation}
The optimum working point is found to be
\begin{equation}
\label{eq:varphi_OPT_homodyne_sqz_coh_plus_sqz_coh}
\varphi_\mathrm{opt}
=2\arctan\left(\frac{\vert\alpha\vert\Upsilon^-\left(\alpha,\zeta\right)}{\vert\beta\vert\cos(\theta_\beta-\theta_\alpha)\Upsilon^-\left(\alpha,\xi\right)}\right)
\end{equation}
yielding the best phase sensitivity
\begin{equation}
\label{eq:Delta_varphi_sqz_coh_sqz_coh_Homodyne_BEST_general}
\Delta\tilde{\varphi}_\mathrm{hom}
=\frac{\sqrt{\Upsilon^-\left(\alpha,\xi\right)\Upsilon^-\left(\alpha,\zeta\right)}}
{\vert\alpha\vert\sqrt{\vert\beta\vert^2\cos^2(\theta_\alpha-\theta_\beta)\Upsilon^-\left(\alpha,\xi\right)
+\vert\alpha\vert^2\Upsilon^-\left(\alpha,\zeta\right)}}
\end{equation}
Assuming now that we are in the high-coherent regime, we impose (PMC2) from equation \eqref{eq:phase_matching_array_0_0_0} and find 
%the optimum angle $\varphi_\textrm{opt}=2\arctan\left(e^{2(r-z)}\vert\alpha\vert/\vert\beta\vert\right)$ yielding 
the optimal phase sensitivity
\begin{equation}
\label{eq:Delta_varphi_sqz_coh_sqz_coh_Homodyne_BEST}
\Delta\tilde{\varphi}_\mathrm{hom}
=\frac{1}
{\sqrt{\vert\alpha\vert^2e^{2r}+\vert\beta\vert^2e^{2z}}}.
\end{equation}

% ---------------------------------------------------------
% ---------------------- DISCUSSION -----------------------
% ---------------------------------------------------------
\subsection{Discussion}

% ------------------- FIGURE ---------- LOSSES --- PMC3 -----
\begin{figure}
\centering
\includegraphics[scale=0.48]{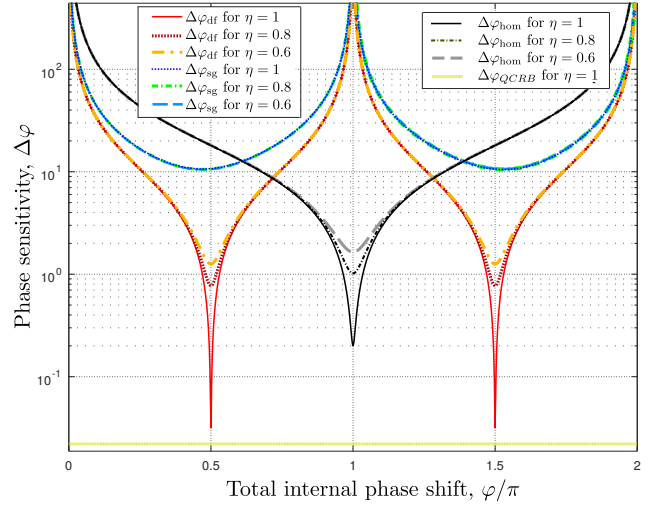}
\caption{The effect of non-unit photo-detection efficiency on the phase sensitivity for a squeezed-coherent plus squeezed-coherent input. A sensible degradation in performance is noticeable especially at the peaks of sensitivity. Being in the low-coherent region, (PMC3) is employed. Parameters used: $\vert\alpha\vert=0.5$, $\vert\beta\vert=0.25$, $r=2.3$, and $z=2.2$.}
\label{fig:Delta_varphi_versus_varphi_alpha05_beta025_LOSSES}
\end{figure}

% ---------------- HEISENBERG SCALING -----------------------
\subsubsection{Analysis of the obtained results}

In Section \ref{subsec:sqz_coherent_sqz_coh_input_Fisher} we concluded that the phase-matching condition (PMC3) given by equation \eqref{eq:phase_matching_array_0_0_pi_over_2} yields the maximum QFI in the low coherent amplitude regime. We depict this scenario in Fig.~\ref{fig:Delta_varphi_versus_varphi_alpha1_beta05} both for the QCRB and realistic detection schemes. Indeed, one notes that the best performance for a difference-intensity detection scheme is obtained using (PMC3) from equation \eqref{eq:phase_matching_array_0_0_pi_over_2}. For a single-mode intensity detection scheme, though, all input phase-matching conditions yield poor results. Equally noteworthy is the substantial sub-optimality of the homodyne detection scheme  w.r.t. the QCRB.

In Section \ref{subsec:sqz_coherent_sqz_coh_input_Fisher} we concluded, too, that the best performance in the high-coherent regime is obtained by employing (PMC2) from equation \eqref{eq:phase_matching_array_0_0_0}. This also applies for realistic detection schemes, as depicted in Fig.~\ref{fig:Delta_varphi_versus_varphi_alpha1000_beta800_noPMC3}. This time, the best performance is given by the homodyne detection technique. Noteworthy, each detection scheme yields its best sensitivity at a different optimal phase shift, $\varphi_\mathrm{opt}$.

From equations \eqref{eq:Fisher_sqzcoh_sqz_coh_PMC_theta_0_phi_0_thetabeta_0} and \eqref{eq:Delta_varphi_sqz_coh_sqz_coh_Homodyne_BEST} we see that in the high-coherent regime, the homodyne can actually reach the QCRB if $r=z$ and we have
\begin{equation}
\label{eq:Delta_varphi_sqz_coh_sqz_coh_Homodyne_BEST_r_equal_z}
\Delta\tilde{\varphi}_\mathrm{hom}
=\frac{e^{-r}}
{\sqrt{\vert\alpha\vert^2+\vert\beta\vert^2}}=\Delta\varphi_{QCRB}.
\end{equation}
We conclude that for (PMC2) the optimal input state is the one with equal squeezing factors in both input ports.

We ask now the question: could we have an experimental advantage if we start from a squeezed-coherent plus squeezed vacuum input as discussed in Section \ref{sec:sqz_coh_plus_squeezed_vacuum} and add some limited displacement to the squeezed vacuum from port $1$ (i. e. we have $\vert\beta\vert\ll\vert\alpha\vert$)? The answer is affirmative and we depict this scenario in Fig.~\ref{fig:Delta_varphi_versus_varphi_sqz-coh_sqz-coh_alpha1000_beta100_phi_pi}. Indeed, for $\beta=0$ we find the result discussed in Section \ref{sec:sqz_coh_plus_squeezed_vacuum} and depicted in Fig.~\ref{fig:Delta_phi_coh_sqz_sqz_vac_vs_theta_alpha1k_phi_0_pi} (solid red curve). The difference-intensity detection scenario has a very peaked optimum at $\varphi_\textrm{opt}=\pi/2$. As $\vert\beta\vert$ starts to grow, the shape of the phase sensitivity is simply translated. Therefore, instead of having a very good phase sensitivity only around $\varphi\approx\pi/2$, we can scan other internal phase shifts by simply manipulating  $\vert\beta\vert$. Please note the we are in the (PMC1) regime and we assume $\vert\beta\vert\ll\vert\alpha\vert$. The addition of the second coherent source negligibly degrades the performance, as seen in the inset of Fig.~\ref{fig:Delta_varphi_versus_varphi_sqz-coh_sqz-coh_alpha1000_beta100_phi_pi}.

%----------------------------------------------------------
% --------------------- EFFECT LOSSES ---------------------
\subsubsection{Non-unit photo detection efficiency}

In this section we use results from Appendix \ref{sec:app:LOSSES} as well as equation \eqref{eq:delta_varphi_LOSSY}.

In Fig.~\ref{fig:Delta_varphi_versus_varphi_alpha05_beta025_LOSSES} we single out (PMC3) from Fig.~\ref{fig:Delta_varphi_versus_varphi_alpha1_beta05} and evaluate the effect of non-unit photo-detection efficiencies. The most noticeable effect is the swift degradation of the peaks of sensitivity for the difference-intensity and homodyne detection schemes. The single-mode intensity detection performance is the least affected by the effect of losses, however this is also due to the poor performance of this detection strategy, given the parameters used in Fig.~\ref{fig:Delta_varphi_versus_varphi_alpha05_beta025_LOSSES}.

\begin{figure}
\centering
\includegraphics[scale=0.48]{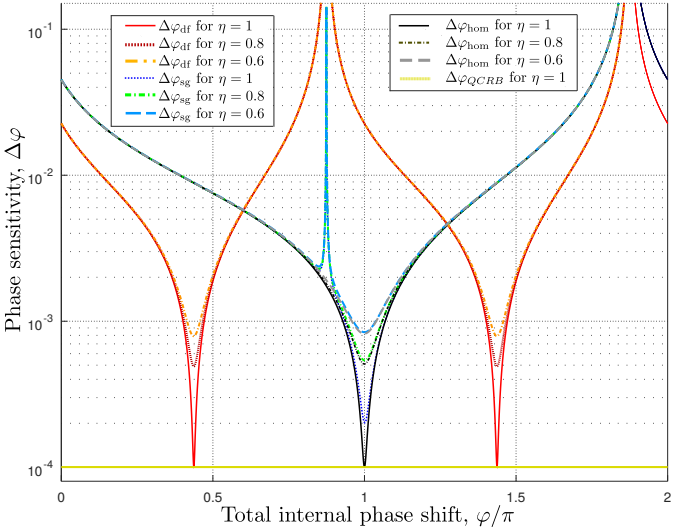}
\caption{The effect of non-unit photo-detection efficiency on the phase sensitivity for a squeezed-coherent plus squeezed-coherent input. A sensible degradation in performance is found for all realistic detection schemes around their respective peak performance. (PMC1) is employed. Parameters used: $\vert\alpha\vert=10^3$, $\vert\beta\vert=200$, $r=2.3$, and $z=2.2$.}
\label{fig:Delta_varphi_versus_varphi_alpha1000_beta200_LOSSES}
\end{figure}

In Fig.~\ref{fig:Delta_varphi_versus_varphi_alpha1000_beta200_LOSSES} a high-$\vert\alpha\vert$  scenario employing (PMC1) is depicted (see also Fig.~\ref{fig:Delta_varphi_versus_varphi_sqz-coh_sqz-coh_alpha1000_beta100_phi_pi} for the loss-less case). This time the effect of non-unit photo-detection efficiency is noticeable for all realistic detection schemes, with a remark similar to the one from Section \ref{subsec:Sqz_coherent_plus_squeezedVac_DISCUSSION} namely that the respective sensitivity peaks are the most impacted by the losses.

% ------------------- FIGURE ---------- LOSSES --- PMC1 -----

% -------------- PHASE-SPACE REPRESENTATION ----------------
\subsubsection{A physical insight on the obtained phase-matching conditions}

We give now some physical insights on the obtained results. For (PMC1) we point the reader to the discussion from Section \ref{subsec:Sqz_coherent_plus_squeezedVac_DISCUSSION}. As mentioned before, $\beta$ is mainly a degrading factor of the overall performance, thus its interest lies only in the regime $\vert\beta\vert\ll\vert\alpha\vert$.

\begin{figure}
\includegraphics[scale=0.7]{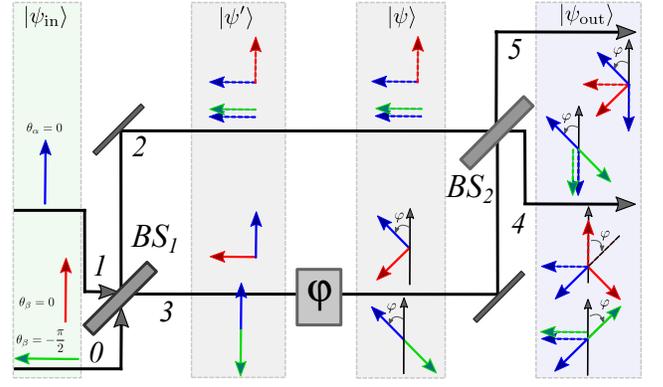}
\caption{A phasor representation of a double-coherent input MZI in two scenarios: (i) both coherent inputs are in phase (red and blue arrows) and (ii) they are $\pi/2$ phase shifted (red and magenta arrows). For equal input amplitudes and (ii), ones notes that the outputs do not depend anymore on the angle $\varphi$.}
\label{fig:BS_MZI_phase_space_phasors}
\end{figure}

The experimentally interesting high-coherent regime setup relies on (PMC2). In references \cite{API18,Pre19} it has been shown that maximum performance from a dual coherent input implies $\theta_\alpha=\theta_\beta$ (see also Fig.~\ref{fig:BS_MZI_phase_space_phasors} for a graphical representation). In our setup, we have two extra squeezings, one in each port. At a careful look, in the case of (PMC2) we actually have twice a coherent plus squeezed vacuum input, namely $\alpha-\xi$ and $\beta-\zeta$. The optimal PMC for each one implies a relation of the type \eqref{eq:phase_matching_cond_coh_plus_sqz_vac}. Indeed, setting $\alpha=\beta$ and $\xi=\zeta$ we get $\mathcal{F}=2\vert\alpha\vert^2e^{2r}$ which is twice the Fisher information for the coherent plus squeezed vacuum input (in the high-$\vert\alpha\vert$ approximation). These PMCs do minimize the term $\Upsilon^-\left(\alpha,\xi\right)$ from the homodyne sensitivity \eqref{eq:Delta_varphi_sqz_coh_sqz_coh_Homodyne} and the term $\Upsilon^-\left(\beta,\zeta\right)$ from the difference-intensity detection sensitivity \eqref{eq:VARIANCE_N_d_FINAL_sqz_coh_sqz_coh_02_Upsilon}. A supplementary justification for the QCRB-optimality of this state can be found in the work of Hofmann (``path-symmetric states can achieve their quantum Cramer-Rao bound'') \cite{Hof09}.

One can argue that in (PMC2) we have $\xi=\zeta$ instead of $\xi=-\zeta$, thus a sub-optimality should be expected from this scheme (see also the discussion from Appendix \ref{sec:app:optimization_two_squeezers}). The argument is valid, however, (PMC2) is a high-coherent scheme i. e. $\{\vert\alpha\vert^2,\:\vert\beta\vert^2\}\gg\{\sinh^2r,\:\sinh^2z\}$, thus the non-optimality from the interaction of the squeezed vacuums should be marginal.

% --------------------- FIGURE ------------

Finally, (PMC3) starts by insisting on the squeezers being in anti-phase. Assuming again $\zeta=-\xi$ we have the state after the beam splitter $\vert\psi'\rangle=\hat{D}_2((i\alpha+\beta/\sqrt{2})\hat{D}_3((\alpha+i\beta)/\sqrt{2})\hat{S}_2(\xi)_2\hat{S}_3(-\xi)\vert0\rangle$ (see Appendix \ref{sec:app:optimization_two_squeezers}). Since we have now $\theta_\alpha-\theta_\beta=\pi/2$ we obtain $\hat{D}_3((\alpha+i\beta)/\sqrt{2})=\hat{D}_3(e^{i\theta_\alpha}(\vert\alpha\vert-\vert\beta\vert)/\sqrt{2})$ with a total annihilation of the coherent amplitude the mode $3$ for $\vert\alpha\vert=\vert\beta\vert$. This phenomenon can be easily represented graphically, as depicted in Fig.~\ref{fig:BS_MZI_phase_space_phasors}. The arrows represent the two input coherent states ($\alpha$ and $\beta$) and for $\theta_\alpha-\theta_\beta=\pi/2$ the amplitude in mode $3$ inside the MZI destructively interferes while at the photo-detectors, none of the outputs depends on the phase $\varphi$ if $\vert\alpha\vert=\vert\beta\vert$.

One can use this argument to point to the sub-optimality of (PMC3). As remarked in the case of (PMC2), one must also pay attention where the given PMCs apply. The destructive interference of the  coherent sources inside the interferometer is a limited nuisance because in the case of (PMC3) we are in the low-coherent regime, $\{\vert\alpha\vert^2,\:\vert\beta\vert^2\}\ll\{\sinh^2r,\:\sinh^2z\}$. We actually rely here on the optimality of the squeezed vacuums. We can approximate the wavefunction after $BS_1$ with  $\vert\psi'\rangle\approx\hat{S}_2(\xi)_2\hat{S}_3(-\xi)\vert0\rangle$ and, as discussed before, this is an optimal state \cite{Lan14}.

% ---------------------------------------------------------
% ---------------- HEISENBERG SCALING ---------------------
\subsubsection{Heisenberg scaling}
We end this work by investigating the Heisenberg scaling \eqref{eq:delta_varphi_HL_coh_sqz_vac} in the case of an input state given by equation \eqref{eq:psi_in_squeezed_coherent_plus_squeezed_coherent}. The total average number of input photons in this scenario is $\langle{N}_\textrm{tot}\rangle=\vert\alpha\vert^2+\vert\beta\vert^2+\sinh^2r+\sinh^2z$. We define $f_\alpha=\vert\alpha\vert^2/\langle{N}_\textrm{tot}\rangle$, $f_\beta=\vert\beta\vert^2/\langle{N}_\textrm{tot}\rangle$, $f_r=\sinh^2r/\langle{N}_\textrm{tot}\rangle$ and $f_z=\sinh^2z/\langle{N}_\textrm{tot}\rangle$. We again assume $\{\vert\alpha\vert^2,\:\vert\beta\vert^2,\:\sinh^2r,\:\sinh^2z\}\gg1$. Since we did not in any way specify the relation among the squeezing factors and the coherent amplitudes, any among the phase-matching conditions discussed before could be optimal. We thus discuss them all and first consider (PMC1) and the QFI given by equation \eqref{eq:Fisher_sqzcoh_sqz_coh_PMC_theta_0_phi_pi_thetabeta_0}. We find $\mathcal{F}\approx4\langle{N}_\textrm{tot}\rangle^2f_r(f_\alpha+f_z)$, a result formally identical to the one obtained in Section \ref{subsec:Sqz_coherent_plus_squeezedVac_DISCUSSION}. However, $\langle{N}_\textrm{tot}\rangle$ is now different and rewriting $\mathcal{F}\approx4\langle{N}_\textrm{tot}\rangle^2f_r(1-f_r-f_\beta)$ we see that the optimum implies $f_\beta\to0$ and $f_r\to1/2$. This scenario does not exclude the constraint $f_\alpha\to0$, thus also two squeezed vacuums can yield the scaling \eqref{eq:delta_varphi_HL_coh_sqz_vac}. By two ``squeezed vacuums'' here and in the following discussion we mean $\{\vert\alpha\vert^2,\:\vert\beta\vert^2\}\ll\{\sinh^2r,\:\sinh^2r\}$ thus, although we assumed $\{\vert\alpha\vert^2,\:\vert\beta\vert^2,\:\sinh^2r,\:\sinh^2z\}\gg1$ we can safely approximate our input state with two squeezed vacuums.

For (PMC2) we have the QFI from equation \eqref{eq:Fisher_sqzcoh_sqz_coh_PMC_theta_0_phi_0_thetabeta_0}. Assuming again $\{\vert\alpha\vert^2,\:\vert\beta\vert^2,\:\sinh^2r,\:\sinh^2z\}\gg1$, we get 
\begin{equation}
\label{eq:Fisher_Heisenberg_limit_csqz_csqz_PMC2}
\mathcal{F}\approx4\langle{N}_\textrm{tot}\rangle^2\left(f_\alpha f_r+f_\beta f_z\right).
\end{equation}
The optimum solution $\mathcal{F}\approx\langle{N}_\textrm{tot}\rangle^2$ is obtained in two scenarios: if $f_r\to1/2$, $f_z\to0$ and $f_\alpha\to1/2$ (thus $f_\beta\to0$)  or if $f_r\to0$, $f_z\to1/2$ and $f_\beta\to1/2$ (thus $f_\alpha\to0$). This time the optimum involves either a coherent source in port $1$ and squeezed vacuum in port $0$ or a coherent source in port $0$ and squeezed vacuum in port $1$, thus excluding the two squeezed vacuums scenario from (PMC1). Why this is so, boils down to the insistence of (PMC2) on having the constraint $\theta-\phi=0$. 

For (PMC3) we have the QFI from equation \eqref{eq:Fisher_sqzcoh_sqz_coh_PMC_theta_0_phi_pi_thetabeta_pi2}. Assuming again $\{\vert\alpha\vert^2,\:\vert\beta\vert^2,\:\sinh^2r,\:\sinh^2z\}\gg1$, we get
\begin{eqnarray}
\label{eq:Fisher_Heisenberg_limit_csqz_csqz_PMC3}
\mathcal{F}\approx4\langle{N}_\textrm{tot}\rangle^2
%\nonumber\\
%\times
\Bigg(f_\alpha f_r+f_\beta f_z
+f_rf_z
\qquad\qquad\qquad
% ---------
\nonumber\\
\qquad\qquad
-\frac{f_\alpha f_\beta(f_r+f_z)^2}
{\frac{f_r^2}{2}
+\frac{f_z^2}{2}
+f_\alpha f_z
+f_\beta f_r}
\Bigg)
\end{eqnarray}
Somehow not surprisingly, the optimum $\mathcal{F}\approx\langle{N}_\textrm{tot}\rangle^2$ is obtained for $f_r=f_z\to1/2$, thus the best scaling is obtained with two squeezed vacuums.

In conclusion, if we impose (PMC1) and $\vert\beta\vert\neq0$, for a Heisenberg scaling the optimum involves (i) either a squeezed vacuum having half the input power in one port and a squeezed coherent state  in the other one or (ii) two squeezed vacuums and no coherent sources. This last result can be put in relation with a similar findings reported in the literature \cite{Pin12,Lan14}. If we impose (PMC2) the optimal Heisenberg scaling is obtained by applying a coherent plus squeezed vacuum input. This result agrees with the conclusions from reference \cite{Lan13}. Finally, imposing (PMC3) and a Heisenberg scaling takes us again to the solution involving two squeezed vacuums.

We can also compare our results with the ones reported in refs. \cite{Spa15,Spa16}. For a squeezed-coherent plus squeezed coherent input state, Sparaciari, Olivares \& Paris found the optimum QFI when $f_r=f_z\approx1/3$ and $f_\alpha=f_\beta$ yielding a scaling $\mathcal{F}=8/9\langle{N}_\textrm{tot}\rangle^2$. If we introduce  these values in equation \eqref{eq:Fisher_Heisenberg_limit_csqz_csqz_PMC2} we find $\mathcal{F}=4/9\langle{N}_\textrm{tot}\rangle^2$. This discrepancy should come as no surprise: while in references \cite{Spa15,Spa16} the authors started from a single-parameter Fisher information, we started from a two-parameter Fisher information approach.

% ---------------------------------------------------------
% ---------------- IMPACT of BS types ------------------
% ---------------------------------------------------------
\section{The impact of the BS type employed on the phase-matching conditions}
\label{sec:impact_BS_on_PMC}
Up to this point we discussed the field operator transformations \eqref{eq:field_op_transf_BS_sym} characterizing a so-called balanced symmetrical or thin-film BS \cite{GerryKnight}. If we introduce the Jordan-Schwinger angular momentum operators \cite{Sch65,Yur86,Dem15} $\hat{J}_x=(\hat{a}_1^\dagger\hat{a}_2+\hat{a}_1\hat{a}_2^\dagger)/2$, $\hat{J}_y=(\hat{a}_1^\dagger\hat{a}_2-\hat{a}_1\hat{a}_2^\dagger)/2i$ and $\hat{J}_z=(\hat{a}_1^\dagger\hat{a}_1-\hat{a}_2^\dagger\hat{a}_2)/2$, the transformation from equation \eqref{eq:field_op_transf_MZI} corresponds to the unitary transformation $\hat{U}_x=e^{i\pi/2\hat{J}_x}$. For example we have $\hat{a}_2=\hat{U}_x^\dagger\hat{a}_0\hat{U}_x=1/\sqrt{2}\hat{a}_0+i/\sqrt{2}\hat{a}_1$. The same initial convention \eqref{eq:field_op_transf_BS_sym}  determines the QFI matrix elements calculated in Appendix \ref{sec:app:Fisher_information} and also the output operator transformations from equation \eqref{eq:field_op_transf_MZI} leading to the observables for the realistic schemes discussed in Appendix \ref{sec:app:variance_calculation}.

However, there are non-symmetric beam splitters (usually called ``cube beam splitters'') that are described by the field operator transformations \cite{Yur86}
\begin{equation}
\label{eq:field_op_transf_MZI_balanced_CUBE_a}
\left\{
\begin{array}{l}
\hat{a}_3=\frac{1}{\sqrt{2}}\left({\hat{a}_1}-{\hat{a}_0}\right)\\
\hat{a}_2=\frac{1}{\sqrt{2}}\left({\hat{a}_1}+{\hat{a}_0}\right)
\end{array}
\right.
\end{equation}
These field operator transformations correspond to the unitary operator $\hat{U}_y=e^{i\pi/2\hat{J}_y}$.  They imply the input-output field operator transformations
\begin{equation}
\label{eq:n4_operator_DEFINITION_BALANCED_cube_BS}
\hat{n}_4
=\cos^2\left(\frac{\varphi}{2}\right){\hat{n}_1}
+\sin^2\left(\frac{\varphi}{2}\right){\hat{n}_0}
+\sin\varphi\Im\left({\hat{a}_1}{\hat{a}_0^\dagger}\right)
\end{equation}
where $\Im$ denotes the imaginary part and
\begin{equation}
\label{eq:N_d_operator_DEFINITION_BALANCED_cube_BS}
\hat{N}_d
=\cos\varphi\left({\hat{n}_1}-{\hat{n}_0}\right)
+2\sin\varphi\Im\left({\hat{a}_1}{\hat{a}_0^\dagger}
\right)
\end{equation}
If we compute now the Fisher matrix coefficients using equation \eqref{eq:field_op_transf_MZI_balanced_CUBE_a} we get $\mathcal{F}_{ss}=\Delta^2{\hat{n}_1}+\Delta^2{\hat{n}_0}$,
\begin{eqnarray}
\label{eq:F_dd_FINAL_FORM_balanced_CUBE_BS}
\mathcal{F}_{dd}=\langle{\hat{n}_1}\rangle
+\langle{\hat{n}_0}\rangle
+2\left(\langle{\hat{n}_0}\rangle\langle{\hat{n}_1}\rangle
-\vert\langle{\hat{a}_0}\rangle\vert^2\vert\langle{\hat{a}_1}\rangle\vert^2\right)
\nonumber\\
+2\Re\left\{\langle{\hat{a}_0^2}\rangle\langle{(\hat{a}_1^\dagger)^2}\rangle
-\langle{\hat{a}_0}\rangle^2\langle{\hat{a}_1^\dagger}\rangle^2
\right\}
% ----------------------------
\end{eqnarray}
and the third Fisher matrix coefficient is
\begin{eqnarray}
\label{eq:F_sd_FINAL_FORM_balanced_CUBE_BS_compact}
\mathcal{F}_{sd}=2\Re\left\{\langle{\hat{a}_0}\rangle\langle{\hat{a}_1^\dagger}\rangle
+\langle{\hat{a}_0}\rangle\left(\langle{\hat{a}_1^\dagger}{\hat{n}_1}\rangle
-\langle{\hat{a}_1^\dagger}\rangle\langle{\hat{n}_1}\rangle\right)
\right.
\nonumber\\
\left.
+\left(\langle{\hat{n}_0}{\hat{a}_0}\rangle-\langle{\hat{a}_0}\rangle\langle{\hat{n}_0}\rangle\right)\langle{\hat{a}_1^\dagger}\rangle
\right\}.
\end{eqnarray}
Please note that in the calculation of the Fisher matrix element $\mathcal{F}_{ss}$ the new expressions for the output number operators have to be used e. g. equation \eqref{eq:n4_operator_DEFINITION_BALANCED_cube_BS} etc.
The optimum QFI as well as the best performance for realistic detection scenarios remain unchanged, however a new assessment of the input PMCs has to be done.

For example, if we consider the squeezed-coherent plus squeezed vacuum input state from equation \eqref{eq:psi_in_squeezed_coherent_plus_squeezed_vac} and a BS characterized by the transformation \eqref{eq:field_op_transf_MZI_balanced_CUBE_a} we obtain the QFI (we recall that is in this scenario $\mathcal{F}=\mathcal{F}_{dd}$)
\begin{eqnarray}
\label{eq:F_dd_coh_sqz_plus_sqz_vac_cube_BS}
\mathcal{F}=
\Upsilon^-\left({\alpha},{\xi}\right)
\qquad\qquad\qquad\qquad\qquad\qquad\qquad\qquad\quad
\nonumber\\
+\frac{\cosh2r\cosh2z+\sinh2r\sinh2z\cos(\theta-\phi)-1}{2}
\quad
% ----------------------------
\end{eqnarray}
This time, contrary to the PMCs given by equations \eqref{eq:phase_matching_cond_coh_plus_sqz_vac} and \eqref{eq:phase_matching_cond_sqz_coh_plus_sqz_vac} we find the optimal QFI 
\eqref{eq:Fisher_sqz-coh_plus_sqz_vac_MAXIMAL} if 
\begin{equation}
\label{eq:phase_matching_VECTOR_sqz_coh_plus_sqz_vac_cube_BS}
\left\{
\begin{array}{l}
2\theta_\alpha-\theta=\pm\pi\\
\theta-\phi=0
\end{array}
\right.
\end{equation}
In a similar manner, all results discussed in Sections \ref{sec:sqz_coh_plus_squeezed_vacuum} and \ref{sec:sqz_coh_plus_sqz_coh} can be rederived.

The physical origin of the sign change (wrt the previous sections) in all terms involving fields from both inputs is easy to explain: while the field operator transformations from equation \eqref{eq:field_op_transf_BS_sym} describe a symmetrical BS, the ones from equation \eqref{eq:field_op_transf_MZI_balanced_CUBE_a} do not. Indeed, in a cube beam splitter one mode propagates without phase shifts, while for the second one the reflection acquires a phase delay of $\pi$. One can also redraw the arrows from Fig.~\ref{fig:BS_MZI_phase_space_phasors} by following the rules of the field operator transformations \eqref{eq:field_op_transf_MZI_balanced_CUBE_a} and convince itself of the new optimal PMCs with a cube type BS.

% ---------------------------------------------------------
% ---------------- C O N C L U S I O N S ------------------
% ---------------------------------------------------------
\section{Conclusions}
\label{sec:conclusions}
In this paper we investigated the phase sensitivity of a Mach-Zehnder interferometer fed with the most general Gaussian input states. Both the theoretical quantum Cram\'er-Rao bound and realistic performances were assessed.

The squeezed-coherent plus squeezed vacuum input state scenario yielded unambiguous phase-matching conditions for a theoretical maximum performance. If the phase of the coherent source is taken to be zero, then the squeezing from the opposite input port has to be zero, too. However, the second squeezer must be in anti-phase. We also showed that a second scenario is possible, when all input phases are zero. Although slightly sub-optimal, this scenario has a good sensitivity over a wide range of internal phase shifts.

The paper discussed in detail the rather complicated case of squeezed-coherent plus squeezed-coherent input. We found three input phase-matching scenarios, each optimal in a certain domain. In the low coherent intensity regime, we found that the optimal input phase-matching condition involves the two coherent sources being dephased by $\pi/2$ and the squeezers in anti-phase. In the high-coherent intensity regime, the optimal input phase matching conditions impose the coherent sources as well as the squeezers to be in phase (if the coherent phases are assumed to be zero).

Practical situations have been discussed with realistic detection schemes, where the addition of the second coherent source is able to bring an experimental advantage. We also showed that with the right phase matching conditions and with equal squeezing in both inputs, some realistic detection techniques are able to saturate the quantum Cram\'er-Rao bound.

% -------- LOSSES ------

When considering losses, all realistic detection schemes show a decrease in performance, the peak performance being the most affected. In most scenarios, the least impacted detection scheme in the lossy case is the homodyne detection. A more thorough investigation on the impact of the different types of losses on the interferometric phase sensitivity is postponed for a future work.

% ------- HEISENBERG SCALING -----

For all input states considered we showed that a Heisenberg scaling is possible. We also showed that in the case of a general Gaussian state, different PMCs lead to different input states that optimize the Heisenberg scaling, confirming and extending some previous results.

We also discussed the impact of the type of beam splitter used. We showed that although the optimal phase sensitivity is unaffected by the type of the beam splitter used, the input phase-matching conditions needed to attain this optimum do change.

\begin{acknowledgments}

The author would like to thank Ms. Anca Preda for interesting discussions, help in some calculations and for double-checking a number of results from this paper.

It is also acknowledged that this work has been supported by the Extreme Light Infrastructure Nuclear Physics (ELI-NP) Phase II, a project co-financed by the Romanian Government and the European Union through the European Regional Development Fund and the Competitiveness Operational Programme (1/07.07.2016, COP, ID 1334).

\end{acknowledgments}

% #########################################################
% ################        APPENDIX         ################
% #########################################################

\appendix

% ---------------------------------------------------------
% ------------ SECTION --- FISHER INFORMATION -------------
% ---------------------------------------------------------
\section{Fisher information}
\label{sec:app:Fisher_information}
Since we assume our input to be in a pure state, we do not need to use the Symmetric Logarithmic Derivative \cite{Dem15,Bra94,Par09} and the QFI is directly 
\begin{equation}
\mathcal{F}(\varphi)= 4\left(\langle\partial_\varphi\psi\vert\partial_\varphi\psi\rangle-\vert\langle\partial_\varphi\psi\vert\psi\rangle\vert^2\right),
\end{equation}
where ${\vert\partial_\varphi\psi\rangle=\partial\vert\psi\rangle/\partial\varphi}$ \cite{Par09,Lan13,Lan14, API18}. To give a basic example for readers unfamiliar with this notation, if the wavefunction is $\vert\psi\rangle=\cos\varphi\vert0\rangle+\sin\varphi\vert1\rangle$ then we have $\vert\partial_\varphi\psi\rangle=-\sin\varphi\vert0\rangle+\cos\varphi\vert1\rangle$.

We consider the general case where each arm of the MZI contains a phase-shift ($\varphi_1$ and, respectively, $\varphi_2$). The estimation is treated as a general two parameter problem \cite{Lan13,Lan14,API18,Jar12}. We define the $2\times2$ Fisher information matrix:
\begin{equation}
\label{eq:app: Fisher_matrix}
   \mathcal{F}=
  \left[ {\begin{array}{cc}
  \mathcal{F}_{ss} & \mathcal{F}_{sd} \\
   \mathcal{F}_{ds} & \mathcal{F}_{dd} \\
  \end{array} } \right]
\end{equation}
where 
\begin{equation}
\label{eq:app:Fisher_matrix_elements}
\mathcal{F}_{ij}=4\Re\{\langle\partial_i\psi\vert\partial_j\psi\rangle-\langle\partial_i\psi\vert\psi\rangle
 \langle\psi\vert\partial_j\psi\rangle\}
\end{equation} 
with $i, j\in \{s,d\}$ and ${\varphi_{s/d}=(\varphi_1\pm\varphi_2})/2$. From this matrix we arrive at the QCRB matrix inequality \cite{Lan13} out of which se retain only the difference-difference phase estimator,
\begin{equation}
(\Delta\varphi_d)^2\geq(\mathcal{F}^{-1})_{dd}
\end{equation}
Using the definition from \eqref{eq:app:Fisher_matrix_elements}, the sum-sum Fisher matrix element $\mathcal{F}_{ss}$ can be computed and yields
\begin{eqnarray}
\label{eq:app:F_ss_FINAL_FORM_GENERAL}
\mathcal{F}_{ss}
=\Delta^2\hat{n}_0+\Delta^2\hat{n}_1
\end{eqnarray}
Similarly the element $\mathcal{F}_{dd}$ is computed and yields
\begin{eqnarray}
\label{eq:app:F_dd_FINAL_FORM_GENERAL_balanced}
\mathcal{F}_{dd}
=\langle\hat{n}_1\rangle+\langle\hat{n}_0\rangle
+2\left(
\langle\hat{n}_0\rangle\langle\hat{n}_1\rangle
-\langle\hat{a}_0^\dagger\rangle\langle\hat{a}_0\rangle\langle\hat{a}_1^\dagger\rangle\langle\hat{a}_1\rangle\right)
\nonumber\\
% ---------------------------
%\nonumber\\
% ---------------------------
-\left(
\langle(\hat{a}_0^\dagger)^2\rangle\langle\hat{a}_1^2\rangle
+\langle\hat{a}_0^2\rangle\langle(\hat{a}_1^\dagger)^2\rangle
\right.
%\nonumber\\
\left.
-\langle\hat{a}_0^\dagger\rangle^2\langle\hat{a}_1\rangle^2
-\langle\hat{a}_0\rangle^2\langle\hat{a}_1^\dagger\rangle^2
\right)
\nonumber\\
% ----------------
=\langle\hat{n}_1\rangle+\langle\hat{n}_0\rangle
+2\left(
\langle\hat{n}_0\rangle\langle\hat{n}_1\rangle
-\vert\langle\hat{a}_0\rangle\vert^2\vert\langle\hat{a}_1\rangle\vert^2\right)
\qquad
\nonumber\\
% ---------------------------
% ---------------------------
-2\Re\left(
\langle\hat{a}_0^2\rangle\langle(\hat{a}_1^\dagger)^2\rangle
-\langle\hat{a}_0\rangle^2\langle\hat{a}_1^\dagger\rangle^2
\right)
\qquad
\end{eqnarray}
The last term we need is $\mathcal{F}_{sd}$ since $\mathcal{F}_{sd}=\mathcal{F}_{ds}$ \cite{Lan13}. We have
\begin{eqnarray}
\label{eq:app:F_sd_GENERIC_FINAL_balanced}
\mathcal{F}_{sd}
=i\left(
\langle\hat{a}_0^\dagger\rangle\langle\hat{a}_1\rangle
-\langle\hat{a}_0\rangle\langle\hat{a}_1^\dagger\rangle
\right)
\qquad\qquad\qquad\qquad\qquad\qquad
% ------------------- SECOND
\nonumber\\
+i\left(
\left(\langle\hat{a}_0^\dagger\hat{n}_0\rangle
-\langle\hat{a}_0^\dagger\rangle\langle\hat{n}_0\rangle\right)\langle\hat{a}_1\rangle
\right.
% ---------------------------- THIRD
%\nonumber\\
\left.
-\left(\langle\hat{n}_0\hat{a}_0\rangle
-\langle\hat{n}_0\rangle\langle\hat{a}_0\rangle\right)\langle\hat{a}_1^\dagger\rangle
\right)
% ---------------------------- THIRD
\nonumber\\
+i\left(
\langle\hat{a}_0^\dagger\rangle
\left(\langle\hat{n}_1\hat{a}_1\rangle
-\langle\hat{a}_1\rangle\langle\hat{n}_1\rangle\right)
\right.
% ---------------------------- THIRD
%\nonumber\\
\left.
-\langle\hat{a}_0\rangle
\left(\langle\hat{a}_1^\dagger\hat{n}_1\rangle
-\langle\hat{a}_1^\dagger\rangle
\langle\hat{n}_1\rangle\right)
\right)
\nonumber\\
% --------- written with Im
=2\Im\left(\langle\hat{a}_0\rangle\langle\hat{a}_1^\dagger\rangle
+\left(\langle\hat{n}_0\hat{a}_0\rangle
-\langle\hat{n}_0\rangle\langle\hat{a}_0\rangle\right)\langle\hat{a}_1^\dagger\rangle
\right.
\qquad\qquad\:\:\,
% ------------------- SECOND
\nonumber\\
\left.
+\langle\hat{a}_0\rangle
\left(\langle\hat{a}_1^\dagger\hat{n}_1\rangle
-\langle\hat{a}_1^\dagger\rangle
\langle\hat{n}_1\rangle\right)
\right)
\qquad
\end{eqnarray}

% ---------------------------------------------------------
% --- APPENDIX --- SECTION ---- VARIANCE COMPUTATION ------
% ---------------------------------------------------------

\section{Calculation of the output variances for the generic case}
\label{sec:app:variance_calculation}
In this appendix we compute the averages $\langle \hat{N}^2\rangle$ as well as the variances $\Delta^2\hat{N}$ for a generic input case.

% ---------------------------------------------------------
% --- DIFFERENCE DETECTION ---- VARIANCE COMPUTATION ------
% ---------------------------------------------------------
\subsection{Difference intensity detection}
\label{subsec:app:variance_calculation_diff_det}
For a difference intensity detection scheme, from eqs.~\eqref{eq:field_op_transf_MZI} and \eqref{eq:N_d_operator_DEFINITION} we obtain the expression of $\hat{N}_d^2$ as a function of input operators. After a long but straightforward calculation we obtain the final expression
\begin{eqnarray}
\label{eq:N_d_SQUARED_FINAL0}
% ---- SIMPLIFY 2
\langle\hat{N}_d^2\rangle
=\cos^2\varphi\left(\langle\hat{n}_0^2\rangle
+\langle\hat{n}_1^2\rangle\right)
-2\cos\left(2\varphi\right)\langle\hat{n}_0\hat{n}_1\rangle
% ------ comment below
%\qquad\quad
\nonumber\\
+\sin^2\varphi\left(
\langle\hat{n}_0\rangle
+\langle\hat{n}_1\rangle
+\langle\hat{a}_0^2(\hat{a}_1^\dagger)^2\rangle
+\langle(\hat{a}_0^\dagger)^2\hat{a}_1^2\rangle\right)
% ------
\nonumber\\
+\sin2\varphi\left(
\langle\hat{n}_0\hat{a}_0\hat{a}_1^\dagger\rangle
+\langle\hat{a}_0^\dagger\hat{n}_0\hat{a}_1\rangle
\right.
% -------------------------
\nonumber\\
\left.
-\langle\hat{a}_0\hat{a}_1^\dagger\hat{n}_1\rangle
-\langle\hat{a}_0^\dagger\hat{n}_1\hat{a}_1\rangle
\right)
%\qquad
\end{eqnarray}
Since we expressly assume that the input state is \emph{separable}, we can write
\begin{eqnarray}
\label{eq:N_d_SQUARED_FINAL1}
% ---- SIMPLIFY 2
\langle\hat{N}_d^2\rangle
=\cos^2\varphi\left(\langle{\hat{n}_0^2}\rangle
+\langle{\hat{n}_1^2}\rangle\right)
-2\cos\left(2\varphi\right)\langle{\hat{n}_0}\rangle\langle{\hat{n}_1}\rangle
% ------ comment below
%\qquad\quad
\nonumber\\
+\sin^2\varphi\left(
\langle{\hat{n}_0}\rangle
+\langle{\hat{n}_1}\rangle
+\langle{\hat{a}_0^2}\rangle\langle{(\hat{a}_1^\dagger)^2}\rangle
+\langle{(\hat{a}_0^\dagger)^2}\rangle\langle{\hat{a}_1^2}\rangle\right)
% ------
\nonumber\\
+\sin2\varphi\left(
\langle{\hat{n}_0\hat{a}_0}\rangle\langle{\hat{a}_1^\dagger}\rangle
+\langle{\hat{a}_0^\dagger\hat{n}_0}\rangle\langle{\hat{a}_1}\rangle
\right.
% ------
\nonumber\\
\left.
-\langle{\hat{a}_0}\rangle\langle{\hat{a}_1^\dagger\hat{n}_1}\rangle
-\langle{\hat{a}_0^\dagger}\rangle\langle{\hat{n}_1\hat{a}_1}\rangle
\right)
\quad
\end{eqnarray}
The term $\langle\hat{N}_d\rangle^2$ can be computed from equation \eqref{eq:Nd_average} and we find the variance
\begin{eqnarray}
\label{eq:VARIANCE_N_d_FINAL}
% ---- SIMPLIFY 2
\Delta^2\hat{N}_d
=\cos^2\varphi\left(
\Delta^2\langle{\hat{n}_0}\rangle
+\Delta^2\langle{\hat{n}_1}\rangle
\right)
% ------ 
%\nonumber\\
% ------ LINE
\nonumber\\
+\sin^2\varphi\left(
\langle{\hat{n}_0}\rangle
+\langle{\hat{n}_1}\rangle
+2\langle{\hat{n}_0}\rangle\langle{\hat{n}_1}\rangle
-2\vert\langle{\hat{a}_0}\rangle\vert^2\vert\langle{\hat{a}_1}\rangle\vert^2
\right.
\nonumber\\
\left.
+2\Re\left(\langle{\hat{a}_0^2}\rangle\langle{(\hat{a}_1^\dagger)^2}\rangle
-\langle{\hat{a}_0}\rangle^2\langle{\hat{a}_1^\dagger}\rangle^2
\right)
\right)
% ------ 3 ------ SIN 2 VARPHI -------
\nonumber\\
+2\sin2\varphi\Re\left(
\left(\langle{\hat{a}_0^\dagger\hat{n}_0}\rangle
-\langle{\hat{n}_0}\rangle\langle{\hat{a}_0^\dagger}\rangle\right)
\langle{\hat{a}_1}\rangle
\right.
% ------ 4 ------ FOURTH and LAST ROW -------
\nonumber
\\
\left.
-\langle{\hat{a}_0}\rangle\left(\langle{\hat{a}_1^\dagger\hat{n}_1}\rangle-\langle{\hat{a}_1^\dagger}\rangle\langle{\hat{n}_1}\rangle\right)
\right)
\qquad
\end{eqnarray}
We mention that the same results can be obtained with the help of the Jordan-Schwinger angular momentum operators \cite{Sch65}, see e. g. \cite{Dem15}.

% ---------------------------------------------------------
% ------ SINGLE DETECTOR ---- VARIANCE COMPUTATION --------
% ---------------------------------------------------------
\subsection{Single-mode intensity detection}
\label{subsec:app:single_intensity_detection}
The calculation is similar to the previous one and we obtain in the single-intensity detection scenario,
%
% ----- BEGIN    W I D E T E X T ----- BEGIN WIDETEXT -----
\begin{widetext}
\begin{eqnarray}
\label{eq:Variance_N_4_FINAL}
% --------------------------------------
\Delta^2\hat{N}_4
=\sin^4\left(\frac{\varphi}{2}\right)\Delta^2\langle{\hat{n}_0}\rangle
+\cos^4\left(\frac{\varphi}{2}\right)\Delta^2\langle{\hat{n}_1}\rangle
%\nonumber\\
+\frac{\sin^2\varphi}{4}\left(
\langle{\hat{n}_0}\rangle
+\langle{\hat{n}_1}\rangle
+2\langle{\hat{n}_0}\rangle\langle{\hat{n}_1}\rangle
-2\vert\langle{\hat{a}_0}\rangle\vert^2\vert\langle{\hat{a}_1^\dagger}\rangle\vert^2
\right)
% -----%-----
\nonumber\\
% =============== SECOND LINE
%----- SQUARE terms
+\frac{\sin^2\varphi}{2}\Re\left(\langle{\hat{a}_0^2}\rangle\langle{(\hat{a}_1^\dagger)^2}\rangle
-\langle{\hat{a}_0}\rangle^2\langle{\hat{a}_1^\dagger}\rangle^2
\right)
-\sin\varphi\Re\langle{\hat{a}_0}\rangle\langle{\hat{a}_1^\dagger}\rangle
%\nonumber\\
% =============== THIRD
-2\sin^2\left(\frac{\varphi}{2}\right)\sin\varphi\Re\left(
\left(\langle{\hat{a}_0^\dagger\hat{n}_0}\rangle
-\langle{\hat{n}_0}\rangle\langle{\hat{a}_0^\dagger}\rangle\right)\langle{\hat{a}_1}\rangle\right)
\nonumber\\ 
% ----
-2\cos^2\left(\frac{\varphi}{2}\right)\sin\varphi\Re\left(
\langle{\hat{a}_0^\dagger}\rangle\left(\langle{\hat{n}_1\hat{a}_1}\rangle
-\langle{\hat{a}_1}\rangle\langle{\hat{n}_1}\rangle
\right)
\right)
\qquad
% --------------------------------
\end{eqnarray}

% ---------------------------------------------------------
% --------------- HOMODYNE DETECTION ----------------------
% ---------------------------------------------------------
\subsection{Balanced homodyne detection}
\label{subsec:app:homodyne_detection}
Using equation \eqref{eq:Obervable_Homodyne} we immediately have
\begin{eqnarray}
\label{eq:Homodyne_del_X_del_varphi}
\vert\partial_\varphi\langle\hat{X}_{\phi_L}\rangle\vert
=\frac{1}{2}\Big\vert\cos\left(\frac{\varphi}{2}\right)\frac{e^{-i\phi_L}\langle{\hat{a}_0}\rangle
+e^{i\phi_L}\langle{\hat{a}_0^\dagger}\rangle}{2}
%\nonumber\\
%\quad
+\sin\left(\frac{\varphi}{2}\right)\frac{e^{-i\phi_L}\langle{\hat{a}_1}\rangle
+e^{i\phi_L}\langle{\hat{a}_1^\dagger}\rangle}{2}
\Big\vert
% ----------
\nonumber\\
=\Big\vert\cos\left(\frac{\varphi}{2}\right)\Re\left(e^{-i\phi_L}\langle{\hat{a}_0}\rangle\right)
%\nonumber\\
+\sin\left(\frac{\varphi}{2}\right)\Re\left(e^{-i\phi_L}\langle{\hat{a}_1}\rangle\right)\Big\vert
%\quad\quad
\end{eqnarray}
The variance of the operator $\hat{X}_{\phi_L}$ is found to be
\begin{eqnarray}
\label{eq:Homodyne_Variance_GENERIC}
\Delta^2\hat{X}_{\phi_L}
=\frac{1}{4}
+\sin^2\left(\frac{\varphi}{2}\right)\frac{e^{-i2\phi_L}(\langle{\hat{a}_0^2}\rangle-\langle{\hat{a}_0}\rangle^2)
+e^{i2\phi_L}(\langle{(\hat{a}_0^\dagger)^2}\rangle
-\langle{\hat{a}_0^\dagger}\rangle^2)
+2(\langle{\hat{n}_0}\rangle
-\langle{\hat{a}_0^\dagger}\rangle\langle{\hat{a}_0}\rangle)
}{4}
\qquad
\nonumber\\
+\cos^2\left(\frac{\varphi}{2}\right)\frac{
e^{-i2\phi_L}(\langle{\hat{a}_1^2}\rangle-\langle{\hat{a}_1}\rangle^2)
+e^{i2\phi_L}(\langle{(\hat{a}_1^\dagger)^2}\rangle-\langle{\hat{a}_1^\dagger}\rangle^2)
+2(\langle{\hat{n}_1}\rangle-\langle{\hat{a}_1^\dagger}\rangle\langle{\hat{a}_1}\rangle)}{4}
\qquad
\nonumber\\
=\frac{1}{4}
%\nonumber\\
+\frac{\sin^2\left(\frac{\varphi}{2}\right)}{2}\left(\Re\left(e^{-i2\phi_L}(\langle{\hat{a}_0^2}\rangle-\langle{\hat{a}_0}\rangle^2)\right)
+\langle\hat{n}_0\rangle-\vert\langle\hat{a}_0\rangle\vert^2\right)
%\nonumber\\
+\frac{\cos^2\left(\frac{\varphi}{2}\right)}{2}\left(\Re\left(e^{-i2\phi_L}(\langle{\hat{a}_1^2}\rangle-\langle{\hat{a}_1}\rangle^2)\right)
+\langle\hat{n}_1\rangle-\vert\langle\hat{a}_1\rangle\vert^2\right)
\quad\quad
\end{eqnarray}

\end{widetext}

% ---------------------------------------------------------
% ------ LOSSES --- GENERAL FORMALISM --------------------
% ---------------------------------------------------------
\section{The impact of non-unit photo-detection efficiency}
\label{sec:app:LOSSES}
If we consider photodetectors having a non-unit quantum efficiency, we model the losses by including a ficticious beam splitter of transmission $\sqrt{\eta}$ in front of an ideal photo-detector \cite{Kim99,Spa16,Ono10}.
Assuming such a beam splitter in front of the photo-detector at the output port $k$, we have the  new annihilation operator
\begin{equation}
\label{eq:app:a4_LOSSY}
\hat{a}'_k=\sqrt{\eta}\hat{a}_k+\sqrt{1-\eta}\hat{a}_v
\end{equation}
where $\hat{a}_v$ is the annihilation operator from the ``vacuum port''. As a convention, $\eta=1$ implies ideal photo-detector. We find immediately
\begin{equation}
\label{eq:app:n4_LOSSY}
\langle\hat{n}'_k\rangle=\eta\langle\hat{n}_k\rangle
\end{equation}
because the port $v$ is always in the vacuum state. After some computations we also  have
\begin{equation}
\label{eq:app:Variance_nk_LOSSY}
\Delta^2\hat{n}'_k=\eta^2\Delta^2\hat{n}_k+\eta\left(1-\eta\right)\langle\hat{n}_k\rangle
\end{equation}
If we consider the output port $4$, we arrive at $\Delta\varphi'_\mathrm{sg}$ from equation \eqref{eq:delta_varphi_LOSSY}.

In the case of a difference-intensity detection scenario, equation \eqref{eq:VARIANCE_N_d_FINAL} is modified to
\begin{equation}
\label{eq:app:Variance_Nd_LOSSY}
\Delta^2\hat{N}'_d=\eta^2\Delta^2\hat{N}_d+\eta\left(1-\eta\right)
\left(\langle\hat{n}_4\rangle+\langle\hat{n}_5\rangle\right)
\end{equation}
therefore the phase sensitivity gives
\begin{equation}
\label{eq:delta_varphi_Nd_LOSSY}
\Delta\varphi'_\mathrm{df}=\frac{\sqrt{\Delta^2\hat{N}_d
+\frac{1-\eta}{\eta}\left(\langle\hat{n}_4\rangle+\langle\hat{n}_5\rangle\right)}}{\vert\partial_\varphi\langle\hat{N}_d\rangle\vert}
\end{equation}
A similar calculation can be performed to include losses for a balanced homodyne detection and we obtain
\begin{equation}
\label{eq:delta_varphi_hom_LOSSY}
\Delta\varphi'_\mathrm{hom}=\frac{\sqrt{\Delta^2\hat{X}_L
+\frac{1}{4}\frac{1-\eta}{\eta}}}{\vert\partial_\varphi\langle\hat{X}_L\rangle\vert}
\end{equation}

% ---------------------------------------------------------
% ---------- SQUEEZED-COHERENT plus SQUEEZD VACUUM  -------
% ---------------------------------------------------------
\section{Calculations for a squeezed-coherent plus squeezed vacuum input}
\label{sec:app:sqz_coh_plus_sqz_vac}
The input state from equation  \eqref{eq:psi_in_squeezed_coherent_plus_squeezed_vac} being factorized (separable) allows a separate analysis of the input ports.

For the input port $0$ we have a squeezed vacuum state generated by the squeezing operator \eqref{eq:Squeezing_operator} with the parameter $\xi=re^{i\theta}$. The two basic equations needed in all calculations are \cite{GerryKnight,Loudon}
\begin{equation}
\label{eq:app:Squueezed_basic_relations}
\left\{
\begin{array}{l}
\hat{S}_0^\dagger\left(\xi\right)\hat{a}_0\hat{S}_0\left(\xi\right)=\cosh{r}\hat{a}_0-\sinh{r}e^{i\theta}\hat{a}_0^\dagger\\
\hat{S}_0^\dagger\left(\xi\right)\hat{a}_0^\dagger\hat{S}_0\left(\xi\right)=\cosh{r}\hat{a}_0^\dagger-\sinh{r}e^{-i\theta}\hat{a}_0
\end{array}
\right.
\end{equation}
From equations \eqref{eq:app:Squueezed_basic_relations} and considering the input state  \eqref{eq:psi_in_squeezed_coherent_plus_squeezed_vac} we have
$\langle{\hat{a}_0}\rangle=0=\langle{\hat{a}_0^\dagger}\rangle$. The average number of photons for a squeezed vacuum state is  $\langle{\hat{n}_0}\rangle=\sinh^2r$ and its variance yields
\begin{equation}
\label{eq:app:Variance_squeezed_vacuum}
\Delta^2{\hat{n}_0}=\frac{\sinh^2{2r}}{2}.
\end{equation}

At input port $1$ we have a squeezed-coherent state, thus using equations \eqref{eq:app:Squueezed_basic_relations} and the properties of coherent states, we have $\langle{{\hat{a}}_1}\rangle=\alpha$, $\langle{{\hat{a}}_1^\dagger}\rangle=\alpha^*$. and the average number of photons is found to be $\langle{\hat{n}_1}\rangle=\vert\alpha\vert^2+\sinh^2z$. We find the results
\begin{equation}
\label{eq:app:a0_squared_a0_dag_squared_COH-SQZ}
\left\{
\begin{array}{l}
\displaystyle
\langle{\hat{a}_1^2}\rangle
=\alpha^2-\frac{1}{2}\sinh2ze^{i\phi}
\\
\mbox{ }\\
\displaystyle
\langle{(\hat{a}_1^\dagger)^2}\rangle
=(\alpha^*)^2-\frac{1}{2}\sinh2ze^{-i\phi}
\end{array}
\right.
\end{equation}
In order to compute the variance we first calculate $\langle{\hat{n}_1^2}\rangle=1/2\sinh^22z+\vert\alpha\vert^2+2\vert\alpha\vert^2\sinh^2z$. Using this result and the average squared $\langle{\hat{n}_1}\rangle^2$ we have the variance
\begin{eqnarray}
\label{eq:app:VARIANCE_squeezed-coherent}
\Delta^2{\hat{n}_1}
=\frac{\sinh^22z}{2}+\Upsilon^-\left({\alpha},{\zeta}\right)
\end{eqnarray}

% ---------------------------------------------------------
% ------------ SECTION --- FISHER INFORMATION -------------
% ---------------------------------------------------------
\subsection{Fisher information calculations}
For a squeezed-coherent plus squeezed vacuum input given by equation \eqref{eq:psi_in_squeezed_coherent_plus_squeezed_vac}, using equations \eqref{eq:app:F_ss_FINAL_FORM_GENERAL}, \eqref{eq:app:Variance_squeezed_vacuum} and \eqref{eq:app:VARIANCE_squeezed-coherent}, we get a sum-sum Fisher matrix coefficient
\begin{eqnarray}
\label{eq:F_ss_coh_sqz_vac_sqz_vac_Upsilon}
\mathcal{F}_{ss}=\frac{\sinh^2{{2r}}}{2}
+\frac{\sinh^2{{2z}}}{2}
+\Upsilon^-\left({\alpha},{\zeta}\right)
\end{eqnarray}
We also compute $\mathcal{F}_{sd}$ from \eqref{eq:app:F_sd_GENERIC_FINAL_balanced}
and get $\mathcal{F}_{sd}=0$. Using equation \eqref{eq:app:F_dd_FINAL_FORM_GENERAL_balanced} and the previous results we also calculate $\mathcal{F}_{dd}$ given by equation \eqref{eq:app:F_dd_coh_sqz_vac_sqz_vac_Upsilon}.
%

% -------------------------------------------------
% -------- DIFFERENCE --- DETECTION -------------
% -------------------------------------------------
\subsection{Difference intensity detection}
\label{subsec:app:coh_sqz_plus_sqz_vac_DIFF_DET}
We start from equation \eqref{eq:VARIANCE_N_d_FINAL} and replace the terms with the expressions from equations \eqref{eq:app:Variance_squeezed_vacuum},  \eqref{eq:app:a0_squared_a0_dag_squared_COH-SQZ} and \eqref{eq:app:VARIANCE_squeezed-coherent}. Using the identity $2\sinh^2r+1=\cosh2r$ takes us to the final result from equation \eqref{eq:VARIANCE_N_d_sqz-coh_sqz_vac_FINAL}. The phase sensitivity $\Delta\varphi_\textrm{df}$ is obtained using equations \eqref{eq:Nd_average_sqz-coherent_squeezed-vac} and \eqref{eq:VARIANCE_N_d_sqz-coh_sqz_vac_FINAL}, yielding
% ----- BEGIN    W I D E T E X T ----- BEGIN WIDETEXT -----
\begin{widetext}
\begin{eqnarray}
\label{eq:Delta_pdi_szq-coh_sqz_vac_GENERAL}
\Delta\varphi_\textrm{df}
=\frac{\sqrt{\left(
{\frac{\sinh^2{2r}}{2}}
+\frac{\sinh^2{2z}}{2}+\Upsilon^-\left({\alpha},{\zeta}\right)
\right)\cot^2\varphi
+
\Upsilon^-\left({\alpha},{\xi}\right)
+\frac{\cosh{2r}\cosh{2z}+\sinh{2r}\sinh{2z}
\cos\left({\phi}-{\theta}\right)}{2}-\frac{1}{2}
}}{\vert{\vert\alpha\vert^2+\sinh^2z}-{\sinh^2r}\vert}
\quad
\end{eqnarray}
We impose now the optimum working point $\varphi_\textrm{opt}=\pi/2$ and have the result from equation \eqref{eq:app:Delta_phi_N_d_sqz-coh_sqz_vac_pi_over_2}.

% --------------------------------------------------
% ----------- SINGLE MODE --------------------------
% --------------------------------------------------
\subsection{Single-mode intensity detection}
Starting from equation \eqref{eq:Variance_N_4_FINAL} and using the previous results takes us to  equation \eqref{eq:Variance_N_4_coh_sqz_sqz_vac}. The phase sensitivity for a single-mode intensity detection scenario is given by
\begin{eqnarray}
\label{eq:Delta_phi_sqz_coh__sqz_vac_SINGLE_DET_Upsilon}
% --------------------------------------
\Delta\varphi_\textrm{sg}=
\frac{\sqrt{\cot^2\left(\frac{\varphi}{2}\right)\left(\frac{\sinh^22z}{2}+\Upsilon^-\left({\alpha},{\zeta}\right)\right)
+\tan^2\left(\frac{\varphi}{2}\right){\frac{\sinh^22r}{2}}
+\Upsilon^-\left({\alpha},{\xi}\right)
+\frac{{\cosh2r}
{\cosh2z}
+{\sinh2r}{\sinh2z}\cos\left(\theta-\phi\right)-1}
{2}}}
{\big\vert{\vert\alpha\vert^2+\sinh^2z}-{\sinh^2r}\big\vert}
\quad\quad
\end{eqnarray}

If we impose now the optimum internal phase shift $\varphi_\textrm{opt}$ from equation \eqref{eq:varphi_opt_sqz-coh_plus_sqz_vac_sing_det}, we obtain the result
\begin{eqnarray}
\label{eq:Delta_varphi_Sqz-coh_Sqz_Vac_SING_Best_GENERAL}
% --------------------------------------
\Delta\tilde{\varphi}_\textrm{sg}=
\frac{\sqrt{{{\sinh2r}}\sqrt{\sinh^22z+2\Upsilon^-\left({\alpha},{\zeta}\right)}
+
%\left(
\Upsilon^-\left({\alpha},{\xi}\right)
+\frac{{\cosh2r}
{\cosh2z}
+{\sinh2r}{\sinh2z}\cos\left(\theta-\phi\right)-1}
{2}
%\right)
}}
{\big\vert{\vert\alpha\vert^2+\sinh^2z}-{\sinh^2r}\big\vert}
\quad\quad
\end{eqnarray}
Further imposing the input phase matching conditions \eqref{eq:phase_matching_cond_coh_plus_sqz_vac} and \eqref{eq:phase_matching_cond_sqz_coh_plus_sqz_vac} yields the best achievable sensitivity
\begin{equation}
\label{eq:app:Delta_varphi_Sqz-coh_Sqz_Vac_SING_Best}
\Delta\tilde{\varphi}_\textrm{sg}\big\vert_{\theta-\phi=\pm\pi}=\frac{\sqrt{\sinh2r\sqrt{\sinh^22z+2\vert\alpha\vert^2e^{2z}}+\vert\alpha\vert^2e^{-2r}+\sinh^2(r-z)}}{\vert\vert\alpha\vert^2+\sinh^2z-\sinh^2r\vert}
\end{equation}
Imposing the phase matching conditions \eqref{eq:phase_matching_cond_coh_plus_sqz_vac} and \eqref{eq:phase_matching_cond_sqz_coh_plus_sqz_vac_BIS} we obtain
\begin{equation}
\label{eq:app:Delta_varphi_Sqz-coh_Sqz_Vac_SING_Best_phi_is_zero}
\Delta\tilde{\varphi}_\textrm{sg}\big\vert_{\theta-\phi=0}=\frac{\sqrt{\sinh2r\sqrt{\sinh^22z+2\vert\alpha\vert^2e^{-2z}}+\vert\alpha\vert^2e^{-2r}+\sinh^2(r+z)}}{\vert\vert\alpha\vert^2+\sinh^2z-\sinh^2r\vert}
\end{equation}
The limit value of $\vert\alpha\vert$, where phase sensitivity from equation \eqref{eq:app:Delta_varphi_Sqz-coh_Sqz_Vac_SING_Best_phi_is_zero} outperforms the one from equation \eqref{eq:app:Delta_varphi_Sqz-coh_Sqz_Vac_SING_Best} is given by equation \eqref{eq:alpha_lim_sqz_coh_plus_sqz_vac}.

% ---------------------------------------------------------
% ------- SECTION --- OPTIMIZATION SQUEEZERS --------------
% ---------------------------------------------------------
\section{The optimization of two input squeezers}
\label{sec:app:optimization_two_squeezers}
In the most general case we have the input state from equation \eqref{eq:psi_in_squeezed_coherent_plus_squeezed_coherent}, however we focus here on the squeezing part of this state (the discussion thus applies to Section \ref{sec:sqz_coh_plus_squeezed_vacuum}, too). Consider the input state
\begin{equation}
\label{eq:app:psi_in_squeezers_only}
\vert\psi_{in}\rangle
\approx\hat{S}_1\left(\zeta\right)\hat{S}_0\left(\xi\right)\vert0\rangle=
 e^{[\zeta^*\hat{a}_1^2-\zeta(\hat{a}_1^\dagger)^2]/2}
e^{[\xi^*\hat{a}_0^2-\xi(\hat{a}_0^\dagger)^2]/2}\vert0\rangle
\end{equation}
We use the decomposition \cite{Tru85} (we recall $\chi=se^{i\vartheta}$)
\begin{equation}
\label{eq:app:squeezer_decomposition_exponentials}
e^{[\chi^*\hat{a}_m^2-\chi(\hat{a}_m^\dagger)^2]/2}
=e^{-\tau\chi(\hat{a}_m^\dagger)^2/2}
e^{-\nu\left(\hat{a}_m^\dagger\hat{a}_m+\frac{1}{2}\right)}
e^{\tau\hat{a}_m^2/2}
\end{equation}
where $\tau=e^{i\vartheta}\tanh{s}$ and  $\nu=\ln\cosh{s}$. Now applying equation \eqref{eq:app:squeezer_decomposition_exponentials} to our input state allows a sizable simplification since the annihilation operators and the number operators give no contribution when applied to the vacuum state and we have
\begin{equation}
\label{eq:app:psi_in_squeezers_only_temp0}
\vert\psi_{in}\rangle
\approx \frac{1}{\sqrt{\cosh{r}\cosh{z}}} e^{-\tau_1(\hat{a}_1^\dagger)^2/2}
e^{-\tau_0(\hat{a}_0^\dagger)^2/2}\vert0\rangle
\end{equation}
where $\tau_1=e^{i\phi}\tanh{z}$ and $\tau_0=e^{i\theta}\tanh{r}$. Since the input creation operators commute, we can group together the exponentials. We want to find out the state vector $\vert\psi'\rangle$ after the first beam splitter. Using the field operator transformations \eqref{eq:field_op_transf_BS_sym} we have
\begin{equation}
\label{eq:app:psi_in_squeezers_only_temp2}
\vert\psi'\rangle
\approx\frac{1}{\sqrt{\cosh{r}\cosh{z}}} e^{\frac{\tau_1-\tau_0}{2}[(\hat{a}_2^\dagger)^2-(\hat{a}_3^\dagger)^3]-i(\tau_1+\tau_0)\hat{a}_2^\dagger\hat{a}_3^\dagger}\vert0\rangle
\end{equation}
We want to have inside the interferometer as much as possible two single-mode squeezed vacuums \cite{Lan14} (one acting as a phase reference for the other). This condition in reinforced when $\mathrm{Arg}(\tau_1)=\mathrm{Arg}(\tau_0)+\pi$.  As remarked by Lang \& Caves \cite{Lan14}, if we assume $\zeta=-\xi$ i. e. start with the input state $\vert\psi_{in}\rangle\approx\hat{S}_1\left(-\xi\right)\hat{S}_0\left(\xi\right)\vert0\rangle$ we find after the beam splitter $\vert\psi'\rangle
\approx (\cosh{r})^{-1} e^{-\tau_0[(\hat{a}_2^\dagger)^2-(\hat{a}_3^\dagger)^3]}\vert0\rangle=\hat{S}_3\left(-\xi\right)\hat{S}_2\left(\xi\right)\vert0\rangle$.

The state vector $\vert\psi'\rangle$ is relevant when computing the QFI, however for realistic schemes we might be interested to find $\vert\psi_\mathrm{out}\rangle$. Starting from the input state \eqref{eq:app:psi_in_squeezers_only} and using the field operator transformations \eqref{eq:field_op_transf_MZI} we find
\begin{eqnarray}
\label{eq:app:psi_out_squeezers_only_temp0}
\vert\psi_\mathrm{out}\rangle
\approx\frac{1}{\sqrt{\cosh{r}\cosh{z}}}
%\nonumber\\
 e^{-\frac{1}{2}\left[\left(\tau_0\sin^2\left(\frac{\varphi}{2}\right)+\tau_1\cos^2\left(\frac{\varphi}{2}\right)\right)(\hat{a}_4^\dagger)^2
 +\left(\tau_0\cos^2\left(\frac{\varphi}{2}\right)+\tau_1\sin^2\left(\frac{\varphi}{2}\right)\right)(\hat{a}_5^\dagger)^2
 +(\tau_1-\tau_0)\sin\varphi\hat{a}_4^\dagger\hat{a}_5^\dagger\right]}\vert0\rangle
\end{eqnarray}
If we impose now the constraint $\zeta=-\xi$ equation \eqref{eq:app:psi_out_squeezers_only_temp0} becomes
\begin{eqnarray}
\label{eq:app:psi_out_squeezers_only_temp1}
\vert\psi_\mathrm{out}\rangle
\approx\frac{1}{\cosh{r}}
%\nonumber\\
 e^{-\frac{1}{2}\left[-\tau_0\cos\varphi(\hat{a}_4^\dagger)^2
 +\tau_0\cos\varphi(\hat{a}_5^\dagger)^2
 -2\tau_0\sin\varphi\hat{a}_4^\dagger\hat{a}_5^\dagger\right]}\vert0\rangle
\end{eqnarray}
\end{widetext}
This state has a strong $\varphi$-dependence, therefore applying the observables described in Section \ref{subsec:output_observables} will yield $\varphi$-dependent results. If we now apply at the input two identical squeezings i. e. $\zeta=\xi$, from equation \eqref{eq:app:psi_out_squeezers_only_temp0} we get 
\begin{eqnarray}
\label{eq:app:psi_out_squeezers_only_no_varphi_dependence}
\vert\psi_\mathrm{out}\rangle
\approx\frac{1}{\cosh{r}}
%\nonumber\\
 e^{-\frac{1}{2}\left[-\tau_0(\hat{a}_4^\dagger)^2
 +\tau_0(\hat{a}_5^\dagger)^2\right]}\vert0\rangle
\end{eqnarray}
and this is the worst case scenario since this state has no $\varphi$-dependence whatsoever.

% ---------------------------------------------------------
% ------- SECTION --- SQUEEZED-COH plus SQUEEZED-COH ------
% ---------------------------------------------------------
\section{Calculations for the squeezed-coherent plus squeezed-coherent input}
\label{sec:app:sqz_coh_plus_sqz_sqz_coh}
In this appendix we detail the calculations needed for the scenario discussed in Section \ref{sec:sqz_coh_plus_sqz_coh}. We rely on results already computed in Appendix \ref{sec:app:sqz_coh_plus_sqz_vac}. The new results needed to complete the calculations are
\begin{equation}
\label{eq:app:COH_SQZ-COH-SQZ_n1a1_minus_n1_a1_AND_dagger}
\left\{
\begin{array}{l}
\langle{\hat{n}_1\hat{a}_1}\rangle
-\langle{\hat{n}_1}\rangle\langle{\hat{a}_1}\rangle
=\alpha\sinh^2z-\frac{\alpha^*}{2}\sinh{2z}e^{i\phi}\\
\langle{\hat{a}_1^\dagger\hat{n}_1}\rangle
-\langle{\hat{n}_1}\rangle\langle{\hat{a}_1^\dagger}\rangle
=\alpha^*\sinh^2z-\frac{\alpha}{2}\sinh{2z}e^{-i\phi}
\end{array}
\right.
\end{equation}
and similarly for port $0$,
\begin{equation}
\label{eq:app:COH_SQZ-COH-SQZ_n0a0_minus_n0_a0_AND_dagger}
\left\{
\begin{array}{l}
\langle{\hat{n}_0\hat{a}_0}\rangle
-\langle{\hat{n}_0}\rangle\langle{\hat{a}_0}\rangle
=\beta\sinh^2r-\frac{\beta^*}{2}\sinh{2r}e^{i\theta}\\
\langle{\hat{a}_0^\dagger\hat{n}_0}\rangle
-\langle{\hat{n}_0}\rangle\langle{\hat{a}_0^\dagger}\rangle
=\beta^*\sinh^2r-\frac{\beta}{2}\sinh{2r}e^{-i\theta}
\end{array}
\right.
\end{equation}
We also state the result of a term that appears repeatedly,

\begin{eqnarray}
\label{eq:TERM_sum_a0_sq_a0_dagger_sq_coh_sqz_vac}
\langle{(\hat{a}_0^\dagger)^2}{\hat{a}_1^2}\rangle
+\langle{\hat{a}_0^2}{(\hat{a}_1^\dagger)^2}\rangle
=2\vert{\alpha}{\beta}\vert^2\cos(2\theta_\alpha-2\theta_\beta)
\qquad\quad
\nonumber\\
-{\vert\beta\vert^2}{\sinh2z}\cos(2\theta_\beta-\phi)
%\nonumber\\
-{\vert\alpha\vert^2}{\sinh2r}\cos(2\theta_\alpha-\theta)
\nonumber\\
+\frac{1}{2}{\sinh2r}{\sinh2z}\cos(\theta-\phi)
\qquad
\end{eqnarray}

% ---------------------------------------------------------
% ---------- SUB - SECTION --- FISHER INFORMATION ---------
% ---------------------------------------------------------
\subsection{Fisher information calculations}
\label{subsec:app:fisher_sqz_coh_plus_sqz_coh}
We use the definition of the Fisher matrix element $\mathcal{F}_{ss}$ and the result from equation \eqref{eq:app:VARIANCE_squeezed-coherent} to obtain
\begin{eqnarray}
\label{eq:F_ss_sqz-coh_plus_sqz-coh_FINAL_FORM}
\mathcal{F}_{ss}=\frac{\sinh^2{{2r}}}{2}
+\Upsilon^-\left({\beta},{\xi}\right)
+\frac{\sinh^2{{2z}}}{2}
+\Upsilon^-\left({\alpha},{\zeta}\right)
\quad
\end{eqnarray}
In the calculation of the Fisher information, the most important matrix element is $\mathcal{F}_{dd}$. Applying the input state \eqref{eq:psi_in_squeezed_coherent_plus_squeezed_coherent} to the definition from equation \eqref{eq:app:Fisher_matrix_elements} gives the result
\begin{eqnarray}
\label{eq:app:F_dd_sqz-coh_plus_sqz_coh_FINAL_FORM}
\mathcal{F}_{dd}
=\Upsilon^+\left(\beta,\zeta\right)
+\Upsilon^+\left({\alpha},{\xi}
\right)
\qquad\qquad\qquad\qquad\qquad
\nonumber\\
+\frac{\cosh{2r}\cosh{2z}-\sinh{2r}\sinh{2z}\cos(\theta-\phi)-1}{2}
\quad
\end{eqnarray}
Finally, the last Fisher matrix element yields
\begin{eqnarray}
\label{eq:F_sd_sqz-coh_plus_sqz_coh_FINAL_FORM}
\mathcal{F}_{sd}
=\vert{\alpha}{\beta}\vert
\Big(\sinh{2r}\sin\left({\theta_\alpha}+{\theta_\beta}-{\theta}\right)
%\right.
\qquad\qquad
\nonumber\\
%\left.
-\sinh{2z}\sin\left({\theta_\alpha}+{\theta_\beta}-{\phi}\right)
%\right.
\nonumber\\
%\left.
-2\left(1+\sinh^2{ r}+\sinh^2{ z}\right)
\sin\left(\theta_\alpha-\theta_\beta\right)
\Big)
\end{eqnarray}

% ---------------------------------------------------------

\subsection{Phase-matching conditions for optimum Fisher information}
\label{subsec:app:PMC_for_fisher_sqz_coh_plus_sqz_coh}

As stated in Section \ref{subsec:sqz_coherent_sqz_coh_input_Fisher}, we start from a known scenario and make our way towards this more general case. If $\vert\beta\vert\to0$, we find ourselves in the squeezed-coherent plus squeezed vacuum input scenario from Section \ref{sec:sqz_coh_plus_squeezed_vacuum}. The phase-matching conditions have been discussed and yield the Fisher information from equation \eqref{eq:Fisher_sqz-coh_plus_sqz_vac_MAXIMAL}. Therefore, we now apply the constraints from equations \eqref{eq:phase_matching_cond_coh_plus_sqz_vac} and \eqref{eq:phase_matching_cond_sqz_coh_plus_sqz_vac} on the Fisher matrix elements $\mathcal{F}_{ss}$, $\mathcal{F}_{dd}$ and $\mathcal{F}_{sd}$ from equations \eqref{eq:F_ss_sqz-coh_plus_sqz-coh_FINAL_FORM}, \eqref{eq:app:F_dd_sqz-coh_plus_sqz_coh_FINAL_FORM} and, respectively, \eqref{eq:F_sd_sqz-coh_plus_sqz_coh_FINAL_FORM}. The Fisher information definition from equation \eqref{eq:Fisher_definition} takes us to
\begin{eqnarray}
\label{eq:Fisher_sqzcoh_sqz_coh_PMC_theta_0_phi_pi_thetabeta_free}
\mathcal{F}
=\Upsilon^+\left(\beta,\zeta\right)
+\vert\alpha\vert^2e^{2r}
+\sinh^2\left(r+z\right)
\qquad\qquad\quad
% --------------------
\nonumber\\
-\frac{\vert\alpha\beta\vert^2\sin^2\left(\theta_\alpha-\theta_\beta\right)
\left(e^{2r}+e^{2z}\right)^2}
{\frac{\sinh^22r}{2}
+\Upsilon^-\left(\beta,\xi\right)
+\frac{\sinh^22z}{2}
+\vert\alpha\vert^2e^{2z}}
\quad
\end{eqnarray}
We allowed $\beta\neq0$, and since $\beta=\vert\beta\vert e^{i\theta_\beta}$ we have to define the angle of the second coherent input, $\theta_\beta$. It can be easily shown that the Fisher information from equation \eqref{eq:Fisher_sqzcoh_sqz_coh_PMC_theta_0_phi_pi_thetabeta_free} is maximized only for $\theta_\alpha-\theta_\beta=n\pi/2$ with $n\in\mathbb{Z}$. In reference \cite{API18}, it has been shown that for a double coherent input, the maximum Fisher information is achieved when 
\begin{equation}
\label{eq:phase_matcing_theta_alpha_equal_theta_beta}
\theta_\alpha-\theta_\beta=0. 
\end{equation}
We thus adopt the PMCs given by equation \eqref{eq:phase_matching_array_0_0_phi_is_pi}, impose this constraint
on equation \eqref{eq:Fisher_sqzcoh_sqz_coh_PMC_theta_0_phi_pi_thetabeta_free} and have immediately the Fisher information from \eqref{eq:Fisher_sqzcoh_sqz_coh_PMC_theta_0_phi_pi_thetabeta_0}. This Fisher information is clearly optimal as ${\vert\beta\vert\to0}$, however there is no reason to be so when $\vert\beta\vert$ is comparable with the other parameters. The poor performance comes from the first term of equation \eqref{eq:Fisher_sqzcoh_sqz_coh_PMC_theta_0_phi_pi_thetabeta_free}, namely $\Upsilon^+\left(\beta,\zeta\right)$. Indeed, in equation \eqref{eq:Fisher_sqzcoh_sqz_coh_PMC_theta_0_phi_pi_thetabeta_0} it takes its  minimal value due to the implied phase-matching condition $2\theta_\beta-\phi=\pm\pi$. In order to maximize the term $\Upsilon^+\left(\beta,\zeta\right)$ from equations \eqref{eq:Fisher_sqzcoh_sqz_coh_PMC_theta_0_phi_pi_thetabeta_free} or \eqref{eq:app:F_dd_sqz-coh_plus_sqz_coh_FINAL_FORM} we need to impose the phase-matching condition
\begin{equation}
\label{eq:phase_matching_2theta_beta_minus_phi_is_zero}
2\theta_\beta-\phi=0
\end{equation}
However, it is easy to see that this PMC cannot be satisfied simultaneously with equations \eqref{eq:phase_matching_cond_coh_plus_sqz_vac}, \eqref{eq:phase_matching_cond_sqz_coh_plus_sqz_vac} and \eqref{eq:phase_matcing_theta_alpha_equal_theta_beta}. We have two solutions to this problem:
\begin{enumerate}
	\item[(i)] continue to impose the PMC from equation \eqref{eq:phase_matcing_theta_alpha_equal_theta_beta} and accept that $\Upsilon^+\left(\beta,\zeta\right)=\vert\beta\vert^2e^{-2z}$
	\item[(ii)] impose $\theta_\alpha-\theta_\beta=\pm\pi/2$ and a whole new discussion begins.
\end{enumerate}

Thus, in case (i) we end up with a trade-off situation and we have to choose which two among three terms from equation \eqref{eq:app:F_dd_sqz-coh_plus_sqz_coh_FINAL_FORM} are to be maximized. If the coherent sources are dominant over the contribution from squeezing, it is natural to maximize $\Upsilon^+\left(\beta,\zeta\right)$ and $
\Upsilon^+\left(\alpha,\xi\right)$. This leads to the PMCs given by equation \eqref{eq:phase_matching_array_0_0_0} and to the QFI from equation \eqref{eq:Fisher_sqzcoh_sqz_coh_PMC_theta_0_phi_0_thetabeta_0}.

Up to this point we assumed that the constraint \eqref{eq:phase_matcing_theta_alpha_equal_theta_beta} yields the maximum Fisher information, and this is certainly true in the high-coherent regime $\{\vert\alpha\vert,\vert\beta\vert\}\gg\{\sinh{r},\sinh{z}\}$. However, in the  $\{\vert\alpha\vert,\vert\beta\vert\}\ll\{\sinh{r},\sinh{z}\}$ regime this is not necessarily true. We thus consider the case (ii) now, namely when $\theta_\alpha-\theta_\beta=(2k+1)\pi/2$ with $k\in\mathbb{Z}$.  Returning again to the  Fisher matrix element $\mathcal{F}_{dd}$ from equation \eqref{eq:app:F_dd_sqz-coh_plus_sqz_coh_FINAL_FORM} we note that there is actually a PMC allowing to simultaneously maximize all terms, namely equation \eqref{eq:phase_matching_array_0_0_pi_over_2}. The penalty for this constraint is the fact that $\mathcal{F}_{sd}\neq0$ and we have the QFI given by equation \eqref{eq:Fisher_sqzcoh_sqz_coh_PMC_theta_0_phi_pi_thetabeta_pi2}.

% ----- BEGIN    W I D E T E X T ----- BEGIN WIDETEXT -----
\begin{widetext}
% ---------------------------------------------------------
% --------------- DIFFERENCE INTENSITY DETECTION ----------
% ---------------------------------------------------------
\subsection{Difference-intensity detection}
\label{subsec:app:COH-SQZ_plus_COH-sqz_difference_detection}
For the input state given by equation \eqref{eq:psi_in_squeezed_coherent_plus_squeezed_coherent}, the variance of $\hat{N}_d$ is found to be
\begin{eqnarray}
\label{eq:VARIANCE_N_d_FINAL_sqz_coh_sqz_coh_02_Upsilon}
% ---- SIMPLIFY 2
\Delta^2\hat{N}_d
=\cos^2\varphi\left(
\frac{\sinh^2{2r}}{2}
+\Upsilon^-\left({\beta},{\xi}\right)
+\frac{\sinh^2{2z}}{2}
+\Upsilon^-\left({\alpha},{\zeta}\right)
\right)
\qquad\qquad\qquad\qquad\qquad\qquad\qquad\qquad\qquad\quad\:\:
% ------ SECOND
\nonumber\\
+\sin^2\varphi\left(\Upsilon^-\left(
{\beta},{\zeta}\right)
+\Upsilon^-\left({\alpha},{\xi}\right)
+\frac{\cosh{2r}\cosh{2z}+\sinh{2r}\sinh{2z}\cos(\theta-\phi)}{2}-\frac{1}{2}
\right)
\qquad\qquad\qquad\quad\:\:
% ------ LAST
\nonumber\\
+\sin2\varphi\vert{\beta}{\alpha}\vert
\Big(
2\left(\sinh^2{{r}}-\sinh^2{ z}\right)\cos(\theta_\alpha-\theta_\beta)
-\sinh{2r}\cos(\theta_\alpha+\theta_\beta-\theta)
+\sinh{2z}\cos(\theta_\alpha+\theta_\beta-\phi)
\Big)
\end{eqnarray}
Using the result from equation \eqref{eq:Nd_average_sqz-coherent_sqz-coh} and the one above, allows the phase sensitivity to be written as
\begin{eqnarray}
\label{eq:app:Delta_varphi_sqz-coh_plus_sqz-coh_Diff_det}
\Delta\varphi_\textrm{df}=\frac{\sqrt{\Delta^2\hat{N}_d}}{\big\vert\sin\varphi\left({\vert\alpha\vert^2}-\vert\beta\vert^2+{\sinh^2z}-{\sinh^2r}\right)
+2\cos\varphi\vert\alpha\beta\vert\cos(\theta_\alpha-\theta_\beta)\big\vert}
\end{eqnarray}
Similar to the previous scenarios an optimum total internal phase shift $\varphi_\textrm{opt}$ can be found. We make the following notations:
\begin{equation}
\label{eq:app:sqz-coh_plus_sqz-coh_Diff_det_ABCDEF_notations}
\left\{
\begin{array}{l}
A=\frac{\sinh^2{2r}}{2}
+\Upsilon^-\left({\beta},{\xi}\right)
+\frac{\sinh^2{2z}}{2}
+\Upsilon^-\left({\alpha},{\zeta}\right)\\
% --------- B --------
B=\Upsilon^-\left(
{\beta},{\zeta}\right)
+\Upsilon^-\left({\alpha},{\xi}\right)
+\frac{\cosh{2r}\cosh{2z}+\sinh{2r}\sinh{2z}\cos(\theta-\phi)}{2}-\frac{1}{2}\\
% ---------- C ----------------
C=\vert{\alpha}{\beta}\vert
\left(
2\left(\sinh^2{{r}}-\sinh^2{ z}\right)\cos(\theta_\alpha-\theta_\beta)
-\sinh{2r}\cos(\theta_\alpha+\theta_\beta-\theta)
+\sinh{2z}\cos(\theta_\alpha+\theta_\beta-\phi)
\right)\\
% ------------ D --------------------
D={\vert\alpha\vert^2}-\vert\beta\vert^2+{\sinh^2z}-{\sinh^2r}\\
% ----------- F ----------------------
F=2\vert\alpha\beta\vert\cos(\theta_\alpha-\theta_\beta)
\end{array}
\right.
\end{equation}
With these notations, a simple calculation shows that the optimum phase shift is given by
\begin{equation}
\label{eq:app:_delta_varphi_OPT_sqz-coh_plus_sqz_coh_Diff_det}
\varphi_\textrm{opt}=\arctan\left(\frac{AD-CF}{BF-CD}\right)+k\pi
\end{equation}
with $k\in\mathbb{Z}$. Inserting $\varphi_\textrm{opt}$ into equation \eqref{eq:app:Delta_varphi_sqz-coh_plus_sqz-coh_Diff_det} yields the optimum phase sensitivity $\Delta\tilde{\varphi}_\textrm{df}$.

% -----------------------------------------------
% ---------- SINGLE INTENSITY DETECTION ---------
% -----------------------------------------------
\subsection{Single-mode intensity detection}
\label{subsec:app:COH-SQZ_plus_COH-sqz_single_detection}
From equation \eqref{eq:Variance_N_4_FINAL}, using the input state given by equation \eqref{eq:psi_in_squeezed_coherent_plus_squeezed_coherent} and the results mentioned earlier, we have
\begin{eqnarray}
\label{eq:app:Variance_N_4_sqz_coh_sqz_coh_FINAL}
% --------------------------------------
\Delta^2\hat{N}_4
=\sin^4\left(\frac{\varphi}{2}\right)
\left(
\frac{\sinh^2{2r}}{2}
+\Upsilon^-\left({\beta},{\xi}\right)
\right)
%\nonumber\\
+\cos^4\left(\frac{\varphi}{2}\right)\left(\frac{\sinh^2{2z}}{2}
+\Upsilon^-\left({\alpha},{\zeta}\right)
\right)
% --------- third LINE
\qquad\qquad\qquad\qquad\qquad\qquad
\nonumber\\
+\frac{\sin^2\varphi}{4}\left(
\Upsilon^-\left({\beta},{\zeta}\right)
+\Upsilon^-\left({\alpha},{\xi}\right)
+\frac{\cosh{2r}\cosh{2z}+\sinh{2r}\sinh{2z}\cos(\theta-\phi)}{2}
-\frac{1}{2}
\right)
\qquad\qquad\qquad
\nonumber\\
% =============== 4
-\sin\varphi\vert{\alpha}{\beta}\vert
\bigg(\cos\left(\theta_\alpha-\theta_\beta\right)
+\sin^2\left(\frac{\varphi}{2}\right)\left(
2\sinh^2{{r}}\cos(\theta_\alpha-\theta_\beta)
-\sinh{2r}\cos(\theta_\alpha+\theta_\beta-\theta)
\right)
%\right.
\qquad\qquad\quad
\nonumber\\ 
% ---- LAAAST
%\left.
+\cos^2\left(\frac{\varphi}{2}\right)\left(
2\sinh^2{ z}\cos(\theta_\alpha-\theta_\beta)
-\sinh{2z}\cos(\theta_\alpha+\theta_\beta-\phi)
\right)
\bigg)
\qquad
\end{eqnarray}
Using the previous result and equation \eqref{eq:N4_derivative_average_coh-sqz_plus_coh-sqz}, we find  the phase sensitivity for a single-mode intensity detection,
\begin{eqnarray}
\label{eq:app:Delta_varphi_sqz-coh_plus_sqz-coh_Sing_det}
\Delta\varphi_\textrm{sg}=\frac{\sqrt{\Delta^2\hat{N}_4}}
{\big\vert\frac{1}{2}\left(\vert{\alpha}\vert^2-\vert{\beta}\vert^2
+\sinh^2z-\sinh^2r\right)+\vert{\alpha}{\beta}\vert
\cos(\theta_\alpha-\theta_\beta)\big\vert}
\end{eqnarray}

% --------- END    W I D E T E X T ----- END WIDETEXT -----
\end{widetext}

\twocolumngrid

% #########################################################
% #############    B I B L I O G R A P H Y    #############
% #########################################################
%
% BibTeX users please use
% \bibliographystyle{}
% \bibliography{}
%
% APS style
\bibliographystyle{apsrev4-1}

% now include the BIB file
\bibliography{MZI_phase_sensitivity_bibtex}

\begin{thebibliography}{53}%
\makeatletter
\providecommand \@ifxundefined [1]{%
 \@ifx{#1\undefined}
}%
\providecommand \@ifnum [1]{%
 \ifnum #1\expandafter \@firstoftwo
 \else \expandafter \@secondoftwo
 \fi
}%
\providecommand \@ifx [1]{%
 \ifx #1\expandafter \@firstoftwo
 \else \expandafter \@secondoftwo
 \fi
}%
\providecommand \natexlab [1]{#1}%
\providecommand \enquote  [1]{``#1''}%
\providecommand \bibnamefont  [1]{#1}%
\providecommand \bibfnamefont [1]{#1}%
\providecommand \citenamefont [1]{#1}%
\providecommand \href@noop [0]{\@secondoftwo}%
\providecommand \href [0]{\begingroup \@sanitize@url \@href}%
\providecommand \@href[1]{\@@startlink{#1}\@@href}%
\providecommand \@@href[1]{\endgroup#1\@@endlink}%
\providecommand \@sanitize@url [0]{\catcode `\\12\catcode `\$12\catcode
  `\&12\catcode `\#12\catcode `\^12\catcode `\_12\catcode `\%12\relax}%
\providecommand \@@startlink[1]{}%
\providecommand \@@endlink[0]{}%
\providecommand \url  [0]{\begingroup\@sanitize@url \@url }%
\providecommand \@url [1]{\endgroup\@href {#1}{\urlprefix }}%
\providecommand \urlprefix  [0]{URL }%
\providecommand \Eprint [0]{\href }%
\providecommand \doibase [0]{http://dx.doi.org/}%
\providecommand \selectlanguage [0]{\@gobble}%
\providecommand \bibinfo  [0]{\@secondoftwo}%
\providecommand \bibfield  [0]{\@secondoftwo}%
\providecommand \translation [1]{[#1]}%
\providecommand \BibitemOpen [0]{}%
\providecommand \bibitemStop [0]{}%
\providecommand \bibitemNoStop [0]{.\EOS\space}%
\providecommand \EOS [0]{\spacefactor3000\relax}%
\providecommand \BibitemShut  [1]{\csname bibitem#1\endcsname}%
\let\auto@bib@innerbib\@empty
%</preamble>
\bibitem [{\citenamefont {{The LIGO Scientific Collaboration}}(2013)}]{LIGO13}%
  \BibitemOpen
  \bibfield  {author} {\bibinfo {author} {\bibnamefont {{The LIGO Scientific
  Collaboration}}},\ }\href {\doibase 10.1038/NPHOTON.2013.177} {\bibfield
  {journal} {\bibinfo  {journal} {Nature Photonics}\ }\textbf {\bibinfo
  {volume} {7}},\ \bibinfo {pages} {613} (\bibinfo {year} {2013})}\BibitemShut
  {NoStop}%
\bibitem [{\citenamefont {Taylor}\ \emph {et~al.}(2013)\citenamefont {Taylor},
  \citenamefont {Janousek}, \citenamefont {Daria}, \citenamefont {Knittel},
  \citenamefont {Hage}, \citenamefont {Bachor},\ and\ \citenamefont
  {Bowen}}]{Tay13}%
  \BibitemOpen
  \bibfield  {author} {\bibinfo {author} {\bibfnamefont {M.~A.}\ \bibnamefont
  {Taylor}}, \bibinfo {author} {\bibfnamefont {J.}~\bibnamefont {Janousek}},
  \bibinfo {author} {\bibfnamefont {V.}~\bibnamefont {Daria}}, \bibinfo
  {author} {\bibfnamefont {J.}~\bibnamefont {Knittel}}, \bibinfo {author}
  {\bibfnamefont {B.}~\bibnamefont {Hage}}, \bibinfo {author} {\bibfnamefont
  {H.-A.}\ \bibnamefont {Bachor}}, \ and\ \bibinfo {author} {\bibfnamefont
  {W.~P.}\ \bibnamefont {Bowen}},\ }\href {\doibase 10.1038/nphoton.2012.346}
  {\bibfield  {journal} {\bibinfo  {journal} {Nature Photonics}\ }\textbf
  {\bibinfo {volume} {7}},\ \bibinfo {pages} {229} (\bibinfo {year}
  {2013})}\BibitemShut {NoStop}%
\bibitem [{\citenamefont {Gard}\ \emph {et~al.}(2017)\citenamefont {Gard},
  \citenamefont {You}, \citenamefont {Mishra}, \citenamefont {Singh},
  \citenamefont {Lee}, \citenamefont {Corbitt},\ and\ \citenamefont
  {Dowling}}]{Gar17}%
  \BibitemOpen
  \bibfield  {author} {\bibinfo {author} {\bibfnamefont {B.~T.}\ \bibnamefont
  {Gard}}, \bibinfo {author} {\bibfnamefont {C.}~\bibnamefont {You}}, \bibinfo
  {author} {\bibfnamefont {D.~K.}\ \bibnamefont {Mishra}}, \bibinfo {author}
  {\bibfnamefont {R.}~\bibnamefont {Singh}}, \bibinfo {author} {\bibfnamefont
  {H.}~\bibnamefont {Lee}}, \bibinfo {author} {\bibfnamefont {T.~R.}\
  \bibnamefont {Corbitt}}, \ and\ \bibinfo {author} {\bibfnamefont {J.~P.}\
  \bibnamefont {Dowling}},\ }\href {\doibase 10.1140/epjqt/s40507-017-0058-8}
  {\bibfield  {journal} {\bibinfo  {journal} {EPJ Quantum Technology}\ }\textbf
  {\bibinfo {volume} {4}},\ \bibinfo {pages} {4} (\bibinfo {year}
  {2017})}\BibitemShut {NoStop}%
\bibitem [{\citenamefont {Li}\ \emph {et~al.}(2014)\citenamefont {Li},
  \citenamefont {Yuan}, \citenamefont {Ou},\ and\ \citenamefont
  {Zhang}}]{Li14}%
  \BibitemOpen
  \bibfield  {author} {\bibinfo {author} {\bibfnamefont {D.}~\bibnamefont
  {Li}}, \bibinfo {author} {\bibfnamefont {C.-H.}\ \bibnamefont {Yuan}},
  \bibinfo {author} {\bibfnamefont {Z.~Y.}\ \bibnamefont {Ou}}, \ and\ \bibinfo
  {author} {\bibfnamefont {W.}~\bibnamefont {Zhang}},\ }\href
  {http://stacks.iop.org/1367-2630/16/i=7/a=073020} {\bibfield  {journal}
  {\bibinfo  {journal} {New Journal of Physics}\ }\textbf {\bibinfo {volume}
  {16}},\ \bibinfo {pages} {073020} (\bibinfo {year} {2014})}\BibitemShut
  {NoStop}%
\bibitem [{\citenamefont {Lang}\ and\ \citenamefont {Caves}(2013)}]{Lan13}%
  \BibitemOpen
  \bibfield  {author} {\bibinfo {author} {\bibfnamefont {M.~D.}\ \bibnamefont
  {Lang}}\ and\ \bibinfo {author} {\bibfnamefont {C.~M.}\ \bibnamefont
  {Caves}},\ }\href {\doibase 10.1103/PhysRevLett.111.173601} {\bibfield
  {journal} {\bibinfo  {journal} {Phys. Rev. Lett.}\ }\textbf {\bibinfo
  {volume} {111}},\ \bibinfo {pages} {173601} (\bibinfo {year}
  {2013})}\BibitemShut {NoStop}%
\bibitem [{\citenamefont {Lang}\ and\ \citenamefont {Caves}(2014)}]{Lan14}%
  \BibitemOpen
  \bibfield  {author} {\bibinfo {author} {\bibfnamefont {M.~D.}\ \bibnamefont
  {Lang}}\ and\ \bibinfo {author} {\bibfnamefont {C.~M.}\ \bibnamefont
  {Caves}},\ }\href {\doibase 10.1103/PhysRevA.90.025802} {\bibfield  {journal}
  {\bibinfo  {journal} {Phys. Rev. A}\ }\textbf {\bibinfo {volume} {90}},\
  \bibinfo {pages} {025802} (\bibinfo {year} {2014})}\BibitemShut {NoStop}%
\bibitem [{\citenamefont {Ataman}\ \emph {et~al.}(2018)\citenamefont {Ataman},
  \citenamefont {Preda},\ and\ \citenamefont {Ionicioiu}}]{API18}%
  \BibitemOpen
  \bibfield  {author} {\bibinfo {author} {\bibfnamefont {S.}~\bibnamefont
  {Ataman}}, \bibinfo {author} {\bibfnamefont {A.}~\bibnamefont {Preda}}, \
  and\ \bibinfo {author} {\bibfnamefont {R.}~\bibnamefont {Ionicioiu}},\ }\href
  {\doibase 10.1103/PhysRevA.98.043856} {\bibfield  {journal} {\bibinfo
  {journal} {Phys. Rev. A}\ }\textbf {\bibinfo {volume} {98}},\ \bibinfo
  {pages} {043856} (\bibinfo {year} {2018})}\BibitemShut {NoStop}%
\bibitem [{\citenamefont {Takeoka}\ \emph {et~al.}(2017)\citenamefont
  {Takeoka}, \citenamefont {Seshadreesan}, \citenamefont {You}, \citenamefont
  {Izumi},\ and\ \citenamefont {Dowling}}]{Tak17}%
  \BibitemOpen
  \bibfield  {author} {\bibinfo {author} {\bibfnamefont {M.}~\bibnamefont
  {Takeoka}}, \bibinfo {author} {\bibfnamefont {K.~P.}\ \bibnamefont
  {Seshadreesan}}, \bibinfo {author} {\bibfnamefont {C.}~\bibnamefont {You}},
  \bibinfo {author} {\bibfnamefont {S.}~\bibnamefont {Izumi}}, \ and\ \bibinfo
  {author} {\bibfnamefont {J.~P.}\ \bibnamefont {Dowling}},\ }\href {\doibase
  10.1103/PhysRevA.96.052118} {\bibfield  {journal} {\bibinfo  {journal} {Phys.
  Rev. A}\ }\textbf {\bibinfo {volume} {96}},\ \bibinfo {pages} {052118}
  (\bibinfo {year} {2017})}\BibitemShut {NoStop}%
\bibitem [{\citenamefont {Preda}\ and\ \citenamefont {Ataman}(2019)}]{Pre19}%
  \BibitemOpen
  \bibfield  {author} {\bibinfo {author} {\bibfnamefont {A.}~\bibnamefont
  {Preda}}\ and\ \bibinfo {author} {\bibfnamefont {S.}~\bibnamefont {Ataman}},\
  }\href {\doibase 10.1103/PhysRevA.99.053810} {\bibfield  {journal} {\bibinfo
  {journal} {Phys. Rev. A}\ }\textbf {\bibinfo {volume} {99}},\ \bibinfo
  {pages} {053810} (\bibinfo {year} {2019})}\BibitemShut {NoStop}%
\bibitem [{\citenamefont {Michaud-Belleau}\ \emph {et~al.}(2018)\citenamefont
  {Michaud-Belleau}, \citenamefont {Genest},\ and\ \citenamefont
  {Desch\^enes}}]{Mic18}%
  \BibitemOpen
  \bibfield  {author} {\bibinfo {author} {\bibfnamefont {V.}~\bibnamefont
  {Michaud-Belleau}}, \bibinfo {author} {\bibfnamefont {J.}~\bibnamefont
  {Genest}}, \ and\ \bibinfo {author} {\bibfnamefont {J.-D.}\ \bibnamefont
  {Desch\^enes}},\ }\href {\doibase 10.1103/PhysRevApplied.10.024025}
  {\bibfield  {journal} {\bibinfo  {journal} {Phys. Rev. Applied}\ }\textbf
  {\bibinfo {volume} {10}},\ \bibinfo {pages} {024025} (\bibinfo {year}
  {2018})}\BibitemShut {NoStop}%
\bibitem [{\citenamefont {Demkowicz-Dobrza\'{n}ski}\ \emph
  {et~al.}(2015)\citenamefont {Demkowicz-Dobrza\'{n}ski}, \citenamefont
  {Jarzyna},\ and\ \citenamefont {Ko\l{}ody\'{n}ski}}]{Dem15}%
  \BibitemOpen
  \bibfield  {author} {\bibinfo {author} {\bibfnamefont {R.}~\bibnamefont
  {Demkowicz-Dobrza\'{n}ski}}, \bibinfo {author} {\bibfnamefont
  {M.}~\bibnamefont {Jarzyna}}, \ and\ \bibinfo {author} {\bibfnamefont
  {J.}~\bibnamefont {Ko\l{}ody\'{n}ski}},\ }\href {\doibase
  10.1016/bs.po.2015.02.003} {\bibfield  {journal} {\bibinfo  {journal}
  {Progress in Optics}\ }\textbf {\bibinfo {volume} {60}},\ \bibinfo {pages}
  {345 } (\bibinfo {year} {2015})}\BibitemShut {NoStop}%
\bibitem [{\citenamefont {Acernese}\ \emph {et~al.}(2014)\citenamefont
  {Acernese}, \citenamefont {Agathos}, \citenamefont {Agatsuma}, \citenamefont
  {Aisa}, \citenamefont {Allemandou} \emph {et~al.}}]{Ace14}%
  \BibitemOpen
  \bibfield  {author} {\bibinfo {author} {\bibfnamefont {F.}~\bibnamefont
  {Acernese}}, \bibinfo {author} {\bibfnamefont {M.}~\bibnamefont {Agathos}},
  \bibinfo {author} {\bibfnamefont {K.}~\bibnamefont {Agatsuma}}, \bibinfo
  {author} {\bibfnamefont {D.}~\bibnamefont {Aisa}}, \bibinfo {author}
  {\bibfnamefont {N.}~\bibnamefont {Allemandou}},  \emph {et~al.},\ }\href
  {\doibase 10.1088/0264-9381/32/2/024001} {\bibfield  {journal} {\bibinfo
  {journal} {Classical and Quantum Gravity}\ }\textbf {\bibinfo {volume}
  {32}},\ \bibinfo {pages} {024001} (\bibinfo {year} {2014})}\BibitemShut
  {NoStop}%
\bibitem [{\citenamefont {Oelker}\ \emph {et~al.}(2014)\citenamefont {Oelker},
  \citenamefont {Barsotti}, \citenamefont {Dwyer}, \citenamefont {Sigg},\ and\
  \citenamefont {Mavalvala}}]{Oel14}%
  \BibitemOpen
  \bibfield  {author} {\bibinfo {author} {\bibfnamefont {E.}~\bibnamefont
  {Oelker}}, \bibinfo {author} {\bibfnamefont {L.}~\bibnamefont {Barsotti}},
  \bibinfo {author} {\bibfnamefont {S.}~\bibnamefont {Dwyer}}, \bibinfo
  {author} {\bibfnamefont {D.}~\bibnamefont {Sigg}}, \ and\ \bibinfo {author}
  {\bibfnamefont {N.}~\bibnamefont {Mavalvala}},\ }\href {\doibase
  10.1364/OE.22.021106} {\bibfield  {journal} {\bibinfo  {journal} {Opt.
  Express}\ }\textbf {\bibinfo {volume} {22}},\ \bibinfo {pages} {21106}
  (\bibinfo {year} {2014})}\BibitemShut {NoStop}%
\bibitem [{\citenamefont {Mehmet}\ and\ \citenamefont
  {Vahlbruch}(2018)}]{Meh18}%
  \BibitemOpen
  \bibfield  {author} {\bibinfo {author} {\bibfnamefont {M.}~\bibnamefont
  {Mehmet}}\ and\ \bibinfo {author} {\bibfnamefont {H.}~\bibnamefont
  {Vahlbruch}},\ }\href {\doibase 10.1088/1361-6382/aaf448} {\bibfield
  {journal} {\bibinfo  {journal} {Class. Quantum Grav.}\ }\textbf {\bibinfo
  {volume} {36}},\ \bibinfo {pages} {015014} (\bibinfo {year}
  {2018})}\BibitemShut {NoStop}%
\bibitem [{\citenamefont {Vahlbruch}\ \emph {et~al.}(2018)\citenamefont
  {Vahlbruch}, \citenamefont {Wilken}, \citenamefont {Mehmet},\ and\
  \citenamefont {Willke}}]{Vah18}%
  \BibitemOpen
  \bibfield  {author} {\bibinfo {author} {\bibfnamefont {H.}~\bibnamefont
  {Vahlbruch}}, \bibinfo {author} {\bibfnamefont {D.}~\bibnamefont {Wilken}},
  \bibinfo {author} {\bibfnamefont {M.}~\bibnamefont {Mehmet}}, \ and\ \bibinfo
  {author} {\bibfnamefont {B.}~\bibnamefont {Willke}},\ }\href {\doibase
  10.1103/PhysRevLett.121.173601} {\bibfield  {journal} {\bibinfo  {journal}
  {Phys. Rev. Lett.}\ }\textbf {\bibinfo {volume} {121}},\ \bibinfo {pages}
  {173601} (\bibinfo {year} {2018})}\BibitemShut {NoStop}%
\bibitem [{\citenamefont {Giovannetti}\ and\ \citenamefont
  {Maccone}(2012)}]{Gio12}%
  \BibitemOpen
  \bibfield  {author} {\bibinfo {author} {\bibfnamefont {V.}~\bibnamefont
  {Giovannetti}}\ and\ \bibinfo {author} {\bibfnamefont {L.}~\bibnamefont
  {Maccone}},\ }\href {\doibase 10.1103/PhysRevLett.108.210404} {\bibfield
  {journal} {\bibinfo  {journal} {Phys. Rev. Lett.}\ }\textbf {\bibinfo
  {volume} {108}},\ \bibinfo {pages} {210404} (\bibinfo {year}
  {2012})}\BibitemShut {NoStop}%
\bibitem [{\citenamefont {Caves}(1981)}]{Cav81}%
  \BibitemOpen
  \bibfield  {author} {\bibinfo {author} {\bibfnamefont {C.~M.}\ \bibnamefont
  {Caves}},\ }\href {\doibase 10.1103/PhysRevD.23.1693} {\bibfield  {journal}
  {\bibinfo  {journal} {Phys. Rev. D}\ }\textbf {\bibinfo {volume} {23}},\
  \bibinfo {pages} {1693} (\bibinfo {year} {1981})}\BibitemShut {NoStop}%
\bibitem [{\citenamefont {Xiao}\ \emph {et~al.}(1987)\citenamefont {Xiao},
  \citenamefont {Wu},\ and\ \citenamefont {Kimble}}]{Xia87}%
  \BibitemOpen
  \bibfield  {author} {\bibinfo {author} {\bibfnamefont {M.}~\bibnamefont
  {Xiao}}, \bibinfo {author} {\bibfnamefont {L.-A.}\ \bibnamefont {Wu}}, \ and\
  \bibinfo {author} {\bibfnamefont {H.~J.}\ \bibnamefont {Kimble}},\ }\href
  {\doibase 10.1103/PhysRevLett.59.278} {\bibfield  {journal} {\bibinfo
  {journal} {Phys. Rev. Lett.}\ }\textbf {\bibinfo {volume} {59}},\ \bibinfo
  {pages} {278} (\bibinfo {year} {1987})}\BibitemShut {NoStop}%
\bibitem [{\citenamefont {Giovannetti}\ \emph {et~al.}(2004)\citenamefont
  {Giovannetti}, \citenamefont {Lloyd},\ and\ \citenamefont {Maccone}}]{Gio04}%
  \BibitemOpen
  \bibfield  {author} {\bibinfo {author} {\bibfnamefont {V.}~\bibnamefont
  {Giovannetti}}, \bibinfo {author} {\bibfnamefont {S.}~\bibnamefont {Lloyd}},
  \ and\ \bibinfo {author} {\bibfnamefont {L.}~\bibnamefont {Maccone}},\ }\href
  {\doibase 10.1126/science.1104149} {\bibfield  {journal} {\bibinfo  {journal}
  {Science}\ }\textbf {\bibinfo {volume} {306}},\ \bibinfo {pages} {1330}
  (\bibinfo {year} {2004})}\BibitemShut {NoStop}%
\bibitem [{\citenamefont {Pezz\'e}\ and\ \citenamefont {Smerzi}(2008)}]{Pez08}%
  \BibitemOpen
  \bibfield  {author} {\bibinfo {author} {\bibfnamefont {L.}~\bibnamefont
  {Pezz\'e}}\ and\ \bibinfo {author} {\bibfnamefont {A.}~\bibnamefont
  {Smerzi}},\ }\href {\doibase 10.1103/PhysRevLett.100.073601} {\bibfield
  {journal} {\bibinfo  {journal} {Phys. Rev. Lett.}\ }\textbf {\bibinfo
  {volume} {100}},\ \bibinfo {pages} {073601} (\bibinfo {year}
  {2008})}\BibitemShut {NoStop}%
\bibitem [{\citenamefont {Jarzyna}\ and\ \citenamefont
  {Demkowicz-Dobrza\ifmmode~\acute{n}\else \'{n}\fi{}ski}(2012)}]{Jar12}%
  \BibitemOpen
  \bibfield  {author} {\bibinfo {author} {\bibfnamefont {M.}~\bibnamefont
  {Jarzyna}}\ and\ \bibinfo {author} {\bibfnamefont {R.}~\bibnamefont
  {Demkowicz-Dobrza\ifmmode~\acute{n}\else \'{n}\fi{}ski}},\ }\href {\doibase
  10.1103/PhysRevA.85.011801} {\bibfield  {journal} {\bibinfo  {journal} {Phys.
  Rev. A}\ }\textbf {\bibinfo {volume} {85}},\ \bibinfo {pages} {011801}
  (\bibinfo {year} {2012})}\BibitemShut {NoStop}%
\bibitem [{\citenamefont {Wu}\ \emph {et~al.}(2019)\citenamefont {Wu},
  \citenamefont {Toda},\ and\ \citenamefont {Hofmann}}]{Wu19}%
  \BibitemOpen
  \bibfield  {author} {\bibinfo {author} {\bibfnamefont {J.-Y.}\ \bibnamefont
  {Wu}}, \bibinfo {author} {\bibfnamefont {N.}~\bibnamefont {Toda}}, \ and\
  \bibinfo {author} {\bibfnamefont {H.~F.}\ \bibnamefont {Hofmann}},\ }\href
  {\doibase 10.1103/PhysRevA.100.013814} {\bibfield  {journal} {\bibinfo
  {journal} {Phys. Rev. A}\ }\textbf {\bibinfo {volume} {100}},\ \bibinfo
  {pages} {013814} (\bibinfo {year} {2019})}\BibitemShut {NoStop}%
\bibitem [{\citenamefont {Burd}\ \emph {et~al.}(2019)\citenamefont {Burd},
  \citenamefont {Srinivas}, \citenamefont {Bollinger}, \citenamefont {Wilson},
  \citenamefont {Wineland}, \citenamefont {Leibfried}, \citenamefont
  {Slichter},\ and\ \citenamefont {Allcock}}]{Bur19}%
  \BibitemOpen
  \bibfield  {author} {\bibinfo {author} {\bibfnamefont {S.~C.}\ \bibnamefont
  {Burd}}, \bibinfo {author} {\bibfnamefont {R.}~\bibnamefont {Srinivas}},
  \bibinfo {author} {\bibfnamefont {J.~J.}\ \bibnamefont {Bollinger}}, \bibinfo
  {author} {\bibfnamefont {A.~C.}\ \bibnamefont {Wilson}}, \bibinfo {author}
  {\bibfnamefont {D.~J.}\ \bibnamefont {Wineland}}, \bibinfo {author}
  {\bibfnamefont {D.}~\bibnamefont {Leibfried}}, \bibinfo {author}
  {\bibfnamefont {D.~H.}\ \bibnamefont {Slichter}}, \ and\ \bibinfo {author}
  {\bibfnamefont {D.~T.~C.}\ \bibnamefont {Allcock}},\ }\href {\doibase
  10.1126/science.aaw2884} {\bibfield  {journal} {\bibinfo  {journal}
  {Science}\ }\textbf {\bibinfo {volume} {364}},\ \bibinfo {pages} {1163}
  (\bibinfo {year} {2019})}\BibitemShut {NoStop}%
\bibitem [{\citenamefont {Malnou}\ \emph {et~al.}(2019)\citenamefont {Malnou},
  \citenamefont {Palken}, \citenamefont {Brubaker}, \citenamefont {Vale},
  \citenamefont {Hilton},\ and\ \citenamefont {Lehnert}}]{Mal19}%
  \BibitemOpen
  \bibfield  {author} {\bibinfo {author} {\bibfnamefont {M.}~\bibnamefont
  {Malnou}}, \bibinfo {author} {\bibfnamefont {D.~A.}\ \bibnamefont {Palken}},
  \bibinfo {author} {\bibfnamefont {B.~M.}\ \bibnamefont {Brubaker}}, \bibinfo
  {author} {\bibfnamefont {L.~R.}\ \bibnamefont {Vale}}, \bibinfo {author}
  {\bibfnamefont {G.~C.}\ \bibnamefont {Hilton}}, \ and\ \bibinfo {author}
  {\bibfnamefont {K.~W.}\ \bibnamefont {Lehnert}},\ }\href {\doibase
  10.1103/PhysRevX.9.021023} {\bibfield  {journal} {\bibinfo  {journal} {Phys.
  Rev. X}\ }\textbf {\bibinfo {volume} {9}},\ \bibinfo {pages} {021023}
  (\bibinfo {year} {2019})}\BibitemShut {NoStop}%
\bibitem [{\citenamefont {Xu}\ \emph {et~al.}(2019)\citenamefont {Xu},
  \citenamefont {Zhang}, \citenamefont {Huang}, \citenamefont {Ma},
  \citenamefont {Liu}, \citenamefont {Yonezawa}, \citenamefont {Zhang},\ and\
  \citenamefont {Xiao}}]{Xu19}%
  \BibitemOpen
  \bibfield  {author} {\bibinfo {author} {\bibfnamefont {C.}~\bibnamefont
  {Xu}}, \bibinfo {author} {\bibfnamefont {L.}~\bibnamefont {Zhang}}, \bibinfo
  {author} {\bibfnamefont {S.}~\bibnamefont {Huang}}, \bibinfo {author}
  {\bibfnamefont {T.}~\bibnamefont {Ma}}, \bibinfo {author} {\bibfnamefont
  {F.}~\bibnamefont {Liu}}, \bibinfo {author} {\bibfnamefont {H.}~\bibnamefont
  {Yonezawa}}, \bibinfo {author} {\bibfnamefont {Y.}~\bibnamefont {Zhang}}, \
  and\ \bibinfo {author} {\bibfnamefont {M.}~\bibnamefont {Xiao}},\ }\href
  {\doibase 10.1364/PRJ.7.000A14} {\bibfield  {journal} {\bibinfo  {journal}
  {Photon. Res.}\ }\textbf {\bibinfo {volume} {7}},\ \bibinfo {pages} {A14}
  (\bibinfo {year} {2019})}\BibitemShut {NoStop}%
\bibitem [{\citenamefont {Grote}\ \emph {et~al.}(2013)\citenamefont {Grote},
  \citenamefont {Danzmann}, \citenamefont {Dooley}, \citenamefont {Schnabel},
  \citenamefont {Slutsky},\ and\ \citenamefont {Vahlbruch}}]{Gro13}%
  \BibitemOpen
  \bibfield  {author} {\bibinfo {author} {\bibfnamefont {H.}~\bibnamefont
  {Grote}}, \bibinfo {author} {\bibfnamefont {K.}~\bibnamefont {Danzmann}},
  \bibinfo {author} {\bibfnamefont {K.~L.}\ \bibnamefont {Dooley}}, \bibinfo
  {author} {\bibfnamefont {R.}~\bibnamefont {Schnabel}}, \bibinfo {author}
  {\bibfnamefont {J.}~\bibnamefont {Slutsky}}, \ and\ \bibinfo {author}
  {\bibfnamefont {H.}~\bibnamefont {Vahlbruch}},\ }\href {\doibase
  10.1103/PhysRevLett.110.181101} {\bibfield  {journal} {\bibinfo  {journal}
  {Phys. Rev. Lett.}\ }\textbf {\bibinfo {volume} {110}},\ \bibinfo {pages}
  {181101} (\bibinfo {year} {2013})}\BibitemShut {NoStop}%
\bibitem [{\citenamefont {Vahlbruch}\ \emph {et~al.}(2016)\citenamefont
  {Vahlbruch}, \citenamefont {Mehmet}, \citenamefont {Danzmann},\ and\
  \citenamefont {Schnabel}}]{Vah16}%
  \BibitemOpen
  \bibfield  {author} {\bibinfo {author} {\bibfnamefont {H.}~\bibnamefont
  {Vahlbruch}}, \bibinfo {author} {\bibfnamefont {M.}~\bibnamefont {Mehmet}},
  \bibinfo {author} {\bibfnamefont {K.}~\bibnamefont {Danzmann}}, \ and\
  \bibinfo {author} {\bibfnamefont {R.}~\bibnamefont {Schnabel}},\ }\href
  {\doibase 10.1103/PhysRevLett.117.110801} {\bibfield  {journal} {\bibinfo
  {journal} {Phys. Rev. Lett.}\ }\textbf {\bibinfo {volume} {117}},\ \bibinfo
  {pages} {110801} (\bibinfo {year} {2016})}\BibitemShut {NoStop}%
\bibitem [{\citenamefont {Schnabel}(2017)}]{Sch17}%
  \BibitemOpen
  \bibfield  {author} {\bibinfo {author} {\bibfnamefont {R.}~\bibnamefont
  {Schnabel}},\ }\href {\doibase 10.1016/j.physrep.2017.04.001} {\bibfield
  {journal} {\bibinfo  {journal} {Physics Reports}\ }\textbf {\bibinfo {volume}
  {684}},\ \bibinfo {pages} {1 } (\bibinfo {year} {2017})}\BibitemShut
  {NoStop}%
\bibitem [{\citenamefont {Yurke}\ \emph {et~al.}(1986)\citenamefont {Yurke},
  \citenamefont {McCall},\ and\ \citenamefont {Klauder}}]{Yur86}%
  \BibitemOpen
  \bibfield  {author} {\bibinfo {author} {\bibfnamefont {B.}~\bibnamefont
  {Yurke}}, \bibinfo {author} {\bibfnamefont {S.~L.}\ \bibnamefont {McCall}}, \
  and\ \bibinfo {author} {\bibfnamefont {J.~R.}\ \bibnamefont {Klauder}},\
  }\href {\doibase 10.1103/PhysRevA.33.4033} {\bibfield  {journal} {\bibinfo
  {journal} {Phys. Rev. A}\ }\textbf {\bibinfo {volume} {33}},\ \bibinfo
  {pages} {4033} (\bibinfo {year} {1986})}\BibitemShut {NoStop}%
\bibitem [{\citenamefont {Pezz\'e}\ \emph {et~al.}(2007)\citenamefont
  {Pezz\'e}, \citenamefont {Smerzi}, \citenamefont {Khoury}, \citenamefont
  {Hodelin},\ and\ \citenamefont {Bouwmeester}}]{Pez07}%
  \BibitemOpen
  \bibfield  {author} {\bibinfo {author} {\bibfnamefont {L.}~\bibnamefont
  {Pezz\'e}}, \bibinfo {author} {\bibfnamefont {A.}~\bibnamefont {Smerzi}},
  \bibinfo {author} {\bibfnamefont {G.}~\bibnamefont {Khoury}}, \bibinfo
  {author} {\bibfnamefont {J.~F.}\ \bibnamefont {Hodelin}}, \ and\ \bibinfo
  {author} {\bibfnamefont {D.}~\bibnamefont {Bouwmeester}},\ }\href {\doibase
  10.1103/PhysRevLett.99.223602} {\bibfield  {journal} {\bibinfo  {journal}
  {Phys. Rev. Lett.}\ }\textbf {\bibinfo {volume} {99}},\ \bibinfo {pages}
  {223602} (\bibinfo {year} {2007})}\BibitemShut {NoStop}%
\bibitem [{\citenamefont {Paris}(1995)}]{Par95}%
  \BibitemOpen
  \bibfield  {author} {\bibinfo {author} {\bibfnamefont {M.~G.}\ \bibnamefont
  {Paris}},\ }\href {\doibase 10.1016/0375-9601(95)00235-U} {\bibfield
  {journal} {\bibinfo  {journal} {Physics Letters A}\ }\textbf {\bibinfo
  {volume} {201}},\ \bibinfo {pages} {132 } (\bibinfo {year}
  {1995})}\BibitemShut {NoStop}%
\bibitem [{\citenamefont {Braunstein}\ and\ \citenamefont
  {Caves}(1994)}]{Bra94}%
  \BibitemOpen
  \bibfield  {author} {\bibinfo {author} {\bibfnamefont {S.~L.}\ \bibnamefont
  {Braunstein}}\ and\ \bibinfo {author} {\bibfnamefont {C.~M.}\ \bibnamefont
  {Caves}},\ }\href {\doibase 10.1103/PhysRevLett.72.3439} {\bibfield
  {journal} {\bibinfo  {journal} {Phys. Rev. Lett.}\ }\textbf {\bibinfo
  {volume} {72}},\ \bibinfo {pages} {3439} (\bibinfo {year}
  {1994})}\BibitemShut {NoStop}%
\bibitem [{\citenamefont {Demkowicz-Dobrza\'{n}ski}\ \emph
  {et~al.}(2012)\citenamefont {Demkowicz-Dobrza\'{n}ski}, \citenamefont
  {Ko\l{}ody\'{n}ski},\ and\ \citenamefont {Gu\c{t}\u{a}}}]{Dem12}%
  \BibitemOpen
  \bibfield  {author} {\bibinfo {author} {\bibfnamefont {R.}~\bibnamefont
  {Demkowicz-Dobrza\'{n}ski}}, \bibinfo {author} {\bibfnamefont
  {J.}~\bibnamefont {Ko\l{}ody\'{n}ski}}, \ and\ \bibinfo {author}
  {\bibfnamefont {M.}~\bibnamefont {Gu\c{t}\u{a}}},\ }\href {\doibase
  10.1038/ncomms2067} {\bibfield  {journal} {\bibinfo  {journal} {Nature
  Communications}\ }\textbf {\bibinfo {volume} {3}},\ \bibinfo {pages} {1063}
  (\bibinfo {year} {2012})}\BibitemShut {NoStop}%
\bibitem [{\citenamefont {Pezz\`e}\ \emph {et~al.}(2015)\citenamefont
  {Pezz\`e}, \citenamefont {Hyllus},\ and\ \citenamefont {Smerzi}}]{Pez15}%
  \BibitemOpen
  \bibfield  {author} {\bibinfo {author} {\bibfnamefont {L.}~\bibnamefont
  {Pezz\`e}}, \bibinfo {author} {\bibfnamefont {P.}~\bibnamefont {Hyllus}}, \
  and\ \bibinfo {author} {\bibfnamefont {A.}~\bibnamefont {Smerzi}},\ }\href
  {\doibase 10.1103/PhysRevA.91.032103} {\bibfield  {journal} {\bibinfo
  {journal} {Phys. Rev. A}\ }\textbf {\bibinfo {volume} {91}},\ \bibinfo
  {pages} {032103} (\bibinfo {year} {2015})}\BibitemShut {NoStop}%
\bibitem [{\citenamefont {Sparaciari}\ \emph {et~al.}(2015)\citenamefont
  {Sparaciari}, \citenamefont {Olivares},\ and\ \citenamefont {Paris}}]{Spa15}%
  \BibitemOpen
  \bibfield  {author} {\bibinfo {author} {\bibfnamefont {C.}~\bibnamefont
  {Sparaciari}}, \bibinfo {author} {\bibfnamefont {S.}~\bibnamefont
  {Olivares}}, \ and\ \bibinfo {author} {\bibfnamefont {M.~G.~A.}\ \bibnamefont
  {Paris}},\ }\href {\doibase 10.1364/JOSAB.32.001354} {\bibfield  {journal}
  {\bibinfo  {journal} {J. Opt. Soc. Am. B}\ }\textbf {\bibinfo {volume}
  {32}},\ \bibinfo {pages} {1354} (\bibinfo {year} {2015})}\BibitemShut
  {NoStop}%
\bibitem [{\citenamefont {Sparaciari}\ \emph {et~al.}(2016)\citenamefont
  {Sparaciari}, \citenamefont {Olivares},\ and\ \citenamefont {Paris}}]{Spa16}%
  \BibitemOpen
  \bibfield  {author} {\bibinfo {author} {\bibfnamefont {C.}~\bibnamefont
  {Sparaciari}}, \bibinfo {author} {\bibfnamefont {S.}~\bibnamefont
  {Olivares}}, \ and\ \bibinfo {author} {\bibfnamefont {M.~G.~A.}\ \bibnamefont
  {Paris}},\ }\href {\doibase 10.1103/PhysRevA.93.023810} {\bibfield  {journal}
  {\bibinfo  {journal} {Phys. Rev. A}\ }\textbf {\bibinfo {volume} {93}},\
  \bibinfo {pages} {023810} (\bibinfo {year} {2016})}\BibitemShut {NoStop}%
\bibitem [{\citenamefont {Liu}\ \emph {et~al.}(2013)\citenamefont {Liu},
  \citenamefont {Jing},\ and\ \citenamefont {Wang}}]{Liu13}%
  \BibitemOpen
  \bibfield  {author} {\bibinfo {author} {\bibfnamefont {J.}~\bibnamefont
  {Liu}}, \bibinfo {author} {\bibfnamefont {X.}~\bibnamefont {Jing}}, \ and\
  \bibinfo {author} {\bibfnamefont {X.}~\bibnamefont {Wang}},\ }\href {\doibase
  10.1103/PhysRevA.88.042316} {\bibfield  {journal} {\bibinfo  {journal} {Phys.
  Rev. A}\ }\textbf {\bibinfo {volume} {88}},\ \bibinfo {pages} {042316}
  (\bibinfo {year} {2013})}\BibitemShut {NoStop}%
\bibitem [{\citenamefont {Dorner}\ \emph {et~al.}(2009)\citenamefont {Dorner},
  \citenamefont {Demkowicz-Dobrzanski}, \citenamefont {Smith}, \citenamefont
  {Lundeen}, \citenamefont {Wasilewski}, \citenamefont {Banaszek},\ and\
  \citenamefont {Walmsley}}]{Dor09}%
  \BibitemOpen
  \bibfield  {author} {\bibinfo {author} {\bibfnamefont {U.}~\bibnamefont
  {Dorner}}, \bibinfo {author} {\bibfnamefont {R.}~\bibnamefont
  {Demkowicz-Dobrzanski}}, \bibinfo {author} {\bibfnamefont {B.~J.}\
  \bibnamefont {Smith}}, \bibinfo {author} {\bibfnamefont {J.~S.}\ \bibnamefont
  {Lundeen}}, \bibinfo {author} {\bibfnamefont {W.}~\bibnamefont {Wasilewski}},
  \bibinfo {author} {\bibfnamefont {K.}~\bibnamefont {Banaszek}}, \ and\
  \bibinfo {author} {\bibfnamefont {I.~A.}\ \bibnamefont {Walmsley}},\ }\href
  {\doibase 10.1103/PhysRevLett.102.040403} {\bibfield  {journal} {\bibinfo
  {journal} {Phys. Rev. Lett.}\ }\textbf {\bibinfo {volume} {102}},\ \bibinfo
  {pages} {040403} (\bibinfo {year} {2009})}\BibitemShut {NoStop}%
\bibitem [{\citenamefont {Demkowicz-Dobrzanski}\ \emph
  {et~al.}(2009)\citenamefont {Demkowicz-Dobrzanski}, \citenamefont {Dorner},
  \citenamefont {Smith}, \citenamefont {Lundeen}, \citenamefont {Wasilewski},
  \citenamefont {Banaszek},\ and\ \citenamefont {Walmsley}}]{Dem09}%
  \BibitemOpen
  \bibfield  {author} {\bibinfo {author} {\bibfnamefont {R.}~\bibnamefont
  {Demkowicz-Dobrzanski}}, \bibinfo {author} {\bibfnamefont {U.}~\bibnamefont
  {Dorner}}, \bibinfo {author} {\bibfnamefont {B.~J.}\ \bibnamefont {Smith}},
  \bibinfo {author} {\bibfnamefont {J.~S.}\ \bibnamefont {Lundeen}}, \bibinfo
  {author} {\bibfnamefont {W.}~\bibnamefont {Wasilewski}}, \bibinfo {author}
  {\bibfnamefont {K.}~\bibnamefont {Banaszek}}, \ and\ \bibinfo {author}
  {\bibfnamefont {I.~A.}\ \bibnamefont {Walmsley}},\ }\href {\doibase
  10.1103/PhysRevA.80.013825} {\bibfield  {journal} {\bibinfo  {journal} {Phys.
  Rev. A}\ }\textbf {\bibinfo {volume} {80}},\ \bibinfo {pages} {013825}
  (\bibinfo {year} {2009})}\BibitemShut {NoStop}%
\bibitem [{\citenamefont {Ono}\ and\ \citenamefont {Hofmann}(2010)}]{Ono10}%
  \BibitemOpen
  \bibfield  {author} {\bibinfo {author} {\bibfnamefont {T.}~\bibnamefont
  {Ono}}\ and\ \bibinfo {author} {\bibfnamefont {H.~F.}\ \bibnamefont
  {Hofmann}},\ }\href {\doibase 10.1103/PhysRevA.81.033819} {\bibfield
  {journal} {\bibinfo  {journal} {Phys. Rev. A}\ }\textbf {\bibinfo {volume}
  {81}},\ \bibinfo {pages} {033819} (\bibinfo {year} {2010})}\BibitemShut
  {NoStop}%
\bibitem [{\citenamefont {Kim}\ \emph {et~al.}(1999)\citenamefont {Kim},
  \citenamefont {Ha}, \citenamefont {Shin}, \citenamefont {Kim}, \citenamefont
  {Park}, \citenamefont {Kim}, \citenamefont {Noh},\ and\ \citenamefont
  {Hong}}]{Kim99}%
  \BibitemOpen
  \bibfield  {author} {\bibinfo {author} {\bibfnamefont {T.}~\bibnamefont
  {Kim}}, \bibinfo {author} {\bibfnamefont {Y.}~\bibnamefont {Ha}}, \bibinfo
  {author} {\bibfnamefont {J.}~\bibnamefont {Shin}}, \bibinfo {author}
  {\bibfnamefont {H.}~\bibnamefont {Kim}}, \bibinfo {author} {\bibfnamefont
  {G.}~\bibnamefont {Park}}, \bibinfo {author} {\bibfnamefont {K.}~\bibnamefont
  {Kim}}, \bibinfo {author} {\bibfnamefont {T.-G.}\ \bibnamefont {Noh}}, \ and\
  \bibinfo {author} {\bibfnamefont {C.~K.}\ \bibnamefont {Hong}},\ }\href
  {\doibase 10.1103/PhysRevA.60.708} {\bibfield  {journal} {\bibinfo  {journal}
  {Phys. Rev. A}\ }\textbf {\bibinfo {volume} {60}},\ \bibinfo {pages} {708}
  (\bibinfo {year} {1999})}\BibitemShut {NoStop}%
\bibitem [{\citenamefont {Paris}(2009)}]{Par09}%
  \BibitemOpen
  \bibfield  {author} {\bibinfo {author} {\bibfnamefont {M.~G.~A.}\
  \bibnamefont {Paris}},\ }\href {\doibase 10.1142/S0219749909004839}
  {\bibfield  {journal} {\bibinfo  {journal} {Int. J. of Quantum Info.}\
  }\textbf {\bibinfo {volume} {07}},\ \bibinfo {pages} {125} (\bibinfo {year}
  {2009})}\BibitemShut {NoStop}%
\bibitem [{\citenamefont {Gerry}\ and\ \citenamefont
  {Knight}(2005)}]{GerryKnight}%
  \BibitemOpen
  \bibfield  {author} {\bibinfo {author} {\bibfnamefont {C.}~\bibnamefont
  {Gerry}}\ and\ \bibinfo {author} {\bibfnamefont {P.}~\bibnamefont {Knight}},\
  }\href {\doibase 10.1017/CBO9780511791239} {\emph {\bibinfo {title}
  {Introductory Quantum Optics}}}\ (\bibinfo  {publisher} {Cambridge University
  Press},\ \bibinfo {year} {2005})\BibitemShut {NoStop}%
\bibitem [{\citenamefont {Rodney}(2000)}]{Loudon}%
  \BibitemOpen
  \bibfield  {author} {\bibinfo {author} {\bibfnamefont {L.}~\bibnamefont
  {Rodney}},\ }\href@noop {} {\emph {\bibinfo {title} {The Quantum Theory of
  Light (Third Edition)}}}\ (\bibinfo  {publisher} {Oxford University Press},\
  \bibinfo {year} {2000})\BibitemShut {NoStop}%
\bibitem [{\citenamefont {Yuen}(1976)}]{Yue76}%
  \BibitemOpen
  \bibfield  {author} {\bibinfo {author} {\bibfnamefont {H.~P.}\ \bibnamefont
  {Yuen}},\ }\href {\doibase 10.1103/PhysRevA.13.2226} {\bibfield  {journal}
  {\bibinfo  {journal} {Phys. Rev. A}\ }\textbf {\bibinfo {volume} {13}},\
  \bibinfo {pages} {2226} (\bibinfo {year} {1976})}\BibitemShut {NoStop}%
\bibitem [{\citenamefont {Mandel}\ and\ \citenamefont
  {Wolf}(1995)}]{MandelWolf}%
  \BibitemOpen
  \bibfield  {author} {\bibinfo {author} {\bibfnamefont {L.}~\bibnamefont
  {Mandel}}\ and\ \bibinfo {author} {\bibfnamefont {E.}~\bibnamefont {Wolf}},\
  }\href {\doibase 10.1017/CBO9781139644105} {\emph {\bibinfo {title} {Optical
  Coherence and Quantum Optics}}}\ (\bibinfo  {publisher} {Cambridge University
  Press},\ \bibinfo {year} {1995})\BibitemShut {NoStop}%
\bibitem [{\citenamefont {Agarwal}(2012)}]{Aga12}%
  \BibitemOpen
  \bibfield  {author} {\bibinfo {author} {\bibfnamefont {G.~S.}\ \bibnamefont
  {Agarwal}},\ }\href {\doibase 10.1017/CBO9781139035170} {\emph {\bibinfo
  {title} {Quantum Optics}}}\ (\bibinfo  {publisher} {Cambridge University
  Press},\ \bibinfo {year} {2012})\BibitemShut {NoStop}%
\bibitem [{\citenamefont {Kim}\ and\ \citenamefont {Sanders}(1996)}]{Kim96}%
  \BibitemOpen
  \bibfield  {author} {\bibinfo {author} {\bibfnamefont {M.~S.}\ \bibnamefont
  {Kim}}\ and\ \bibinfo {author} {\bibfnamefont {B.~C.}\ \bibnamefont
  {Sanders}},\ }\href {\doibase 10.1103/PhysRevA.53.3694} {\bibfield  {journal}
  {\bibinfo  {journal} {Phys. Rev. A}\ }\textbf {\bibinfo {volume} {53}},\
  \bibinfo {pages} {3694} (\bibinfo {year} {1996})}\BibitemShut {NoStop}%
\bibitem [{\citenamefont {Ou}(1996)}]{Ou96}%
  \BibitemOpen
  \bibfield  {author} {\bibinfo {author} {\bibfnamefont {Z.~Y.}\ \bibnamefont
  {Ou}},\ }\href {\doibase 10.1103/PhysRevLett.77.2352} {\bibfield  {journal}
  {\bibinfo  {journal} {Phys. Rev. Lett.}\ }\textbf {\bibinfo {volume} {77}},\
  \bibinfo {pages} {2352} (\bibinfo {year} {1996})}\BibitemShut {NoStop}%
\bibitem [{\citenamefont {Hofmann}(2009)}]{Hof09}%
  \BibitemOpen
  \bibfield  {author} {\bibinfo {author} {\bibfnamefont {H.~F.}\ \bibnamefont
  {Hofmann}},\ }\href {\doibase 10.1103/PhysRevA.79.033822} {\bibfield
  {journal} {\bibinfo  {journal} {Phys. Rev. A}\ }\textbf {\bibinfo {volume}
  {79}},\ \bibinfo {pages} {033822} (\bibinfo {year} {2009})}\BibitemShut
  {NoStop}%
\bibitem [{\citenamefont {Pinel}\ \emph {et~al.}(2012)\citenamefont {Pinel},
  \citenamefont {Fade}, \citenamefont {Braun}, \citenamefont {Jian},
  \citenamefont {Treps},\ and\ \citenamefont {Fabre}}]{Pin12}%
  \BibitemOpen
  \bibfield  {author} {\bibinfo {author} {\bibfnamefont {O.}~\bibnamefont
  {Pinel}}, \bibinfo {author} {\bibfnamefont {J.}~\bibnamefont {Fade}},
  \bibinfo {author} {\bibfnamefont {D.}~\bibnamefont {Braun}}, \bibinfo
  {author} {\bibfnamefont {P.}~\bibnamefont {Jian}}, \bibinfo {author}
  {\bibfnamefont {N.}~\bibnamefont {Treps}}, \ and\ \bibinfo {author}
  {\bibfnamefont {C.}~\bibnamefont {Fabre}},\ }\href {\doibase
  10.1103/PhysRevA.85.010101} {\bibfield  {journal} {\bibinfo  {journal} {Phys.
  Rev. A}\ }\textbf {\bibinfo {volume} {85}},\ \bibinfo {pages} {010101}
  (\bibinfo {year} {2012})}\BibitemShut {NoStop}%
\bibitem [{\citenamefont {Schwinger}(1965)}]{Sch65}%
  \BibitemOpen
  \bibfield  {author} {\bibinfo {author} {\bibfnamefont {J.}~\bibnamefont
  {Schwinger}},\ }in\ \href {\doibase 10.2172/4389568} {\emph {\bibinfo
  {booktitle} {Quantum Theory of Angular Momentum}}},\ \bibinfo {editor}
  {edited by\ \bibinfo {editor} {\bibfnamefont {L.}~\bibnamefont {Biedenharn}}\
  and\ \bibinfo {editor} {\bibfnamefont {H.~V.}\ \bibnamefont {Dam}}}\
  (\bibinfo  {publisher} {Academic, New York},\ \bibinfo {year} {1965})\ Chap.\
  \bibinfo {chapter} {On Angular Momentum}\BibitemShut {NoStop}%
\bibitem [{\citenamefont {Truax}(1985)}]{Tru85}%
  \BibitemOpen
  \bibfield  {author} {\bibinfo {author} {\bibfnamefont {D.~R.}\ \bibnamefont
  {Truax}},\ }\href {\doibase 10.1103/PhysRevD.31.1988} {\bibfield  {journal}
  {\bibinfo  {journal} {Phys. Rev. D}\ }\textbf {\bibinfo {volume} {31}},\
  \bibinfo {pages} {1988} (\bibinfo {year} {1985})}\BibitemShut {NoStop}%
\end{thebibliography}%

\end{document}